\documentclass[a4paper,11pt]{article}
\usepackage{jcappub}

\pdfoutput=1

\usepackage{tabularx,booktabs}
\newcolumntype{Y}{>{\centering\arraybackslash}X}

\usepackage{fontawesome}
\usepackage[utf8]{inputenc}
\usepackage[font=small,labelfont=bf,tableposition=top]{caption}
\usepackage{float}
\usepackage{adjustbox}
\usepackage{arydshln}
\usepackage{empheq}

\usepackage[percent]{overpic}
\usepackage{amsmath,amssymb,amsbsy,amstext, amsthm, simplewick, amsfonts, bm}
\usepackage{hyperref}
\usepackage[normalem]{ulem}
\usepackage{graphicx}
\usepackage{caption}
\usepackage{subcaption}
\usepackage{siunitx}
\usepackage{upgreek}
\usepackage{framed}
\usepackage{wrapfig}
\usepackage{multirow}
\usepackage{bbm}
\usepackage[nameinlink, capitalise]{cleveref}
\usepackage[svgnames,dvipsnames,x11names]{xcolor}
\usepackage{wrapfig}
\usepackage{soul}
\usepackage[htt]{hyphenat}
\allowdisplaybreaks
\usepackage{empheq}
\usepackage[T1]{fontenc}
\include{fontawsome5}

\usepackage[top=2.5cm,bottom=3.cm,right=2.5cm,left=2.5cm]{geometry}
\def\la{\langle}
\def\ra{\rangle}

\def\ln{{\rm ln}}

\renewcommand*{\vec}{\textbf}

\def\be{\begin{equation}}
\def\ee{\end{equation}}
\def\bea{\begin{eqnarray}}
\def\eea{\end{eqnarray}}
\def\ba{\begin{align}}

\def\bi{\begin{itemize}}
\def\ei{\end{itemize}}

\newcommand{\nn}{ \nonumber }

\newcommand{\gnl}{g_{\rm NL}}

\def\bx{{\bf x}}

\def\bq{{\bf q}}

\def\bB{{\bf B}}

\def\cM{{\mathcal M}}

\renewcommand*{\vec}{\textbf}

\newcommand*{\non}  {\nonumber}
\newcommand*{\lb}  {\left(}
\newcommand*{\rb}  {\right)}

\newcommand{\eq}[1]{\begin{align}#1\end{align}}
\newcommand{\eeq}[1]{\begin{equation}#1\eend{equation}}

\newlength{\labelgutter}
\newcommand{\rowlabel}[2]{%
  \put(-16,#1){\makebox[0pt][l]{\footnotesize #2\hspace{0.25em}}}%
}

\def\v#1{\vec{#1}}
\def\vk{\v{k}}
\def\vq{\v{q}}

\def\cP{\mathcal P}
\def\cS{\mathcal S}
\def\cK{\mathcal K}

\def\be{\begin{equation}}
\def\ee{\end{equation}}
\def\bea{\begin{align}}
\def\eea{\end{align}}

\subheader{
\footnotesize
\vspace*{-3.7em}
\begin{flushright}
RBI-ThPhys-2025-nn
\end{flushright}
}

\title{\fontsize{20}{31}\selectfont{Parity in Composite-Field Galaxy Correlators}}

\author[a]{Zucheng Gao,}
\author[a]{Azadeh Moradinezhad Dizgah,}
\author[b,c,d]{Zvonimir Vlah}
\makeatletter
\renewcommand\afterAuthorSpace{\vskip 6pt}
\makeatother

\affiliation[a]{Laboratoire d’Annecy de Physique Theorique (LAPTh), CNRS/USMB, 9 Chemin de Bellevue BP110 - Annecy -  F-74941 - ANNECY CEDEX - FRANCE}
\affiliation[b]{Division of Theoretical Physics, Ru\dj er Bo${\rm \check{s}}$kovi\' c Institute, Zagreb HR-10000, Croatia}
\affiliation[c]{Kavli Institute for Cosmology Cambridge, Madingley Road, Cambridge, CB3 0HA, UK}
\affiliation[d]{DAMTP, Centre for Mathematical Sciences, Wilberforce Road, Cambridge CB3 0WA, UK}

\emailAdd{zucheng.gao@lapth.cnrs.fr}
\emailAdd{azadeh.moradinezhad@lapth.cnrs.fr}
\emailAdd{zvlah@irb.hr}

\abstract{Detecting parity violation on cosmological scales would provide a striking clue to new physics. Large-scale structure offers the raw statistical power---many three-dimensional modes---to make such tests. However, for scalar observables, like galaxy clustering, the leading parity-sensitive observable is the trispectrum, whose high dimensionality makes the measurement and noise estimation challenging. We present two late-time parity-odd kurto spectra that compress the parity-odd scalar trispectrum into one-dimensional, power–spectrum–like observables. They are built by correlating (i) two appropriately weighted quadratic composite fields, or (ii) a linear and cubic composite field, constructed from dark matter (DM) or galaxy overdensity fields. We develop an FFTLog pipeline for efficient theoretical predictions of the two observables. We then validate the estimators for a specific parity-odd primordial template on perturbative DM field, and on DM and halo fields in full N-body \texttt{Quijote} simulations, with and without parity-odd initial conditions, in real and redshift space. For DM, the variance is dominated by the parity-even contribution---i.e., the gravitationally induced parity-even trispectrum---and is efficiently suppressed by phase-matched fiducial subtraction. For halos, discreteness-driven stochasticity dominates and is not appreciably reduced by subtraction; however, optimal weighting and halo–matter cross kurto spectra considerably mitigate this noise and enhance the signal. Using controlled down-sampling of the matter field, we empirically calibrate how the parity-even variance scales with number density and volume, and provide an illustrative forecast for the detectability of parity-odd kurto spectra in a Euclid-like spectroscopic galaxy survey.}

\begin{document}

\maketitle

\section{Introduction}

Observations of the large-scale structure (LSS) of the Universe provide a unique window into fundamental physics. With the advent of Stage-IV cosmological surveys such as the Dark Energy Spectroscopic Instrument (DESI)~\cite{DESI:2016fyo}, Euclid~\cite{Euclid:2024yrr}, SPHEREx~\cite{SPHEREx:2014bgr}, the Nancy Grace Roman Space Telescope~\cite{Wang:2021oec}, the Prime-Focus Spectrograph (PFS)~\cite{PFSTeam:2012fqu}, and the Rubin Observatory LSST~\cite{LSSTDarkEnergyScience:2012kar}, we are in an exciting precision era, where detailed statistical measurements of the matter distribution via galaxy clustering and lensing can test a wide range of new physics scenarios. These include the nature of dark energy (DE) and dark matter (DM), the properties of primordial fluctuations, and possible violations of fundamental spacetime symmetries.

Parity, the transformation property of a system under point-like reflection, is one such symmetry. Although parity violation is well established in the weak nuclear interactions \cite{Wu:1957my,Lee:1956qn}, its status on cosmological scales is not fully explored.\footnote{See \cite{Kamada:2022nyt} for a broad overview of how microscopic chirality feeds macroscopic parity-odd phenomena in astrophysics and cosmology.} Detection of a parity-odd cosmological signal could indicate new physics in the early or late Universe---e.g., exotic inflationary dynamics \cite{Cabass:2022oap} or modifications of gravity \cite{Creque-Sarbinowski:2023wmb}---or could arise from astrophysical processes such as helical large-scale magnetic fields generated by rotating, turbulent plasma \cite{Brandenburg:2004jv}.\footnote{Such dynamo-generated helicity is largely localized to galactic scales.} For scalar observables, such as CMB temperature anisotropies, weak gravitational lensing fields, and galaxy overdensity, parity violation does not manifest at the level of two- or three-point correlation functions, which are invariant under parity transformations. Consequently, the four-point function (trispectrum) is the leading-order statistic capable of capturing parity-odd signatures in scalar fields. Several recent works have explored the possibility of using the CMB~\cite{Philcox:2023ffy,Philcox:2025wts,Greco:2025xtt} and galaxy trispectra to probe such effects~\cite{Cahn:2023L,Hou:2022rcd,Philcox:2022,Cabass:2022oap,Philcox:2024mmz,Slepian:2025kbb}. However, trispectrum (and its covariance~\cite{Hou:2021ncj}) estimation and interpretation remain computationally formidable due to the high dimensionality of the observable, complicated by survey geometry, noise, and nonlinear gravitational evolution \cite{Philcox:2024mmz,Krolewski:2024paz,Adari:2024vkf}. These challenges motivate the development of lower-dimensional observables that efficiently retain the essential information encoded in the trispectrum without requiring its full measurement and analysis.

A powerful and flexible framework for compressing information from N-point correlation functions is provided by \textit{composite-field spectra} (CF), defined as two-point correlation functions constructed from cross-correlations of appropriately weighted products of the observed overdensity fields evaluated at the same spatial location. These CF-spectra offer computationally efficient and interpretable alternatives to direct polyspectrum estimation, enabling targeted extraction of specific physical signatures—such as nonlinear bias, primordial non-Gaussianity, or parity violation—by carefully designing their weights. At lowest order, this approach underpins the galaxy \textit{skew spectrum}~~\cite{Schmittfull:2014tca,Schmittfull:2020hoi}, which cross-correlates the galaxy field with a weighted square of itself, efficiently capturing the full tree-level bispectrum information. The idea has been further extended to approximate trispectrum information through \textit{kurto spectra}~~\cite{Lazeyras:2017hxw,Abidi:2018eyd}, constructed from correlations between pairs of quadratic CFs or between a cubic CF and the linear field.\footnote{Optimally weighted skew spectra in harmonic space, which saturate the Cramer-Rao limit of the bispectrum in the weakly non-Gaussian limit, were first introduced in the context of constraining primordial non-Gaussianity from the CMB data \cite{Komatsu:2003iq,Munshi:2009ik}. More recent works in the context of harmonic-space on the topic include \cite{Munshi:2020ofi,Munshi:2021uwn,Chakraborty:2022aok}.} In constructing skew and kurto spectra, the mode-coupling kernels that define CFs are chosen to mirror those in perturbative description of DM and its biased tracers, enabling the resulting estimators to closely approximate maximum-likelihood estimators for amplitudes of primordial polyspectra, the growth rate of structure, and galaxy biases.

While skew and kurto spectra have thus far primarily targeted gravitational nonlinearities \cite{Hou:2022rcd,Hou:2024blc} and primordial non-Gaussianities \cite{MoradinezhadDizgah:2019xun}, the CF framework is flexible enough to accommodate other types of physical signatures. In particular, it can be readily extended to isolate parity-violating signals. This can be achieved by suitably modifying the mode-coupling kernels and filtering the input fields to isolate the parity-odd components of the underlying trispectrum. Such parity-sensitive CFs yield two-point spectra that vanish under parity-even conditions, providing targeted and efficient estimators designed explicitly for probing parity. 

Building on our previous work on skew spectra, we extend the CF–spectra construction to capture parity-odd trispectrum that may arise from early- or late-Universe physics. The general construction is similar to what has been presented in \cite{Jamieson:2024mau}, but here we develop and apply the estimators to dark-matter and halo fields rather than primordial fluctuations alone. We introduce an FFTLog-based algorithm that enables rapid and accurate numerical evaluation of parity-odd CF spectra. We then rigorously test these estimators against synthetic density fields generated with Eulerian perturbation theory and validate their robustness with detailed measurements from fully non-linear $N$-body simulations of matter and halos. This allows us to systematically quantify the impact of non-linear gravitational evolution, halo/galaxy bias, and stochastic noise on estimator performance. Furthermore, we identify the dominant sources of variance in the measurements, assess the detectability of parity-violating signatures on simulations, and make forecast for their detectability in a Euclid-like spectroscopic galaxy clustering data. 

The rest of the paper is organized as follows: in section \ref{sec:th-parity-eu}, we review the theoretical motivations for considering parity-odd signatures by briefly discussing example of early-universe models which produce parity-odd primordial trispectrum and how the primordial trispectrum is imprinted on LSS. In section \ref{sec:parity-CFs}, we discuss the general construction of CF-spectra as efficient estimators of N-point functions, specializing to the case of parity-odd kurto spectra. In section \ref{sec:th-FFTLog}, we describe the fast computational algorithm based on FFTLog approach for evaluating the parity-odd kurto spectra. In section \ref{sec:sims} we present the measurements of parity-odd kurto spectra on simulated DM and halo fields with parity violating initial conditions, and discuss the expected detectability of the signal and the main limitations. Finally, in section \ref{sec:conclusions} we draw our conclusions.

\section{Parity Violation in Scalar Sector and the Physics of Inflation}\label{sec:th-parity-eu}

We briefly survey inflationary mechanisms that generate an \emph{imaginary} (parity-odd) component of the primordial trispectrum.\footnote{For a recent  review of mechanisms generating inflationary three- and four-point functions, see~\cite{Philcox:2025bvj}.} Since our goal is to test CF-spectra—specifically, parity-odd kurto-spectra—we will ultimately adopt a single phenomenological template (a separable pseudoscalar contact form) and defer a broader exploration of shapes to future work.

Four illustrative inflationary scenarios known to yield a parity-odd \emph{scalar} trispectrum are the following.
\emph{First}, single-field models with a non-Bunch–Davies vacuum and a nonlinear dispersion relation, exemplified by ghost inflation \cite{Arkani-Hamed:2003juy}, in which the inflaton develops a time-dependent background and higher-derivative self-interactions containing a Levi-Civita tensor. These operators produce a parity-odd, \emph{separable} contact trispectrum that is suppressed in the fully squeezed limit and peaks when all four wavenumbers are comparable (equilateral to mildly folded shapes) \cite{Cabass:2022rhr,Cabass:2022oap}.
\footnote{The same higher-derivative terms generate an equilateral bispectrum \cite{Senatore:2009gt}, tightly constrained by Planck \cite{Planck:2019kim}; A viable ghost-inflation scenario therefore requires a mechanism—such as a symmetry or tuning—that suppresses the cubic coefficients relative to the quartic ones  \cite{Cabass:2022rhr,Philcox:2025bvj}.}
\emph{Second}, cosmological-collider models with tree-level exchange of a massive spinning field \cite{Arkani-Hamed:2015bza}, where a pseudo-scalar coupling yields a parity-odd, \emph{non-separable} trispectrum that depends on the intermediate mass and spin and peaks in collapsed configurations \cite{Liu:2019fag,Cabass:2022rhr,Cabass:2022oap}. A high-fidelity separable approximation to the collapsed limit is given in \cite{Philcox:2025bvj}.
\emph{Third}, models with a chiral $U(1)$ gauge source during inflation. The resulting \emph{scalar} trispectrum’s shape and amplitude depend sensitively on how the axion and gauge sector are realized, and they typically do not produce fully separable trispectra (see, e.g., \cite{Bartolo:2015dga,Niu:2022fki,Fujita:2023inz,Stefanyszyn:2023qov}). As a benchmark, in the standard axion–inflation setup where the \emph{inflaton itself is an axion} coupled via a Chern–Simons term, the trispectrum acquires a parity-odd component that is \emph{enhanced in squeezed (collapsed) configurations} \cite{Bartolo:2015dga,Shiraishi:2016mok,Cho:2025rvg}.\footnote{Two other representative axion scenarios are: (i) models with a \emph{massive} $U(1)$ gauge field, for which the parity-odd trispectrum arises from spin-1 exchange, factorizes into bispectrum-like building blocks, and exhibits collapsed-limit enhancement with a characteristic chiral angular dependence \cite{Stefanyszyn:2023qov}; and (ii) ``rolling axion'' scenarios where a \emph{spectator} axion has time-dependent velocity, which instead produce parity-odd trispectra peaking in (quasi) equilateral configurations, with amplitudes controlled by the axion roll, gauge coupling, and related parameters \cite{Niu:2022fki,Fujita:2023inz}.}
\emph{Fourth}, helical primordial magnetic fields (PMFs) \cite{Shaw:2009nf,Caprini:2009pr,Caprini:2014mja,Caprini:2017vnn,Bamba:2021wyx} furnish another scalar-sector mechanism: the helical component of the PMF sources a parity-odd (imaginary) curvature trispectrum at leading order \cite{Yura:2025mus}. The shape is generically \emph{non-separable}, controlled by the magnetic spectral index and helicity fraction; the parity-odd signal is enhanced in collapsed configurations and can exceed the even component in parts of parameter space \cite{Yura:2025mus}.

Instead of modeling the exact shapes of primordial scalar trispectra from specific early Universe scenarios, we adopt the simple phenomenological template for a contact-type parity-odd trispectrum introduced in \cite{Coulton:2024}, for which there are readily available $N$-body simulations (the \texttt{Quijote-ODD} suite \faGlobe\footnote{\url{https://quijote-simulations.readthedocs.io/en/latest/odd.html}}) to test our estimators. This form captures the contact-type parity-odd trispectrum generated by quartic couplings in ghost inflation and, being separable, enables both fast theory evaluation and a clear demonstration of CF-spectra performance.

In this template, the parity-odd primordial trispectrum originates from a purely imaginary cubic correction to Gaussian primordial fluctuations,
\begin{align}\label{eq:phi_ng}
\phi(\vk) &= \phi_{\rm G}(\vk) + \phi^{(3)}_{\rm NG}(\vk) \non \\
&= \phi_{\rm G}(\vk) + i \, g_{\rm NL}^{\rm PO} \!\int_{\bq_1,\bq_2,\bq_3} (2\pi)^3 \delta_D(\vk-\bq_{123}) \, K_{\rm NL,PO}^{(3)}(\bq_1,\bq_2,\bq_3)\,\phi_{\rm G}(\bq_1)\phi_{\rm G}(\bq_2)\phi_{\rm G}(\bq_3),
\end{align}
where we defined $\int_\vq = \int d^3q/(2\pi)^3$. The overall amplitude \(g_{\rm NL}^{\rm PO}\) is real and the cubic kernel is a pseudo-scalar,
\begin{equation}
K_{\rm NL,PO}^{(3)}(\vk_1,\vk_2,\vk_3) \;=\; \vk_1\!\cdot\!\big(\vk_2\times\vk_3\big)\, S(k_1,k_2,k_3).
\end{equation}
Following \cite{Coulton:2024}, we adopt the separable choice
\be\label{eq:K3_Coulton}
K_{\rm NL,PO}^{(3)}(\vk_1,\vk_2,\vk_3) \;=\; \vk_1\!\cdot\!\big(\vk_2\times\vk_3\big)\, k_1^\alpha k_2^\beta k_3^\gamma .
\ee
To leading order in $g_{\rm NL}^{\rm PO}$, one insertion of $\phi^{(3)}_{\rm NG}$ contracted with three Gaussian fields yields the connected four-point function
\be
\langle \phi(\vk_1)\phi(\vk_2)\phi(\vk_3)\phi(\vk_4)\big\rangle_c
=(2\pi)^3\,\delta_D(\vk_1+\vk_2+\vk_3+\vk_4)\,T_\phi(\vk_1,\vk_2,\vk_3,\vk_4),
\ee
\be \label{eq:Tphi_contact}
T_\phi(\vk_1,\vk_2,\vk_3,\vk_4)
=\; i\, g_{\rm NL}^{\rm PO}\;
K^{(3)}_{\rm NL,PO}(\vk_1,\vk_2,\vk_3)\;
P_\phi(k_1)\,P_\phi(k_2)\,P_\phi(k_3)
\;+\;23~\mathrm{perms},
\ee
where $P_\phi$ is the Gaussian power spectrum of $\phi$. The $24$ ordered terms correspond to the $4$ choices of which external leg carries the cubic field and, for each choice, the \(3!\) Wick pairings that contract the three Gaussian fields with the three legs of $K_{\rm NL,PO}^{(3)}$. The Fourier-space trispectrum in eq.~\eqref{eq:Tphi_contact} is purely imaginary and flips sign under spatial inversion since the scalar triple product is odd under parity. 

Although our focus in this paper is on the scalar sector, parity can also be broken in the \emph{vector} and \emph{tensor} sector---either via chiral gravity (e.g., dynamical Chern–Simons and chiral scalar–tensor extensions) \cite{Lue:1998mq,Jackiw:2003pm,Alexander:2007kv,Nojiri:2020pqr,Cai:2022lec,Crisostomi:2017ugk,Bartolo:2020gsh,Moretti:2024fzb} or through chiral gauge dynamics during inflation that preferentially excite one GW helicity \cite{Anber:2012du,Adshead:2013qp,Bastero-Gil:2022fme}. The most direct probes are \emph{vectorial} and \emph{tensorial} observables: galaxy spin power spectra~\cite{Shim:2024tue}, parity-odd CMB polarization two-point functions ($TB,\,EB$) \cite{Lue:1998mq,Gerbino:2016mqb} and parity-odd tensor/mixed higher-point functions (bispectra, trispectra) \cite{Bartolo:2017szm,Bartolo:2018elp,Philcox:2023xxk}, with analogous tests in LSS cosmic shear (nonzero $EB$ and parity-odd shear bispectra) \cite{Biagetti:2020lpx,Philcox:2023uor,Kurita:2025hmp}. Tensor-sector parity violation also induces \emph{scalar} signatures via (i) primordial graviton exchange, which yields a parity-odd scalar trispectrum—the collapsed/squeezed ``cosmological fossil" of a helical long mode \cite{Masui:2010cz,Jeong:2012df,Schmidt:2013gwa,Masui:2017fzw,Creque-Sarbinowski:2023wmb}---and (ii) late-time lensing by a chiral GW background, generating a parity-breaking galaxy four-point through gradient–curl couplings \cite{Inomata:2024ald}. Finally, parity can be violated along photon propagation via \emph{cosmic birefringence}, which rotates $E\!\to\!B$ and produces nonzero $TB/EB$ even without chiral gravitons \cite{Minami:2020odp,Fujita:2020ecn,Cosmoglobe:2023pgf}. Let us note that the estimator of the trispectrum we introduce in this paper and describe in details in section \ref{sec:parity-CFs} is applicable in searching for imprints of vector/tensor chirality if it produces non-zero trispectrum of scalar fluctuations (in early or late-time). For scenarios which generate chiral higher-order ($N>2$) correlation functions of tensor fluctuations, the efficient estimators introduced here for bispectrum and trispectrum (the skew and kurto spectra) should be extendable. We leave investigation of such scenarios to a future work.

The primordial (parity-odd) trispectrum, like that produced by the template in eq. \eqref{eq:phi_ng}, impacts galaxy statistics in two ways. First, most directly, it seeds a nonzero connected trispectrum of the late-time matter field (and hence of galaxies). At linear level in perturbation theory (neglecting redshift-space distortions), we have
\begin{align} 
\delta_m(\vk,z)&=\mathcal M(k,z)\,\phi(\vk), \label{eq:deltam}\\
T_m^{\rm PNG}(\vk_1,\vk_2,\vk_3,\vk_4,z) 
&=\prod_{i=1}^4 \mathcal M(k_i,z) T_\phi(\vk_1,\vk_2,\vk_3,\vk_4), \label{eq:Tm_lin}
\end{align}
where $\cM$ is the linear transfer function of matter including the linear growth. Nonlinear gravitational evolution induces additional non-zero trispectrum contribution. This is why we have used the superscript ``PNG'' for the trispectrum. The induced galaxy trispectrum is given by 
\begin{align} 
\delta_g(\vk,z) &= b_1(z) \delta_m(\vk,z), \label{eq:deltag}\\
T_g^{\rm PNG}(\vk_1,\vk_2,\vk_3,\vk_4,z) &= b_1^4(z) T_m(\vk_1,\vk_2,\vk_3,\vk_4,z). \label{eq:Tg_lin}
\end{align}
Again, we have used the superscript ``PNG'' to distinguish the primordial origin of the galaxy trispectrum from contributions due to nonlinear gravitational evolution and galaxy biases. Therefore, the parity-odd property of the primordial trispectrum gets directly imprinted on the galaxy trispectrum. The second imprint of non-zero primordial trispectrum (in general any type of primordial non-Gaussianity) is that it alters the galaxy–matter biasing relation, necessitating new operators in the perturbative renormalized bias expansion \cite{Assassi:2014fva, Assassi:2015fma, Desjacques:2016bnm}. Parity-odd imprints on LSS may also be generated at \emph{late} times, e.g., through astrophysical couplings to a pseudo-scalar sector (such as axion–photon interactions) or via effective chiral-gravity phenomena at low redshift. The framework of effective field theory of LSS (EFTofLSS) \cite{McDonald:2006mx,Baumann:2010tm,Carrasco:2012cv,Cabass:2022avo}, provides a unified language to incorporate both the primordial and late-time parity-odd imprints on galaxy biases, by consistently including relevant bias and stochastic operators allowed by symmetries and degrees of freedom of the early and late Universe. In this work we focus on the direct imprints of parity-odd primordial trispectrum on matter and galaxy fields and construct efficient estimators to isolate it. We refer the interested reader to Ref. \cite{Cabass:2022oap} about the characterization of late-time parity-breaking bias operators, and leave the investigation of the galaxy bias and stochasticity in the presence of (parity-odd) primordial scalar trispectrum to a future work.

Let us close this section by noting that when analyzing real data, any apparent parity‑odd trispectrum (or compressed statistic such as the kurto‑spectra) must survive stringent systematic checks. Galaxy‑survey systematics—mask geometry, stellar contamination, seeing variations, depth fluctuations, redshift‑dependent selection, fiber collisions---may either leak even‑parity power into odd modes or render determination of error budget challenging \cite{Hou:2022wfj,Hou:2025wuc}. Since in this paper our focus is on testing the performance of parity-odd kurto spectra on simulated matter and halo density field in a periodic box, we will not further discuss these challenges.

\section{Efficient Estimators to Extract Trispectrum Information} \label{sec:parity-CFs}

In this section, we begin by briefly describing the general construction of weighted CF-spectra as proxies for N-point correlation functions. Focusing on 3- and 4-point correlation functions, we then review the construction of weighted skew and kurto spectra, which serve as optimal estimators of the bispectrum and trispectrum in the semi-linear regime, where the separability of N-point functions holds. We then demonstrate how the weights of the two kurto spectra can be tuned to isolate parity-odd contributions to the galaxy trispectrum, while simultaneously nulling the parity-even component. Finally, we relate the parity-odd kurto spectra to the general irreducible decomposition of tensor fields.

\subsection{Composite-field spectra as proxies for N-point correlation functions}\label{subsec:CF}

In general, to capture the information encoded in the N-point statistics of a non-Gaussian field, one can construct pseudo two-point correlation functions by correlating CFs---built from multiple copies of the original field evaluated at the same spatial location---with either each other or the original field. As mentioned earlier, in contrast to N-point CFs, the CF-spectra offer a crucial advantage of being a lower dimensional observable. This implies that  the computational cost of measuring them from the data is comparable to that of the power spectrum, and if estimating the covariance matrices from mocks, their analysis requires significantly smaller number of mocks compared to that needed for N-point functions. Clearly, one can construct the CF-spectra from the same observed field, e.g., galaxy overdensity, or from multiple fields, e.g., galaxy over density and CMB lensing. In the discussion below, we assume that only a single observed field is used. Later, when discussing the parity-odd kurto spectra, we generalize the definition to multiple observed fields. 

Composite-fields, denoted as ${\mathcal D}_n[\delta](\bx)$ below, are defined as products of filtered density fields in real space,
\be\label{eq:CF_def_config}
{\mathcal D}_n[\delta](\bx) = \prod_{i=1}^{n} {\mathcal D}_i(\bx)\delta(\bx),
\ee
where ${\mathcal D}_i$ are arbitrary filters applied on the observed field. Therefore, in Fourier space, an $n$-th order CF is defined as a convolution involving $n$ copies of the observed overdensity field and a mode-coupling kernel $D_n$, which corresponds to the Fourier transform of the product of real-space filters:
\begin{align}\label{eq:CF_def_Fourier}
    D_n[\delta](\vk) &= \int_{\vq_1} \cdots \int_{\vq_{n-1}} D_n(\vq_1,\cdots,\vq_{n-1},\vk-\vq_1-\cdots-\vq_{n-1})  \notag \\ & \hspace{0.8in}  \times \delta_R(\vq_1)  \cdots\delta_R(\vq_{n-1})\delta_R(\vk-\vq_1-\cdots-\vq_{n-1}), 
\end{align}
where the subscript $R$ indicates that the observed density field is smoothed on scale $R$, i.e., $\delta_R(\vk) = W_R(\vk) \delta(\vk)$. The smoothing suppresses contributions from fluctuations on scales smaller than $R$, in analogy with imposing a small-scale cutoff in the analysis of $N$-point CFs. Several previous works on composite fields (e.g., \cite{Schmittfull:2014tca,MoradinezhadDizgah:2019xun,Hou:2022rcd,Hou:2024blc}) have adopted a Gaussian filter instead of a Fourier-space top-hat, in order to avoid ringing effects when applied to data. In this paper, our main results are obtained with a Gaussian filter, $W_R(k) ={\rm exp}(-k^2 R^2)/2$. However, in section~\ref{sec:sims} we also explore the impact of the filter choice, in particular using a smooth hyperbolic tangent (tanh) filter that avoids ringing while preserving a closer correspondence with a sharp $k_{\rm max}$ cutoff.

The scalar pseudo two-point statistics of the CFs (with one another and the original observed field) is given by
\be\label{eq:CF_spec}
\langle D_n[\delta](\vk) D_m[\delta](\vk')\rangle = (2\pi)^3 \delta_D(\vk+\vk') \cP_{D_n[\delta],D_m[\delta]}(\vk).
\ee
As for the standard power spectrum, in real space, the above two-point function is isotropic (depends only on the magnitude of wavevector), while in redshift-space, it will also depend on line-of-sight direction. One can further perform angular averaging or more generally define multipoles of this pseudo-power spectrum. 

In defining the composite-field spectra, the filters $D_n$ can be tailored to target specific physical signatures in the $N$-point statistics. For example, as will be described in section \ref{subsec:skew_kurto}, in the weakly non-Gaussian regime, when the theoretical $N$-point template is (approximately) separable, one can choose $D_n$ such that the resulting CF spectra reproduce---up to the usual disconnected (mean-field/Wick) subtraction---the maximum-likelihood estimator for the template amplitude. This principle underlies the weighted skew and kurto spectra (discussed in the next subsection) \cite{Schmittfull:2014tca,MoradinezhadDizgah:2019xun,Schmittfull:2020hoi,Abidi:2018eyd} to constrain galaxy-bias parameters, the growth rate of structure, and amplitudes of primordial non-Gaussianity. As another example of tailored kernels, one can design $D_n$ to isolate parity-odd contributions to the trispectrum, which is the focus of this work and is detailed in section~\ref{subsec:kurto_PO}.

Note that while we have defined the CFs in eqs. \eqref{eq:CF_def_config} and \eqref{eq:CF_def_Fourier} and the corresponding pseudo-power spectra in eq. \eqref{eq:CF_spec} as scalar quantities, these definitions can be extended to tensorial quantities. As we will discuss in section \ref{subsec:kurto_PO}, the CF spectrum estimators that capture the parity-odd trispectrum necessarily have to be constructed from vector fields. We will only focus on scalar CF-spectra and leave the consideration of the tensorial spectra to a future work.

\subsection{Connection to maximum-likelihood estimators} \label{subsec:skew_kurto}

\noindent In this subsection we connect composite-field (CF) spectra to maximum-likelihood (ML) estimators: we show how \emph{skew} (quadratic–linear) and \emph{kurto} (cubic–linear and quadratic–quadratic) spectra reproduce, in the weakly non-Gaussian limit, the ML estimators for separable bispectrum and trispectrum amplitudes (up to the usual disconnected subtraction). We first derive the skew–ML correspondence with explicit $k$ cuts (no smoothing in the ML form), then reorganize the trispectrum ML estimator into the two kurto spectra. In our baseline results, we do not use the optimal weighting described here, but only for a subset of our results, we show the improvements in the detectability of parity-odd trispectrum when these weights are used. 
 
\paragraph{Skew spectra and the ML bispectrum estimator.} Weighted skew spectra, introduced in \cite{Schmittfull:2014tca,Schmittfull:2020hoi}, are the lowest–order CF spectra defined in eq.~\eqref{eq:CF_spec}: the angle–averaged cross–power of a filtered \emph{quadratic} composite field with a filtered \emph{linear} copy of the field. In the weakly non-linear regime, when the galaxy bispectrum can be written as a sum of \emph{product-separable} terms, one can choose the quadratic mode-coupling kernel and the linear filter so that the resulting skew spectra coincide with the \emph{cubic part} of the maximum-likelihood (ML) estimator for the corresponding amplitudes. To make this correspondence clear, we briefly review the ML estimator for a noisy data set,\footnote{We largely follow the notation of previous skew-spectra works \cite{Schmittfull:2014tca,MoradinezhadDizgah:2019xun,Schmittfull:2020hoi,Hou:2022rcd,Hou:2024blc}, with minor changes for consistency across this paper.} then reorganize it into skew spectra.

For a theoretical bispectrum model
\be
\big\langle \delta(\vk_1)\delta(\vk_2)\delta(\vk_3)\big\rangle_{\rm th}
=(2\pi)^3\,\delta_D(\vk_1+\vk_2+\vk_3)\,B_{\rm th}(\vk_1,\vk_2,\vk_3),
\ee
in the nearly Gaussian limit, the maximum-likelihood estimator for its amplitude reads \cite{Taylor:2000hq,Fergusson:2010ia}
\begin{align}
\label{eq:ML_bis}
\hat A
=\frac{1}{N_{\rm th}^B}\int_{\vk_1}\!\int_{\vk_2}\!\int_{\vk_3}
\big\langle \delta(\vk_1)\delta(\vk_2)\delta(\vk_3)\big\rangle_{\rm th}
&\Big[
C^{-1}\!\big(\delta^{\mathrm{obs}}_{\vk_1}\big)\,
C^{-1}\!\big(\delta^{\mathrm{obs}}_{\vk_2}\big)\,
C^{-1}\!\big(\delta^{\mathrm{obs}}_{\vk_3}\big)
 \non \\
&\hspace{2cm} -3\,C^{-1}\!\big(\delta^{\mathrm{obs}}_{\vk_1}\delta^{\mathrm{obs}}_{\vk_2}\big)\,
C^{-1}\!\big(\delta^{\mathrm{obs}}_{\vk_3}\big)
\Big],
\end{align}
where $N_{\rm th}^B$ normalizes the estimator, $\delta^{\mathrm{obs}}_{\vk}$ is the observed (noisy) field, and $C$ is its covariance. The integrals are restricted to the selected range of $k\in[k_{\min},k_{\max}]$. Assuming a diagonal covariance on the scales of interest, we have $C^{-1}(\delta^{\mathrm{obs}}_{\vk})\simeq \delta_{\mathrm{obs}}(\vk)/P_{\rm obs}(k)$. The second (linear/mean-field) term, which we drop in subsequent discussion, is required if the field is statistically inhomogeneous (e.g., due to survey mask and anisotropic noise).

For a bispectrum template expanded as a sum of $m_B$ product-separable terms, 
\be
\label{eq:Bsep}
B_{\rm th}(\vk_1,\vk_2,\vk_3) \;=\; \sum_{\alpha=1}^{m_B} A_\alpha\, \prod_{j=1}^{3} g_j^\alpha(\vk_j),
\ee
the estimator for the amplitude of a single separable term (for one chosen ordering) is
\be
\hat A_\alpha \;=\; \frac{(2\pi)^3}{N_{\rm th}}\int_{\vk} \frac{g_3^\alpha(\vk)\, \delta_{\rm obs}(-\vk)}{P_{\rm obs}(k)}
\int_{\bq} \frac{g_1^\alpha(\vq)\, \delta_{\rm obs}(\bq)}{P_{\rm obs}(q)} \,
\frac{g_2^\alpha(\vk-\bq)\, \delta_{\rm obs}(\vk-\bq)}{P_{\rm obs}(|\vk-\bq|)}.
\ee
Comparing with eq.~\eqref{eq:CF_def_Fourier}, the inner integral defines the quadratic composite
\be\label{eq:CF_2nd}
\cS_\alpha(\vk) \;=\; \int_{\bq} D_2^{(\alpha)}(\bq,\vk-\bq)\, \delta_{\rm obs}(\bq)\,\delta_{\rm obs}(\vk-\bq),
~~
D_2^{(\alpha)}(\bq,\vk-\bq) \;=\; \frac{g_1^\alpha(\vq)\, g_2^\alpha(\vk-\bq)}{P_{\rm obs}(q)\, P_{\rm obs}(|\vk-\bq|)},
\ee
and the linear filtered field is given by
\be
\tilde \delta_{\rm obs}^{(\alpha)}(\vk) \;=\;  \frac{g_3^\alpha(\vk)}{P_{\rm obs}(k)}\, \delta_{\rm obs}(\vk).
\ee
Thus the expectation value of the estimator is the $k$-integral of the isotropic skew spectrum,
\be
\big\langle \hat A_\alpha \big\rangle
\;=\; 4\pi \int_0^\infty k^2\,dk \;\cP_{\cS_\alpha,\tilde \delta_{\rm obs}^{(\alpha)}}(k),
\qquad
\cP_{\cS_\alpha,\tilde \delta}(k)\;\equiv\;\int \frac{d\Omega_{\hat\vk}}{4\pi}\,
\big\langle \cS_\alpha(\vk)\, \tilde \delta(-\vk)\big\rangle .
\ee
Note that while in the ML estimator scale-cuts are explicitly enforced by restricting triangle legs to $k_{\rm min}\leq k\leq k_{\max}$, in CF constructions like eq.~\eqref{eq:CF_2nd}, the convolution sums over \emph{all} internal momenta and to suppress contributions from $k>k_{\max}$ and avoid ringing/aliasing in FFTs, we smooth the input field
$\delta_R(\vk)\;\equiv\;W_R(k)\,\delta_{\rm obs}(\vk)$ in the composite fields. 

Assuming Gaussian initial conditions, it was shown in Refs. \cite{Schmittfull:2014tca, Schmittfull:2020hoi} that a set of 3 (14) skew spectra fully capture the information of the real- (redshift-) space bispectrum (on large scales) for parameters that appear as overall amplitudes of separable contributions to the bispectrum. This include amplitudes of galaxy bias and stochastic operators, amplitude of primordial power spectrum, and growth rate of structure. In the presence of primordial non-Gaussianity, additional skew spectra are needed to optimally estimate amplitude of primordial bispectrum \cite{Schmittfull:2014tca,MoradinezhadDizgah:2019xun}. 

\paragraph{Kurto spectra and the ML trispectrum estimator.} 
The next-order CF-spectra, referred to as kurto spectra \cite{Abidi:2018eyd,Lazeyras:2017hxw}, capture the information in trispectrum on semi-linear regime with specific choice of the mode coupling kernels. The correspondence to ML estimator of trispectrum  is a generalization of what we described for skew spectra. The ML estimator for amplitude of a theoretical trispectrum is given by \cite{Fergusson:2010ia} 
\begin{align}
\label{eq:ML_tri}
\hat A
=\frac{1}{N_{\rm th}^T}
\int_{\vk_1}\!\int_{\vk_2}\!\int_{\vk_3}\!\int_{\vk_4}
& \big\langle \delta(\vk_1)\delta(\vk_2)\delta(\vk_3)\delta(\vk_4)\big\rangle_{\rm th}
\Big[
\,C^{-1}\!\big(\delta^{\mathrm{obs}}_{\vk_1}\big)\,
  C^{-1}\!\big(\delta^{\mathrm{obs}}_{\vk_2}\big)\,
  C^{-1}\!\big(\delta^{\mathrm{obs}}_{\vk_3}\big)\,
  C^{-1}\!\big(\delta^{\mathrm{obs}}_{\vk_4}\big)\nonumber\\
&-6\,C^{-1}\!\big(\delta^{\mathrm{obs}}_{\vk_1}\delta^{\mathrm{obs}}_{\vk_2}\big)\,
        C^{-1}\!\big(\delta^{\mathrm{obs}}_{\vk_3}\big)\,
        C^{-1}\!\big(\delta^{\mathrm{obs}}_{\vk_4}\big) +3\,C^{-1}\!\big(\delta^{\mathrm{obs}}_{\vk_1}\delta^{\mathrm{obs}}_{\vk_2}\big)\,
      C^{-1}\!\big(\delta^{\mathrm{obs}}_{\vk_3}\delta^{\mathrm{obs}}_{\vk_4}\big)
\Big].
\end{align}
As in the bispectrum case, the second and third terms on the right-hand side of eq. \eqref{eq:ML_tri} are the Gaussian–disconnected (quadratic and constant) contributions. They vanish in expectation for a homogeneous, periodic box, but in the presence of a survey mask or anisotropic noise they must be kept and estimated (e.g.\ from simulations or randomized catalogs) to ensure an unbiased estimator. In analogy to the bispectrum case, assuming a product-separable template of trispectrum,
\be
\label{eq:tri_separable_form_general}
T_{\rm th}(\vk_1,\vk_2,\vk_3,\vk_4)=\sum_{\beta=1}^{m_T} A_\beta \prod_{j=1}^{4} g_j^\beta(\vk_j),
\ee
the quartic piece of eq. \eqref{eq:ML_tri} can be reorganized into two power–spectrum–like \emph{kurto spectra} by choosing appropriate filters. As for skew spectra, we enforce the scale cuts by smoothing the fields entering the composite fields. We note that the separability of trispectrum means that it can be factorised by multiplication of separate functions depending on k-vectors. In the case of non-separable theoretical trispectrum template, we can approximate it with separable forms, as discussed in~\cite{Philcox:2025bvj}

We define the cubic composite and linear filtered field
\begin{align}
\mathcal K_\beta(\vk) &= \int_{\bq_1}\!\int_{\bq_2}\,
D^{(\beta)}_3(\bq_1,\bq_2,\vk-\bq_1-\bq_2)\,
\delta_{\rm obs}(\bq_1)\,\delta_{\rm obs}(\bq_2)\,\delta_{\rm obs}(\vk-\bq_1-\bq_2), \\
\tilde\delta_{\rm obs}^{(\beta)}(\vk) &= \frac{g_4^\beta(\vk)}{P_{\rm obs}(k)}\,\delta_{\rm obs}(\vk),
\end{align}
together with two quadratic composites built from disjoint pairs,
\begin{align}
\mathcal S_{\beta,12}(\vk) &= \int_{\bq}\,
D^{(\beta)}_{2,12}(\bq,\vk-\bq)\,\delta_{\rm obs}(\bq)\,\delta_{\rm obs}(\vk-\bq), \\
\mathcal S_{\beta,34}(\vk) &= \int_{\bq}\,
D^{(\beta)}_{2,34}(\bq,\vk-\bq)\,\delta_{\rm obs}(\bq)\,\delta_{\rm obs}(\vk-\bq),
\end{align}
with kernels
\begin{align}
\label{eq:D_3_beta}
D^{(\beta)}_3(\bq_1,\bq_2,\vk-\bq_1-\bq_2)
&=\frac{g_1^\beta(\vq_1)\,g_2^\beta(\vq_2)\,g_3^\beta(\vk-\bq_1-\bq_2)}
{P_{\rm obs}(q_1)\,P_{\rm obs}(q_2)\,P_{\rm obs}(|\vk-\bq_1-\bq_2|)}, \\
\label{eq:D_2_betas}
D^{(\beta)}_{2,12}(\bq,\vk-\bq)
&=\frac{g_1^\beta(\vq)\,g_2^\beta(\vk-\bq)}{P_{\rm obs}(q)\,P_{\rm obs}(|\vk-\bq|)},\qquad
D^{(\beta)}_{2,34}(\bq,\vk-\bq)
=\frac{g_3^\beta(\vq)\,g_4^\beta(\vk-\bq)}{P_{\rm obs}(q)\,P_{\rm obs}(|\vk-\bq|)} \nonumber.
\end{align}
The two types of kurto spectra that capture the trispectrum information on amplitude-like parameters are then defined as 
\begin{align} 
\cP_{3\times 1}^{(\beta)}(k) & = \cP_{\cK_\alpha\tilde \delta}^{(\beta)}(k)\;\equiv\;
\int \frac{d\Omega_{\hat\vk}}{4\pi}\,\big\langle \mathcal K_\beta(\vk)\,\tilde\delta_{\rm obs}^{(\beta)}(-\vk)\big\rangle,
\\
\cP_{2\times 2}^{(\beta)}(k) &= \cP_{\cS_\alpha \cS_\beta}^{(\beta)}(k)\;\equiv\;
\int \frac{d\Omega_{\hat\vk}}{4\pi}\,\big\langle \mathcal S_{\beta,12}(\vk)\,\mathcal S_{\beta,34}(-\vk)\big\rangle.
\end{align}

As with skew spectra, to capture the information of galaxy trispectrum on galaxy biases, amplitudes of primordial power spectrum and trispectrum, and growth rate of structure, a large set of distinct kurto spectra are needed. In this work our aim is more specific: we target the \emph{parity-odd} trispectrum generated by a particular cubic interaction of the inflaton. Accordingly, we restrict attention to the $\cP_{3\times 1}$ and $\cP_{2\times 2}$ whose filters are tailored to the primordial (parity-odd) trispectrum and its propagation to late-time galaxy fields via the linear transfer function. The explicit, parity-sensitive filter choices—and the way they null parity-even contributions by construction—are presented next.

\subsection{Isolating parity-odd trispectrum with kurto-spectra} \label{subsec:kurto_PO}

The recipe that guided construction of CF-spectra in section \ref{subsec:skew_kurto} is simple: build a composite field whose nonlinear kernel $D^{(\alpha)}_n$ matches the mode couplings appearing in the targeted $n$-point function, then compress that information by cross-correlating the CFs with a suitable (filtered) copy of the density and with one another.  For the bispectrum we used scalar quadratic kernels and obtained the skew spectra; for the parity-even trispectrum we extended the same idea to correlations of a cubic CF and original field and to two quadratic CFs, giving the kurto spectra. For noisy data, as reviewed above, matching the ML estimators of bispectrum and trispectrum further required inverse variance weighting of the input observed field. 

We discuss here briefly regarding the loop counting of the kurto spectrum. All the loop counting referred in this paper is in the sense of the power-spectrum-like quantity rather than trispectrum-like. Following~\cite{Abidi:2018eyd}, the leading-order kurto spectrum is a one-loop contribution involving two linear matter power spectra, which does not corresponds to a trispectrum. The next-to-leading order kurto spectrum is a two-loop contribution (three linear power spectra), which corresponds to a tree-level trispectrum.

As described in section \ref{sec:th-parity-eu}, the parity-odd trispectrum is proportional to a pseudo-scalar built from three wave-vectors, e.g.,
$\widehat T^{\rm odd}\!\propto\!(\boldsymbol{q}_1\times\boldsymbol{q}_2)\!\cdot\!\boldsymbol{q}_3$. Therefore, the kurto-spectra estimators sensitive to this parity-odd trispectrum must itself supply another pseudo-scalar so their product is an ordinary scalar and survives angular averaging. One option to achieve this is to consider  
\be
\cP_{2\times 2}^{\rm PO}(k) = \int \frac{d\Omega_k}{4\pi}\langle {\bf V}^{ab}({\bf k}) \cdot {\bf A}^{cd}(-{\bf k}) \rangle,
\ee
that is defined as the scalar correlator of an ordinary vector quadratic field, ${\bf V}^{ab}$, with a pseudo-vector ${\bf A}^{cd}$. An example of such composite fields, which we will use in our analysis, was constructed in \cite{Jamieson:2024mau}
\be\label{eq:VA_conf}
V_i^{ab}(\bx) = \delta^a(\bx)\partial_{i}\delta^b(\bx), 
\qquad
A_i^{cd}(\bx)  = \epsilon_{ijk} \partial_j \partial^{-2} V_k^{cd}(\bx). 
\ee
with their Fourier transforms given by
\be
\label{eq:VA_Fourier}
V^{ab}_i(\mathbf{k})=-i \int_{\mathbf q} (\mathbf{k}-\mathbf q)_i\,f_a(q)\,f_b(|\mathbf k-\mathbf q|)
\,\delta(\mathbf q)\,\delta(\mathbf k-\mathbf q),
\qquad
A^{cd}_i(\mathbf{k})=-\frac{i}{k^{2}}\,[\mathbf k\times {\bf V}^{cd}(\mathbf{k})]_i .
\ee
It is essential to note that the longitudinal part of ${\bf V}_{ab}$ cancels unless the two radial windows inside the same convolution are different: $f_a\neq f_b$ (and similarly $f_c\neq f_d$ for $A^{cd}$).  If we set $f_a=f_b$ the integrand becomes symmetric, $V^{aa}_i\propto k_i$, the curl vanishes, $A^{aa}=0$, and the parity-odd kurto spectrum disappears.  The filters across the two quadratic CFs can be identical, however, as they do not affect the transverse projection. 

In contrast to Ref. \cite{Jamieson:2024mau}, in this work, we consider the matter or halo overdensities as input fields in construction of parity-odd kurto spectra instead of primordial gravitational potential, $\phi$. When relating the halo/matter kurto spectra to primordial parity-odd mode coupling kernel in eq.  \eqref{eq:phi_ng}, we relate the matter density field to $\phi$ via matter transfer function.

To construct the second type of kurto-spectra, i.e., correlation of a pseudo-scalar cubic field (thus sensitive to the parity-odd signature) and the original field (or its filtered version),
\be
\mathcal{P}^{\rm PO}_{3\times1}(k) = \int \frac{d\Omega_k}{4\pi}\langle \delta(\vk_2) \Psi^{abc}(\vk_1)\rangle
\ee

Ref. \cite{Jamieson:2024mau} proposed using the following cubic field
\be\label{eq:cubic_psudoscalar}
\Psi^{abc}(\bx) = \nabla \delta^a(\bx) \cdot \bB^{bc}(\bx) 
\ee
where $\bB^{bc}$ is a quadratic pseudo-vector field, 
\be
\bB^{bc}(\bx) = \nabla \delta^b(\bx)\times \nabla \delta^c(\bx) 
\ee
with their Fourier transforms given by\footnote{Ref. \cite{Jamieson:2024mau} allowed for an additional overall filter for $\bf B$, which in their numerical evaluation of the estimator was set to unity. We therefore, do not write this extra filter in the expression below.}
\begin{align}
\Psi^{abc}(\vk) &= -i \int_{\bq_1} (\vk-\bq_1)_i f_a(|\vk-\bq_1|)\delta(\vk-\bq_1)\bB^{bc}_i(\bq_1),\\
\bB^{bc}_i(\bq_1) &= - \epsilon_{ijk} \int_{\bq_2} \bq_{2,j} (\bq_1-\bq_2)_k f_b(q_2) f_c(|\bq_1-\bq_2|) \delta(\bq_2)\delta(\bq_1-\bq_2).
\end{align}

In what follows, we will use the same scalar filters as Ref. \cite{Jamieson:2024mau}, 
\be 
f_a(k) = k^2, \qquad f_b(k) = f_d(k) = 1, \qquad f_c(k) = k^{-2},
\ee 
and additionally smooth each of the overdensity fields entering the composite fields with a Gaussian filter to limit the contribution of small-scale fluctuations as described in eq.~\eqref{eq:CF_def_Fourier}.

\subsection{Connection to irreducible tensor decomposition}

\noindent The constructions of section ~\ref{subsec:kurto_PO} can be understood more generally in the language of irreducible tensor decomposition. Using the framework developed for the study of galaxy shapes in \cite{Vlah:2019}, we apply it here to the constructed vector fields.  This perspective clarifies why parity-odd kurto spectra vanish in parity-symmetric theories, and shows explicitly which helicity channels of vector and tensor correlators are isolated by our estimators.

\paragraph{Vector case}Consider a vector field $V_i(\vk)$. Expanding in the helicity basis $\{\hat k,\, e^\pm\}$ gives
\be
V_i(\vk) = v^{(0)}(\vk)\,\hat k_i + v^{(+1)}(\vk)\,e^+_i + v^{(-1)}(\vk)\,e^-_i,
\ee
where $m=0,\pm1$ label the longitudinal and transverse helicities.  
Statistical isotropy enforces diagonal correlators in $m$, so the most general two-point function reads
\be
\langle V_i(\vk)V_j(\vk')\rangle = (2\pi)^3 \delta_D(\vk+\vk') 
\Big[ -P_0(k)\hat k_i\hat k_j + P_+(k)(\delta^K_{ij}-\hat k_i\hat k_j) + iP_\times(k)\epsilon_{ijl}\hat k_l\Big].
\ee
Here $P_0$ and $P_+$ are parity-even spectra, while $P_\times$ multiplies the antisymmetric tensor $\epsilon_{ijl}$ and is therefore parity-odd.  

Now consider the composite fields of eq.~\eqref{eq:VA_Fourier}, $V^{ab}$ and its curl-like partner $A^{cd}$. Their scalar contraction,
\be
\cP^{\rm PO}_{2\times2}(k) = \int \frac{d\Omega_k}{4\pi}\,
\langle {\bf V}^{ab}(\vk)\cdot{\bf A}^{cd}(-\vk)\rangle,
\ee
is proportional to $P_\times(k)$; explicitly one finds
\be
\cP^{\rm PO}_{2\times2}(k) \;=\; -\frac{2}{k}\,P_\times(k).
\ee
Thus the parity-odd $2\times2$ kurto spectrum cleanly isolates the antisymmetric, helicity-odd part of the vector correlator, and vanishes identically in parity-symmetric theories.

\paragraph{Tensor case}The same logic applies to higher-rank tensors \cite{Vlah:2019}. A general symmetric rank-2 tensor field $\Pi_{ij}(\vk)$ can be decomposed into irreducible SO(3) components:  
\be
\Pi_{ij}(\vk) = \tfrac{1}{3}\delta^K_{ij}\,\Pi^{(0)}_0(\vk)
\,+\, \sum_{m=-2}^2 \Pi^{(m)}_2(\vk)\,Y^{(m)}_{ij}(\hat k),
\ee
where $\Pi^{(0)}_0$ is the scalar trace ($\ell=0$) and $\Pi^{(m)}_2$ are the traceless rank-2 pieces ($\ell=2$) expanded in a tensor harmonic basis $Y^{(m)}_{ij}$.  

Statistical isotropy implies that different helicity modes do not mix,
\be
\langle \Pi^{(m)}_\ell(\vk)\,\Pi^{(m')}_{\ell'}(\vk')\rangle
= (2\pi)^3\delta_D(\vk+\vk')\,\delta^K_{mm'}\,P^{(m)}_{\ell\ell'}(k).
\ee
Under parity, helicities transform as $m\mapsto -m$, which enforces the equality $P^{(+m)}_{\ell\ell'}(k)=P^{(-m)}_{\ell\ell'}(k)$ in parity-invariant theories. Departures from this equality define parity-odd spectra. 

In the rank-2 case, two independent antisymmetric combinations can appear: the $m=\pm1$ and $m=\pm2$ sectors, yielding two parity-odd spectra $P_d^\times(k)$ and $P_e^\times(k)$. These are the tensorial analogs of the $P_\times(k)$ component identified in the vector case. Physically, they correspond to the helical parts of spin-1 and spin-2 modes, respectively. For the associated tensorial basis operators, we refer the reader to Appendix A of \cite{Vlah:2019}. Explicit expressions for the tensor harmonics $Y^{(m)}_{ij}$ and their transformation properties are also provided in Appendix~\ref{append:helicity_basis}.

From this perspective, the parity-odd kurto spectra are scalar contractions that project onto the helicity-odd sectors. The $\cP_{2\times2}$ estimator isolates the antisymmetric $P_\times$ channel of a vector correlator, while the $\cP_{3\times1}$ estimator, built from the pseudoscalar cubic field $\Psi^{abc}$, provides a direct scalar probe of parity-odd pseudoscalar contractions. Together, they form complementary observables for isolating parity violation.

Although in this paper we focus on scalar observables (density fields and their composite spectra), the irreducible-tensor decomposition provides a general framework within which all possible parity-odd two-point spectra can be systematically classified by their helicity structure. In this language, the parity-odd kurto spectra appear as the simplest scalar examples.

\section{Fast Theory Calculation using FFTLog Algorithm}\label{sec:th-FFTLog}

For a separable cubic mode coupling kernel of the form in eq. \eqref{eq:K3_Coulton}, the parity-odd kurto spectra can be efficiently computed using the FFTLog algorithm \cite{Hamilton2000,Schmittfull:2016jsw,McEwen:2016fjn,Simonovic:2017mhp}. We present elements of this fast computation in this section. After setting up the generalities, we discuss the parity-odd 2x2 and 3x1 kurto spectra separately in two subsections.

Given the relation of matter fluctuations to primordial gravitational potential through the transfer function, the parity-odd cubic mode coupling in eq.~\eqref{eq:cubic_phi} produces a parity-odd matter density field, which we denote by $\delta_p$, and is given by $\delta_p(\vk,z) = \cM(k,z) \phi_{\rm NG}(\vk)$. Assuming this primordial parity-odd non-Gaussianity does not modify biasing relation of the tracers like halos and galaxies (i.e., no new bias operator is introduced), the induced parity-odd trispectrum of matter fluctuations also gets directly imprinted on halo/galaxy overdensities since (at linear order) we have $\delta_h^{\rm PO}(\vk, z) = b_1(z) \delta_p(\vk,z)$. To make the notation more concise, from here on, we drop the explicit redshift dependencies. We note that the calculation done in this section does not use the optimal estimator, discussed in Sec.~\ref{subsec:skew_kurto}.

\subsection{Separable non-linear kernels} 

Describing a general separable cubic mode coupling kernel that involves a pseudo-scalar as
\be
\label{eq:K3_separable_general}
	K^{(3)}_{\rm NL, PO}(\vk_1,\vk_2,\vk_3) = \epsilon_{ijk} A_i(\vk_1)B_j(\vk_2)C_k(\vk_3),
\ee
and using the Fourier integral representation of Dirac delta 
\be
\delta_D(\vk) = \frac{1}{(2\pi)^3} \int d^3 x \  e^{i\vk\cdot{\vec x}}
\ee
the cubic-order parity-violating non-Gaussianity of primordial fluctuations is given by    
\begin{align} \label{eq:cubic_phi}
\phi_{\rm NG}(\vk) & =i \gnl^{\rm PO} \epsilon_{ijk} \int d^3 x \ e^{-i\vk\cdot{\vec x}} \notag \\
&\hspace{1in} \times \int_{\bq_1}e^{i{\vec x}\cdot \vq_1} A_i(\vq_1) \phi_{\rm G}(\bq_1)\int_{\bq_2} e^{i{\vec x}\cdot \vq_2} B_j(\vq_2) \phi_{\rm G}(\bq_2) \int_{\bq_3} e^{i{\vec x}\cdot \vq_3} C_k(\vq_3)   \phi_{\rm G}(\bq_3) \notag \\
& =   i \gnl^{\rm PO} \epsilon_{ijk} \int d^3 x \ e^{-i\vk\cdot{\vec x}} {\tilde A}_i({\vec x)} {\tilde B}_j({\vec x)} {\tilde C}_k({\vec x)}
\end{align}
where we defined the tilde fields as 
\be
\{{\tilde A}, {\tilde B}, {\tilde C}\}_i({\vec x})=  \int_{\bq}e^{i{\vec x}\cdot \vq} \ \{A,B,C\}_i(\vq) \phi_{\rm G}(\vq)
\ee

The form of the cubic kernel in eq. \eqref{eq:K3_Coulton} has an additional simplification, which is that $\{A_i,B_j,C_k\}$ have collinear forms, i.e.,
\be\label{eq:colinear}
    \{A, B, C\}_i(\vec q) = \{a(q),b(q),c(q)\} q_i \, .
\ee
This gives us 
\be
   {\{\tilde A, \tilde B, \tilde C\}}_i(\vec x) 
    = - i \partial_i \int_{\vec q} e^{i \vec x \cdot \vec q}  \{a(q),b(q),c(q)\}  \phi_{\rm G}(\vec q) \non\\
    = - i \partial_i \{a(\vec x),b(\vec x),c(\vec x)\}.
\ee
Therefore, eq. \eqref{eq:cubic_phi} reduces to

\be
\phi_{\rm NG}(\vk) =  - \gnl^{\rm PO} \epsilon_{ijk} \int d^3 x \ e^{-i\vk\cdot{\vec x}} \partial_i a({\vec x}) \partial_j b({\vec x}) \partial_kc({\vec x}).
\ee

We note that the $A_i$ defined here is different from the composed vector field $A_i^{cd}$ defined in eq.~\eqref{eq:VA_conf}.

\subsection{The $\mathbf{2\times 2}$ parity-odd kurto-spectrum}\label{subsec:2x2}

As discussed in section \ref{subsec:kurto_PO}, the first estimator sensitive to parity-odd component of the trispectrum, the $2\times 2$ kurto spectrum, is obtained from correlation of a vector and a pseudo-vector fields. Using the two composite-fields in eq. \eqref{eq:VA_Fourier}, their cross correlation in Fourier space is given by

\eq{\label{eq:VA_original}
    \langle V^{ab}(\vk) \cdot A^{cd}(\vk')\rangle &= \epsilon_{ijk} \frac{ik'_j}{k'^2}\int d^3 x \ d^3 x' \ e^{-i\vec x \cdot \vec k - i\vec x' \cdot \vec k'} \langle \delta^{a}(\vec x) \partial_i \delta^{b}(\vec x) \delta^{c}(\vec x') \partial_k \delta^{d}(\vec x')  \rangle.
}
As discussed in section \ref{subsec:kurto_PO}, the two fields entering the definition of each of the quadratic fields above, have to be differently filtered in order to have nonzero correlator $\langle V^{ab}_i A^{a'b'}_i\rangle$. Here, we assumed a general case of applying four distinct filters to the four fields. 

To understand the nonzero contributions to the above correlator, it is helpful to use perturbation theory power counting. Lets consider the fields entering the correlator to be matter overdensities to simplify the discussion. Extension to biased tracers is straightforward. Up to 2-loop order, perturbative description of the correlator in eq. \eqref{eq:VA_original} requires expansion of matter density field up to third order. Since parity-odd contributions to matter density field appear at third order and higher, we have
\be\label{eq:deltam_cubic}
\delta_{\rm m}^{\rm NL}(\vk) = \delta_{\rm m}^{(1)}(\vk) + \delta_{\rm m}^{(2)}(\vk) + \left[\delta_{\rm m, PE}^{(3)}(\vk) + \delta_{\rm m, PO}^{(3)}(\vk)\right],
\ee
where the subscripts PO and PE denotes the parity-odd and parity-even third order matter field. In addition to the 1-loop contribution, there would be two 2-loop contributions which we refer to as $\langle 2211 \rangle$ and $\langle 3111 \rangle$, the former is parity-even and the latter has both parity-odd and parity-even components. Given the choice of the filter in constructing $A_i^{cd}$ field, which involves a Levi-Civita symbol, the parity-even correlators vanish (the correlator is nonzero only if we have odd powers of Levi-Civita symbols in the integrand of eq. \eqref{eq:VA_original}. At 3-loop level, the only parity-odd contribution is $\langle 3221 \rangle$, while at 4-loop order only $\langle 3331 \rangle$ is nonzero. As we will see in section \ref{subsubsec:ept_field}, the 2-loop parity-odd contribution is the dominant one if large displacement contributions are correctly resummed. 

In summary, the leading-order contribution to $\cP_{2\times 2}^{\rm PO}$ involves one factor of cubic parity-odd field, $\delta_p$. Among the four permutations in the correlator on the right-hand of eq. \eqref{eq:VA_original}, $\delta_p$ can appear with no or one derivative. To illustrate our FFT-based implementation of the 2x2 kurto spectra, let us detail the derivation steps for leading-order contributions to one of these permutations.  
\eq{ \label{eq:2x2_fft_firstperm}
    \langle \delta_{p}^{a}(\vec x) &\partial_i \delta_{L}^{b}(\vec x) \delta_{L}^{c}(\vec x') \partial_k \delta_{L}^{d}(\vec x')  \rangle \nn \\
    & = i \gnl^{\rm PO}\epsilon_{lmn} \int_{\vec p} e^{i\vec p\cdot \vec x} f_a(\vec p) \cM(p)  \int d^3 r \  e^{-i\vec r\cdot \vec p} \langle \tilde{A}_l(\vec r) \tilde{B}_m(\vec r) \tilde{C}_n(\vec r) \partial_i \delta_{L}^{b}(\vec x) \delta_{L}^{c}(\vec x') \partial_k \delta_{L}^{d}(\vec x')\rangle \nn \\
    & =  i \gnl^{\rm PO} \epsilon_{lmn} \int d^3 r \  F_a(\vec x - \vec r) \left( \langle \tilde{A}_l(\vec r)\partial_i \delta_{L}^{b}(\vec x)\rangle \langle \tilde{B}_m(\vec r) \delta_{L}^{c}(\vec x')\rangle \langle \tilde{C}_n(\vec r) \partial_k \delta_{L}^{d}(\vec x')\rangle  + \rm perms \right), 
}
where $\delta_L$ is a filtered linear matter overdensity field, and we have defined 
\be \label{eq:F_a}
F_a(\vec u) = \int_{\vec p} e^{i\vec p\cdot \vec u} f_a(p) \cM(p) = \frac{1}{2\pi^2}\int p^2 dp \ j_0(pu)f_a(p)\cM(p).
\ee 
Using eq. \eqref{eq:colinear}, the three correlators on the right side of eq. \eqref{eq:2x2_fft_firstperm} can be represented with a general form of (replacing $\{\tilde A,a\}$ with $\{\tilde B,b\}$ and $\{\tilde C,c\}$),
\eq{
    \langle \tilde{A}_l(\vec r)\partial^n_i \delta_{L}^{b}(\vec x)\rangle &= \int_{\vec k_1, \vec k_2} e^{i\vec k_1\cdot \vec r} e^{i\vec k_2\cdot \vec x} i k_{1,l} (ik_{2,i})^n a(k_1) f_b(k_2) \langle \phi_{\rm G}(\vec k_1)\delta_{L}(\vec k_2) \rangle \non \\
    & = i^{n+1}\int_{\vec k_1} e^{i\vec k_1\cdot (\vec r-\vec x)} k_{1,l} (-k_{1,i})^n a(k_1) f_b(k_1) \frac{P_{L}(k_1)}{\cM(k_1)}, \nn \\
    &\label{eq:corr_tensor_1_2}\equiv 
    \begin{cases}
    \xi_{l}^{{\tilde A}b}(\vec r - \vec x), \qquad n=0 \\
    \xi_{li}^{{\tilde A}b}(\vec r - \vec x), \qquad n=1
    \end{cases}
}
Therefore, the contribution of the permutation in eq. \eqref{eq:2x2_fft_firstperm} to the 2x2 spectrum in eq.~\eqref{eq:VA_original} is given by 
\eq{
    \langle V^{ab}(\vk) \cdot A^{cd}(\vk')\rangle 
    & = -(2\pi)^3\delta_{\rm D}(\vec k + \vec k')  i \gnl^{\rm PO} \epsilon_{ijk}\epsilon_{lmn} \frac{ik_j}{k^2} \nn \\&\times \int d^3 {\tilde x} \ d^3 {\tilde x}' \  e^{-i\tilde{\vec x}  \cdot \vec k + i \tilde{\vec x}' \cdot \vec k'}  F_a(\tilde{\vec x}) \left( \xi_{li}^{{\tilde A}b}(-\tilde{\vec x} ) \xi_{m}^{{\tilde B}c}(- \tilde{\vec x}')  \xi_{nk}^{{\tilde C}d}(-\tilde{\vec x}')  + \rm perms \right).
}
where we have done a change of variable $\tilde{\vec x} = \vec x - \vec r$ and $\tilde{\vec x}' = \vec x' - \vec r$. The Dirac delta is the result of integration over $r$. For simplicity, in the rest of the calculation, we drop the tilde symbols.  
	
In general, since statistical homogeneity enforces all correlators to depend only on the relative separation, rotational invariance constrains the allowed tensor structures: a scalar correlator depends only on the distance $x$; a vector correlator 
with one free index must align with the unit separation vector $\hat x_i$; and a rank-2 tensor 
correlator decomposes uniquely into a trace part proportional to $\delta_{ij}$ and a traceless 
symmetric part built from $\hat x_i \hat x_j - \delta_{ij}/3 $. The above expression can therefore be simplified by decomposing the tensor and vector correlators  into their irreducible components under spatial rotations. In particular, the tensor correlator 
takes the form
\eq{\label{eq:corr_tensor}
    \xi^{{\tilde A}b}_{ij} (\vec x) \;=\;  \delta^{\rm K}_{ij}\,\xi^{{\tilde A}b}_{0}(x) 
    + \Big( \hat x_i \hat x_j - \frac{1}{3}\,\delta^{\rm K}_{ij}\Big)\,\xi^{{\tilde A}b}_{2}(x) \, ,
}
where the monopole and quadrupole components are given by
\eq{
    \xi^{{\tilde A}b}_{0}(x) &= \frac{1}{3}\, \delta^{\rm K}_{ij}\,\xi^{{\tilde A}b}_{ij}(\vec x) 
    \;=\; \frac{1}{3}\int \frac{k^2 dk}{2\pi^2}\, k^2\,a(k)f_b(k)\,\frac{P_{L}(k)}{\cM(k)}\, j_0(kx), \\
    \xi^{{\tilde A}b}_{2}(x) &= \frac{3}{2}\,\Big(\hat x_i \hat x_j - \frac{1}{3}\,\delta^{\rm K}_{ij}\Big)\,\xi^{{\tilde A}b}_{ij}(\vec x) 
    \;=\; -\int \frac{k^2 dk}{2\pi^2}\, k^2\,a(k)f_b(k)\,\frac{P_{L}(k)}{\cM(k)}\, j_2(kx).
}
Similarly, the vector correlator reduces to a single dipole structure along the separation vector,
\eq{
    \xi_{j}^{{\tilde A}b}(\vec x) &= \hat{x}_j\, \xi_1^{{\tilde A}b}(x), \\
    \xi_1^{{\tilde A}b}(x) &= -\int \frac{k^2 dk}{2\pi^2}\, k\,a(k)f_b(k)\,\frac{P_{L}(k)}{\cM(k)}\, j_1(kx).
}

Defining $\xi_{02}^{{\tilde A}b} = \xi_0^{{\tilde A}b} - \frac{1}{3}\xi_2^{{\tilde A}b}$, we write 
\eq{
    &\mathcal{P}_{2\times 2}^{\rm PO, type1}(k) = - i( i \gnl^{\rm PO})\int_{\vec x, \vec x'} e^{-i \vec x \cdot \vec k + i \vec x' \cdot \vec k} F_a(x)  \nn \\
    &\hspace{0.2in} \times \epsilon_{ijk}\epsilon_{lmn} \lb \xi_{02}^{{\tilde A}b}(x) \delta_{li}^{\rm K} + \hat{x}_i\hat{x}_l \xi_2^{{\tilde A}b}(x)  \rb  \frac{k_j}{k^2} \hat{x}'_m \xi_1^{{\tilde B}c} (x') \lb \xi_{02}^{{\tilde C}d}(x') \delta_{nk}^{\rm K} + \hat{x}'_n\hat{x}'_k \xi_2^{{\tilde C}d}(x') \rb + \mathrm{perms} \nn \\
    & \hspace{0.2in} = - i ( i \gnl^{\rm PO}) \int_{\vec x, \vec x'} e^{-i \vec x \cdot \vec k + i \vec x' \cdot \vec k} F_a(x)\non\\
    &\hspace{1cm}\times \Biggl( 2\frac{\vec k \cdot \hat{x}'}{k^2} \xi_{02}^{{\tilde A}b}(x)\xi_1^{{\tilde B}c} (x')\xi_{02}^{{\tilde C}d}(x') + \left[ \frac{\vec k \cdot \hat{x}'}{k^2} - \frac{\vec k \cdot \hat{x}}{k^2}(\hat{x}\cdot \hat{x}') \right] \xi_2^{{\tilde A}b}(x)\xi_1^{{\tilde B}c}(x')\xi_{02}^{{\tilde C}d}(x')\Biggr) + \mathrm{perms} \nn \\
    & \hspace{0.2in} = ( \gnl^{\rm PO}) \frac{(4\pi)^2}{k}\left[\int x^2 dx j_0(kx) F_a(x) (2\xi_{02}^{{\tilde A}b}(x) + \xi_{2}^{{\tilde A}b}(x)) \int  x'^2 dx' j_1(kx') \xi_1^{{\tilde B}c}(x') \xi_{02}^{{\tilde C}d} (x') \right. \nn\\
    & \left. \hspace{0.2in} -\int x^2dx F_a(x)\lb \frac{1}{3}\xi_2^{{\tilde A}b}(x) j_0(kx) - \frac{2}{3} \xi_2^{{\tilde A}b}(x) j_2(kx) \rb \int  x'^2dx' \xi_1^{{\tilde B}c}(x')\xi_{02}^{{\tilde C}d}(x') j_1(kx')\right] + \mathrm{perms} 
}
Therefore, we have 
\begin{empheq}[box=\fbox]{align} \label{eq:P_2x2_type1}\mathcal{P}_{2\times 2}^{\rm PO, type1}(k) & = \frac{(4\pi)^2}{k} (\gnl^{\rm PO}) \int  x^2 dx F_a(x) \lb 2\xi_{0}^{{\tilde A}b}(x)j_0(kx) + \frac{2}{3}\xi_{2}^{{\tilde A}b}(x) j_2(kx) \rb \nn \\
&\times \int x'^2 dx' j_1(kx') \xi_1^{{\tilde B}c}(x') \xi_{02}^{{\tilde C}d}(x') + \mathrm{perms.}
\end{empheq}

\vspace{0.2in}
The derivation of the contribution from the other permutation, in which the gradient acts on the parity-odd field, is given by
\eq{
    &\langle \delta_L^{a}(\vec x)\, \partial_i \delta_p^{b}(\vec x)\, 
      \delta_L^{c}(\vec x')\, \partial_k \delta_L^{d}(\vec x') \rangle  \non \\
    &\hspace{0.2in} = \epsilon_{lmn} \int d^3 r \; 
      F_{b,i}(\vec x - \vec r)\, 
      \Big( \langle \tilde{A}_l(\vec r)\, \delta_L^{a}(\vec x)\rangle \,
            \langle \tilde{B}_m(\vec r)\, \delta_L^{c}(\vec x')\rangle \,
            \langle \tilde{C}_n(\vec r)\, \partial_k \delta_L^{d}(\vec x')\rangle 
      + \mathrm{perms} \Big),
}
where we have defined
\be \label{eq:F_b_i}
    F_{b,i}(\vec u) \;=\; -\hat{u}_i \int \frac{p^2\,dp}{2\pi^2}\, p\, f_b(p)\,\cM(p)\, j_1(pu).
\ee 
The evaluation follows analogously to the previous case, so we do not repeat the details and only quote the final expression:
\begin{empheq}[box=\fbox]{align} 
\label{eq:P_2x2_type2}
    \mathcal{P}_{2\times 2}^{\rm PO, type2}(k)  
    &= \frac{(4\pi)^2}{k} \gnl^{\rm PO} \int x^2 dx \; F_{b,1}(x)\,
        \left[ \frac{2}{3}\,\xi_{1}^{{\tilde A}a}(x)\, j_0(kx) + \frac{2}{3}\,\xi_{1}^{{\tilde A}a}(x)\, j_2(kx) \right] \nn \\
    &\quad \times \int x'^2 dx' \; \xi_{1}^{{\tilde B}c}(x)\, \xi_{02}^{{\tilde C}d}(x')\, j_1(kx') 
      + \mathrm{perms}.
\end{empheq}
The final result for $\mathcal{P}_{2\times 2}^{\rm PO}$ is obtained by summing over all permutations of the two kinds.

\subsection{The $\mathbf{3\times 1}$ parity-odd kurto-spectrum}\label{subsec:3x1}

The second estimator of trispectrum is obtained by correlating a cubic composite-field and another density field. In order to be parity-sensitive, the cubic field has to be a pseudo‑scalar as discussed in section \ref{subsec:kurto_PO}. 
Using the cubic pseudo-scalar given in eq. \eqref{eq:cubic_psudoscalar} its correlator with the input field defining the $3 \times 1$ kurto spectrum is given by
\be\label{eq:3x1}
\langle \Phi^{abc}(\vec{k}) \delta(\vec{k}') \rangle = -i \epsilon_{ijk} \int_{\vec{x}, \vec{x}'} e^{-i\vec{k}\cdot \vec{x} - i \vec{k}'\cdot \vec{x}'} \langle \partial_i \delta^{a} (\vec{x}) \partial_j \delta^b(\vec{x}) \partial_k \delta^c(\vec{x}) \delta(\vec{x}') \rangle.
\ee
As for the $2\times 2$ kurto spectrum, there are two types of contributions to the correlator on the right side of eq. \eqref{eq:3x1} depending on whether or not a derivative is applied on the parity-odd field. In the first case (permutation with $\partial_i \delta_p$), we have
\begin{align} 
    &\langle \partial_i \delta_p^{a} (\vec{x}) \partial_j \delta_L^b(\vec{x}) \partial_k \delta_L^c(\vec{x}) \delta_L(\vec{x}') \rangle  \nn \\
    &\hspace{0.2in} = (i \gnl^{\rm PO}) \epsilon_{lmn}  \partial_i\int_{\vec p} e^{i\vec p\cdot \vec x} f_a(\vec p) \cM(p) \int d^3 r\ e^{-i\vec r\cdot \vec p} \langle \tilde{A}_l(\vec r) \tilde{B}_m(\vec r) \tilde{C}_n(\vec r) \partial_j \delta_L^{b}(\vec x) \partial_k \delta_L^{c}(\vec x) \delta_L(\vec x')\rangle \nn \\ 
    & \hspace{0.2in} = (i \gnl^{\rm PO}) \epsilon_{lmn} \int d^3 r\ F_{a,i}(\vec x - \vec r) \left( \langle \tilde{A}_l(\vec r)\partial_j \delta_L^{b}(\vec x)\rangle \langle \tilde{B}_m(\vec r) \partial_k\delta_L^{c}(\vec x)\rangle \langle \tilde{C}_n(\vec r) \delta_L(\vec x')\rangle  + \rm perms \right). 
\end{align}
where $F_{a,i}$ is given by eq. \eqref{eq:F_b_i}. Therefore, the contribution of this permutation to the $3\times 1$ kurto spectrum is given by
\begin{empheq}[box=\fbox]{align} 
\label{eq:P_3x1_type1}
    &\mathcal{P}_{3\times 1}^{\rm PO, type1} = 2 (4\pi)^2  \gnl^{\rm PO} \int x^2 dx F_{a,1}(x) \xi_{02}^{{\tilde A}b}(x) \xi_{02}^{{\tilde B}c}(x) j_1(kx) \int x'^2 dx' j_1(kx') \xi_1^{\tilde C}(x').
\end{empheq}

For the second type of permutations in which the parity-odd field appears with no derivative, we have 

\begin{align} 
    &\langle \partial_i \delta_L^{a} (\vec{x}) \partial_j \delta_L^b(\vec{x}) \partial_k \delta_L^c(\vec{x}) \delta_p(\vec{x}') \rangle \nn \\
    & \hspace{0.2in} = (i \gnl^{\rm PO}) \epsilon_{lmn} \int_{\vec r} \cM(\vec x' - \vec r)\left( \langle \tilde{A}_l(\vec r)\partial_i \delta_L^{a}(\vec x)\rangle \langle \tilde{B}_m(\vec r) \partial_j\delta_L^{b}(\vec x)\rangle \langle \tilde{C}_n(\vec r) \partial_k\delta_L^c(\vec x)\rangle  + \rm perms \right).
\end{align}
which contributes to the $3\times 1$ kurto spectrum as
	\begin{empheq}[box=\fbox]{align} 
    \label{eq:P_3x1_type2}
		\mathcal{P}_{3\times 1}^{\rm PO, type2} &= (4\pi)^2\gnl^{\rm PO}  \int x'^2 dx' \cM(x') j_0(kx') \int x^2 dx j_0(kx) \left[ 6 \xi_{02}^{{\tilde A}a} (x) \xi_{02}^{{\tilde B}b}(x)\xi_{02}^{{\tilde C}c}(x) \right.   \\
        & \hspace{1cm}\left. + 2 \left(\xi_2^{{\tilde A}a}(x)\xi_{02}^{{\tilde B}b}(x) \xi_{02}^{{\tilde C}c}(x) + \xi_{02}^{Aa}(x)\xi_2^{{\tilde B}b}(x)\xi_{02}^{{\tilde C}c}(x) + \xi_{02}^{{\tilde A}a}(x)\xi_{02}^{{\tilde B}b}(x)\xi_2^{{\tilde C}c}(x)  \right)\right] \nn
	\end{empheq}

\vspace{-0.4in}    
\subsection{Numerical implementation}
\label{subsec:numerical_implementation}

We evaluate the correlation functions entering eqs.~\eqref{eq:P_2x2_type1}, 
\eqref{eq:P_2x2_type2}, \eqref{eq:P_3x1_type1}, and \eqref{eq:P_3x1_type2} 
using the FFTLog algorithm~\cite{Hamilton2000}, which performs fast Hankel 
(Fourier–Bessel) transforms on logarithmically spaced grids. The general 
transform takes the form
\begin{align}
\xi_\ell(x) &= \int \frac{k^2 dk}{2\pi^2} f(k) j_{\ell} (kx), \nn \\
&= \sum_i c_i \int \frac{k^2 dk}{2\pi^2} k^{\nu_i} j_{\ell} (kx), \non \\
&=\pi^{-1.5}\sum_i c_i \frac{1}{x^{3+\nu_i}} 2^{\nu_i} \frac{\Gamma(\frac{1}{2}(\ell+3+\nu_i))}{\Gamma(\frac{1}{2}(\ell-\nu_i))},
\label{eq:fftlog-master}
\end{align}
where $f(k)$ has been expanded on a log grid as a sum of complex power laws 
$k^{\nu_i}$ with $\nu_i=b+i\eta_i$, and $b$ is the FFTLog bias. The coefficients 
$c_i$ are obtained by FFT in $\ln k$ of $k^{-b}f(k)$; the analytic Hankel 
transform of each power law then yields last line of eq.~\eqref{eq:fftlog-master}.

In practice, we proceed as follows. First, we choose a $k$ range 
$[k_{\min},k_{\max}]$, number of samples $N$ (here $N=512$), and a bias $b$ 
chosen so that $k^{-b}f(k)$ is approximately periodic across the interval, i.e. 
$k_{\min}^{-b}f(k_{\min})\approx k_{\max}^{-b}f(k_{\max})$. We then sample 
$f(k)$ on a logarithmic grid 
$k_n=k_{\min}\exp(n\Delta)$ with $\Delta=\ln(k_{\max}/k_{\min})/N$, form 
$g(k)=k^{-b}f(k)$, and compute its FFT to obtain amplitudes $c_i$ at frequencies 
$\eta_i=2\pi i/(N\Delta)$. The correlation functions $\xi_\ell(x)$ are evaluated 
on the corresponding log-$x$ grid using eq.~\eqref{eq:fftlog-master}. We then build the required real-space products (such as the cubic combinations in 
eq.~\eqref{eq:P_3x1_type2}), which we denote by $\Xi_\ell(x)$, and inverse-transform them back to 
$k$ space with the same FFTLog machinery:
\[
F_\ell(k)=4\pi\int x^2\,\mathrm{d}x\,\Xi_\ell(x)\,j_\ell(kx),
\]
again by expanding $x^{-b_x}\Xi_\ell(x)$ on a log grid and applying FFT.

Special care is required to control numerical precision and edge effects. We 
apply smooth tapers at both ends of the $k$-range: a Gaussian at high $k$ to 
suppress ringing and a smooth cutoff at low $k$ when the integrand is 
infrared-divergent. The latter is implemented as
\[
W_{\rm IR}(k)=\frac{1}{2}\Bigl[1+\tanh\!\Bigl(\alpha\,\ln\!\frac{k}{k_{\min}^{\rm IR}}\Bigr)\Bigr],
\]
with steepness $\alpha\gg1$. In particular, the correlation function 
$\xi^{\tilde A c}_1$ that enters eqs.~\eqref{eq:P_2x2_type1} and 
\eqref{eq:P_2x2_type2} receives factors of $k^{-2}$ from $a(k)$ and $f_c(k)$, 
together with $1/\cM(k)$, leading to a divergent integrand at low $k$; 
multiplication by $W_{\rm IR}$ regularizes the integral. The cutoff 
$k_{\min}^{\rm IR}$ is chosen slightly above the simulation fundamental mode 
$k_f$; in practice $k_{\min}^{\rm IR}\approx 1.24\,k_f$ yields stable results. 
After forming the configuration-space products, we lightly taper the edges of 
$\Xi_\ell(x)$ before the inverse transform to further reduce wrap-around.

With $N=512$ and the \texttt{pyfftlog} package \faGithub\footnote{\url{https://github.com/emsig/pyfftlog}}, 
a single FFTLog call takes $\sim 11\,\mathrm{ms}$ on an Apple Silicon M2. The 
$\mathcal{P}_{2\times 2}$ calculation requires 87 Hankel transforms 
($\sim 0.87\,\mathrm{s}$), and $\mathcal{P}_{3\times 1}$ requires 79 
($\sim 0.79\,\mathrm{s}$).

\section{Comparison with Simulations}\label{sec:sims} 

In this section, we present the measurement of parity-odd kurto spectra on simulations. We employ two sets of simulations: (i) a perturbative DM simulation based on Eulerian perturbation theory (EPT), and (ii) \texttt{Quijote-ODD} suite~\cite{Coulton:2024}, an N-body simulation with a specific parity-violating model encoded in the initial conditions and evolved according to Newtonian dynamics. The EPT simulation is designed to study in detail the effects of gravitational non-linearity, disentangling different perturbative contributions to the kurto spectra. In this case, we restrict our analysis to the $\mathcal{P}_{2\times 2}$ kurto spectra, since we expect the main conclusions about the impact of parity-even trispectrum contributions as a source of noise to be qualitatively similar for both observables, with differences arising only from the distinct perturbative terms involved. For the \texttt{Quijote-ODD} suite, we measure kurto spectra on both the DM field and the halo field to assess detectability in conditions closer to real surveys. In this case, we analyze both $\mathcal{P}_{2\times 2}$ and $\mathcal{P}_{3\times 1}$ and compare their respective behaviors. We also contrast the estimators measured in real space and in redshift space, including redshift-space distortions (RSD). Finally, we present the signal-to-noise ratio of the observables measured in the simulations and provide forecasts for their detectability in the Euclid spectroscopic sample. We also note that the $\cP_{2\times 2 }$ and $\cP_{3\times 1}$ has a dimension of $\left(\mathrm{Mpc}/h \right)^{12}$, in all the following figures, for simplicity, we do not explicitly show the dimension.

\subsection{DM field}\label{subsec:dm_field}

We first compare the kurto spectra of DM field on the EPT and \texttt{Quijote-ODD} simulations.

\subsubsection{EPT field}\label{subsubsec:ept_field}

We generate the EPT field using the Python package \textbf{\texttt{nbodykit}} \faGlobe\footnote{\url{https://nbodykit.readthedocs.io/en/latest/}}. The field is constructed in a cubic box of side length $1\ \mathrm{Gpc}/h$ with a $512^3$ mesh grid, matching the resolution of the \texttt{Quijote-ODD} simulations (see next subsection for specifications). We begin by generating the linear matter overdensity field at redshift $z$, $\delta^{(1)} = \delta_L$, from a Gaussian random field whose power spectrum corresponds to that of DM in the cosmology used for the \texttt{Quijote-ODD} suite. Second-order parity-even fields are then generated via the GridSPT algorithm~\cite{Taruya:2018jtk}, including contributions from the linear density field, contraction of displacement and derivative of linear density fields, and the tidal field. We also investigate the second-order field without the displacement contribution as described below. For the third-order field, we retain only the parity-odd cubic contribution in eqs.\eqref{eq:phi_ng}, \eqref{eq:K3_Coulton}, \eqref{eq:deltam}, neglecting the gravitationally induced contribution (i.e., dropping the $\delta_{\rm m, PE}^{(3)}$ term in eq.\eqref{eq:deltam_cubic}). Finally, we extend the skew-spectra package \textbf{\texttt{skewspec}} \faGithub\footnote{\url{https://github.com/mschmittfull/skewspec.git}} to measure the parity-odd kurto spectra. 

We pedagogically vary the second-order kernel to study the impact of its different components on $\mathcal{P}_{2\times 2}$. Specifically, we consider four scenarios (expressed in configuration space)
\begin{align}\label{eq:2nd}
    {\rm Growth}:& \quad \left[\delta^{(1)}\right]^2, \nn \\
     {\rm Tidal}:& \quad \frac{3}{2}\left(\frac{\partial_i\partial_j}{\nabla^2}\delta^{(1)}\right)\left(\frac{\partial_i\partial_j}{\nabla^2}\delta^{(1)}\right) - \frac{1}{2}\left[\delta^{(1)}\right]^2, \nn \\
     {\rm No \ displacement}:& \quad \frac{5}{7}\left[\delta^{(1)}\right]^2 + \frac{2}{7}\left(\frac{\partial_i\partial_j}{\nabla^2}\delta^{(1)}\right)\left(\frac{\partial_i\partial_j}{\nabla^2}\delta^{(1)}\right), \nn \\
     {\rm Full} \ F_2:& \quad \frac{5}{7}\left[\delta^{(1)}\right]^2 -\left(\frac{\partial_i}{\nabla^2}\delta^{(1)}\right)\partial_i \delta^{(1)} + \frac{2}{7}\left(\frac{\partial_i\partial_j}{\nabla^2}\delta^{(1)}\right)\left(\frac{\partial_i\partial_j}{\nabla^2}\delta^{(1)}\right).
\end{align}

We clarify our notation below. For the kurto spectra, we consider correlators of four overdensity fields, $\langle \delta\delta\delta\delta\rangle$. Labeling the linear, non-linear (second-order parity-even), and parity-odd fields as $1$, $2$, and $3$, respectively, to isolate the dominant source of noise and signal, we measure five versions of the kurto spectrum for each second-order kernel scenario:
 \begin{align}\label{eq:correlators}
     {\rm total}:& \quad \la (1+2+3)(1+2+3)(1+2+3)(1+2+3) \ra \nn \\
     {\rm tot}12:& \quad \la(1+2)(1+2)(1+2)(1+2)\ra \nn \\
     {\rm tot}13:& \quad \la(1+3)(1+3)(1+3)(1+3)\ra \nn \\
     \la 3111 \ra:& \quad {\rm expected \ leading \ order \ (LO) \ signal} \nn \\
     \la 3221 \ra:& \quad {\rm expected \ next \ to \ leading \ order \ \ (NLO) \ signal}
 \end{align}

\begin{figure}[t]
  \centering
  \begin{subfigure}{\linewidth}
    \centering
    \makebox[\linewidth][r]{%
      \hspace*{\labelgutter}%
      \begin{overpic}[height=0.42\textheight] 
        {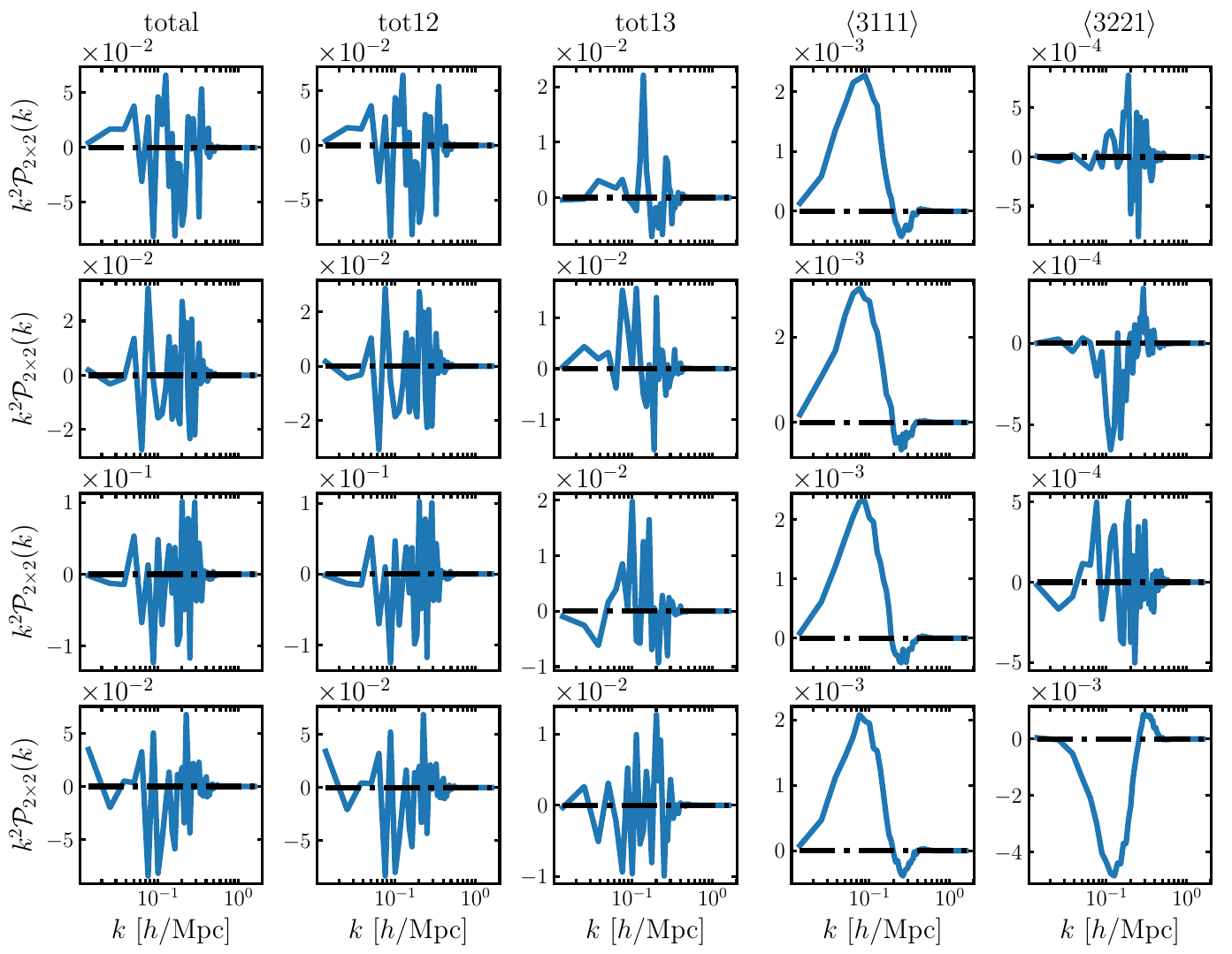}
        \rowlabel{64}{Growth}
        \rowlabel{46}{Tidal}
        \rowlabel{29}{No disp.}
        \rowlabel{12}{Full $F_2$}
      \end{overpic}%
    }
    \caption{$z=0$}   \vspace{0.15in} 
  \end{subfigure}
  \begin{subfigure}{\linewidth}
    \centering
    \makebox[\linewidth][r]{%
      \hspace*{\labelgutter}%
      \begin{overpic}[height=0.42\textheight]
        {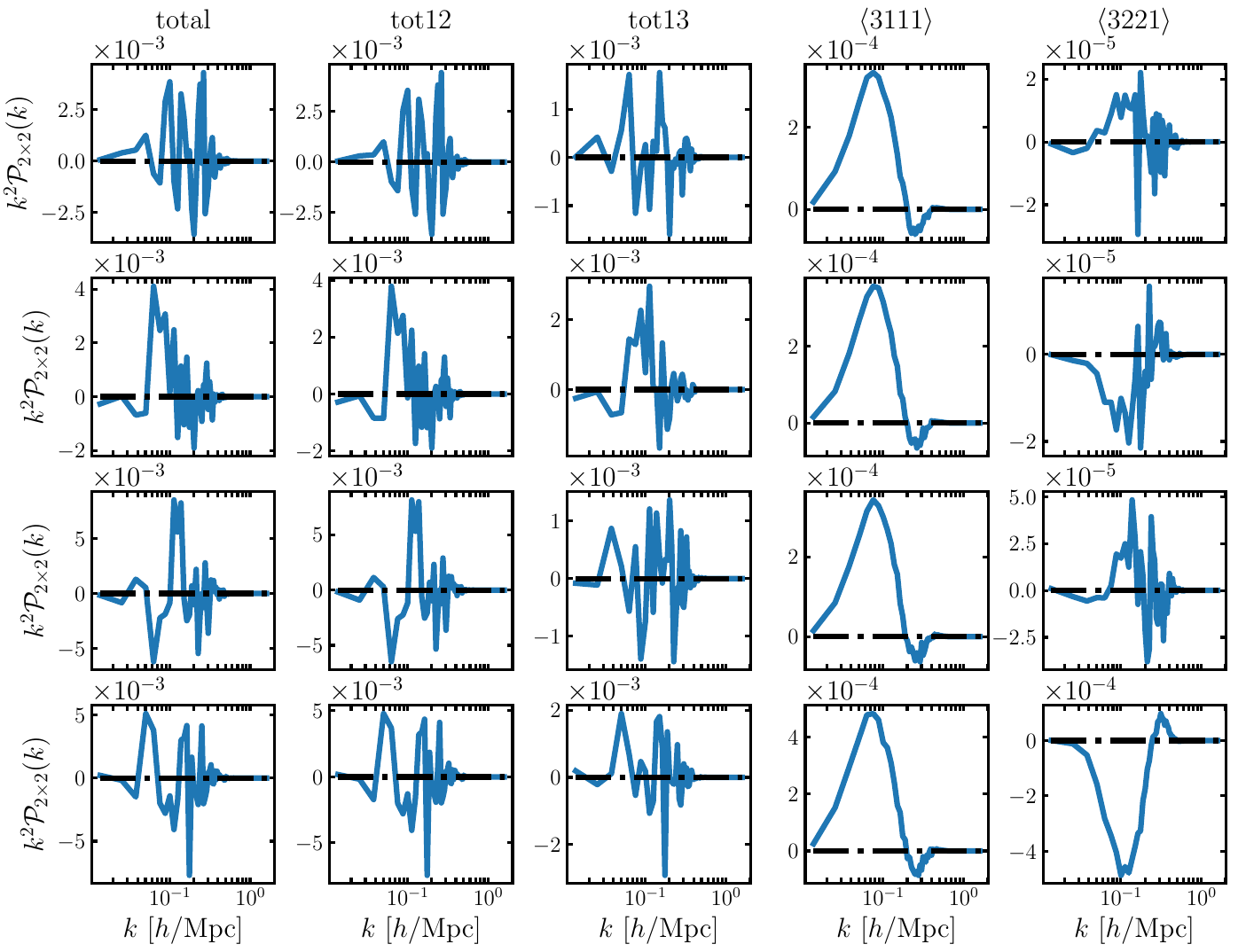}
        \rowlabel{64}{Growth}
        \rowlabel{46}{Tidal}
        \rowlabel{29}{No disp.}
        \rowlabel{12}{Full $F_2$}
      \end{overpic}%
    }
    \caption{$z=1$} \vspace{0.1in} 
  \end{subfigure}
    \caption{The EPT $k^2\mathcal{P}_{2\times 2}$ at redshift $z=0$ (a) and $z=1$ (b), with $R_G=5\, \mathrm{Mpc}/h$. Each column corresponds to the five cases, while each row corresponds to each kernel scenario. Note that each row is a new realization of the linear field, thus there is variation in signals even without the second-order fields.}
  \label{fig:EPT_kurto_twoz} \vspace{-0.2in}
\end{figure}

We show the measurement for a single realization at two redshifts, $z=0$ (a) and $z=1$ (b), in figure~\ref{fig:EPT_kurto_twoz}. For both redshifts, the columns correspond to the correlators defined in eq.~\eqref{eq:correlators}, while the rows correspond to the four assumptions for the second-order field in eq.~\eqref{eq:2nd}. Note that the third and fourth columns do not contain any second-order contributions, and the difference observed in different rows for these two columns is just the differing noise for different realizations. A comparison of the kurto spectra including all contributions (tot), averaged over several EPT realizations, will be presented in section~\ref{subsubsec:QvsEPT}. Throughout, we denote by $R_G = 5\,\mathrm{Mpc}/h$ the scale of the Gaussian filter applied to each overdensity field. The impact of varying the smoothing scale of the Gaussian filter is investigated in section~\ref{subsubsec:QvsEPT}.

At both redshifts, the measurement on the total field (left column) is very noisy. The second column contains no signal and therefore clearly shows the noise introduced by parity-even trispectrum from correlators involving linear and second-order fields. 
The expectation values of all correlators contributing to tot$12$ are zero, yet the variance in one realization is an order of magnitude larger than the expected LO signal (fourth column). The third column, which contains no second-order field, 
has a noise level about four (two) times smaller than the second column at $z=0$ ($z=1$). Since both the second and third columns include the $\la 1111 \ra$ term, the difference in noise arises from cross-correlators between second- and third-order fields with the linear field. The variance of these terms decreases at higher redshift, where the non-linearity due to second-order terms is weaker. From the third column, however, we see that---even without the second-order field---the expected signal is still about five times weaker than the noise (due to the noise of parity-even trispectrum of the linear field, which is effectively the disconnected part of the eight-point correlation function). 

The fifth column shows the next-to-leading-order parity-odd signal. For different choices of the second-order kernel, the contributions of the growth and tidal terms at $z=0$ ($z=1$) are about a factor of five (ten) smaller than the leading-order parity-odd signal shown in the fourth column. 
However, once we add the displacement term—i.e., using the full $F_2$ kernel—the NLO signal becomes comparable to, or even larger than, the LO signal. This large NLO signal is artificial (originates from the lack of proper IR resummation in our EPT field) and is not present in the full $N$-body dark-matter field. 

\subsubsection{\texttt{\textbf{Quijote-ODD}} dark matter vs EPT fields}\label{subsubsec:QvsEPT}

Next, we test our estimators beyond perturbation theory using the \texttt{Quijote-ODD} simulation suite \cite{Coulton:2024}, a variant of the \texttt{Quijote} collisionless $N$-body suite \cite{Villaescusa-Navarro:2019bje} \faGlobe\footnote{\url{https://quijote-simulations.readthedocs.io/en/latest/}} in which parity-odd statistics are injected at the level of the initial conditions. Each run evolves nonlinearly $512^3$ dark-matter particles in a periodic box of volume $(1\,\mathrm{Gpc}/h)^3$, assuming a fiducial $\Lambda$CDM cosmology with $\{\Omega_{\rm m},\Omega_{\rm b},h,n_s,\sigma_8\}=\{0.3175,0.049,0.6711,0.9624,0.834\}$ (setting total neutrino mass to zero). Parity violation is injected through a parity-odd primordial trispectrum (implemented by adding a cubic, parity-odd contribution to the primordial field) characterized by the amplitude $g^{\rm PO}_{\rm NL}$ (see eqs.~\eqref{eq:phi_ng} and~\eqref{eq:K3_Coulton}). Two phased-matched ensembles are provided: \texttt{ODD\_p} ($g^{\rm PO}_{\rm NL}=+10^6$) and \texttt{ODD\_m} ($g^{\rm PO}_{\rm NL}=-10^6$)---each paired with Gaussian fiducial runs sharing the same random seeds. This pairing enables (near) cosmic-variance cancellation by forming, for each realization, the kurto spectrum difference $\Delta\cP(k)\equiv \cP(\texttt{ODD\_p})-\cP(\texttt{fid})$. Initial conditions are generated with a suitably modified version of the \texttt{2LPTIC} code \faGlobe\footnote{\url{https://cosmo.nyu.edu/roman/2LPT/}} for primordial non-Gaussianity \cite{Scoccimarro:2011pz}, which computes second-order Lagrangian displacements and velocities from the non-Gaussian field before standard Newtonian evolution. In this section we use the dark-matter snapshots at $z=1$; results on halo catalogs are presented in section~\ref{subsec:Halo}.

\begin{figure}[htbp!]
    \begin{subfigure}{0.5\textwidth}
        \hspace{-0.1in}\includegraphics[width=\linewidth]{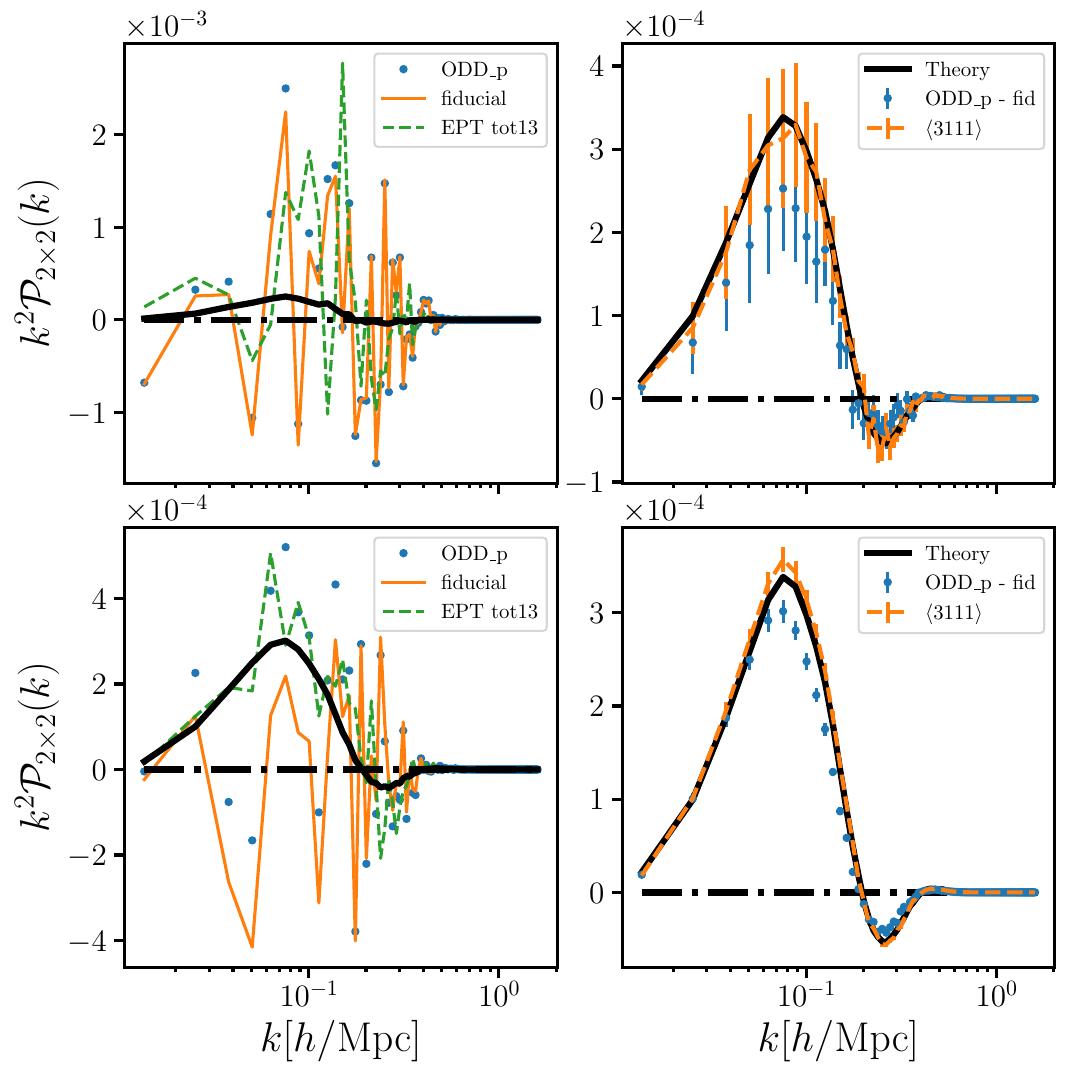}
        \caption{$R_G = 5\,\mathrm{Mpc}/h$}
        \label{fig:matter_2x2_RG5}
    \end{subfigure}\hfill
    \begin{subfigure}{0.5\textwidth}
        \includegraphics[width=\linewidth]{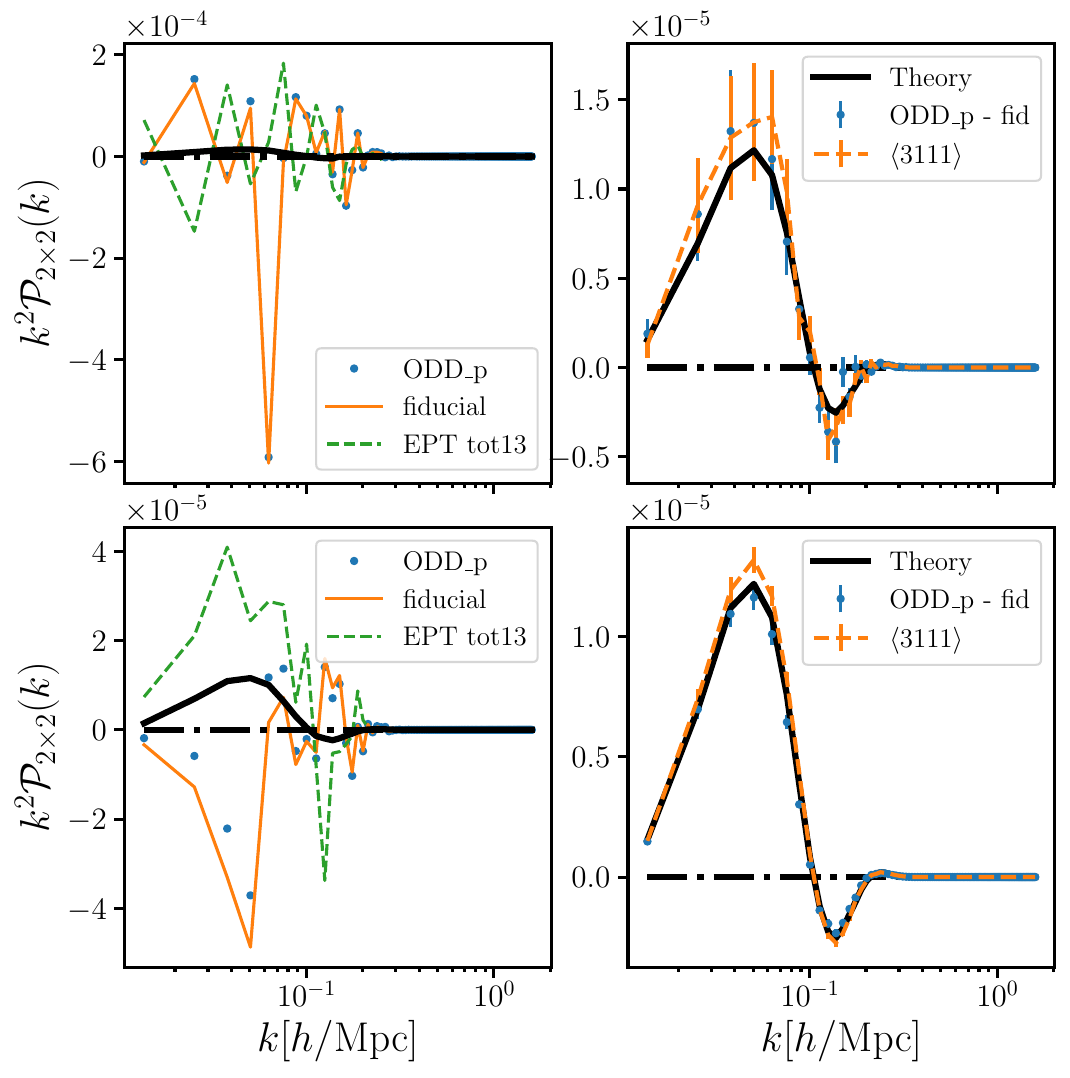}
        \caption{$R_G = 10\,\mathrm{Mpc}/h$}
        \label{fig:matter_2x2_RG10}
    \end{subfigure}
    \caption{Real-space matter kurto spectrum $\mathcal{P}_{2\times2}$ at $z=1$ from \texttt{Quijote} snapshots, compared with EPT-field measurements and theory, for $R_G=5\,\mathrm{Mpc}/h$ (figure~\ref{fig:matter_2x2_RG5}) and $R_G=10\,\mathrm{Mpc}/h$ (figure~\ref{fig:matter_2x2_RG10}). All spectra are multiplied by $k^{2}$ for visual clarity. In each subfigure, top row shows single realization, while bottom row is the average over 40 realizations. Left panels show \texttt{Quijote} \texttt{ODD\_p} (blue dots), matched fiducial simulation (orange solid), EPT \texttt{tot13} (green dashed), while the right panels show the difference of kurto spectra on \texttt{ODD\_p} and fiducial simulations (blue dots), the leading-order $\langle3111\rangle$ contribution on the EPT field (orange dashed). In both panels, the black solid line shows the theoretical prediction of parity-odd signal from eqs. \eqref{eq:P_2x2_type1} and \eqref{eq:P_2x2_type2}. The error bars correspond to the diagonal of the full covariance matrix of the $\cP_{2\times2}^{\rm PO}$ estimated from 500 fiducial simulations.}
    \label{fig:Q_2x2_matter_RG5_RG10}
\end{figure}
 \begin{figure} 
        \begin{subfigure}[b]{0.49\textwidth}
            \hspace{-0.15in}\includegraphics[width=\linewidth]{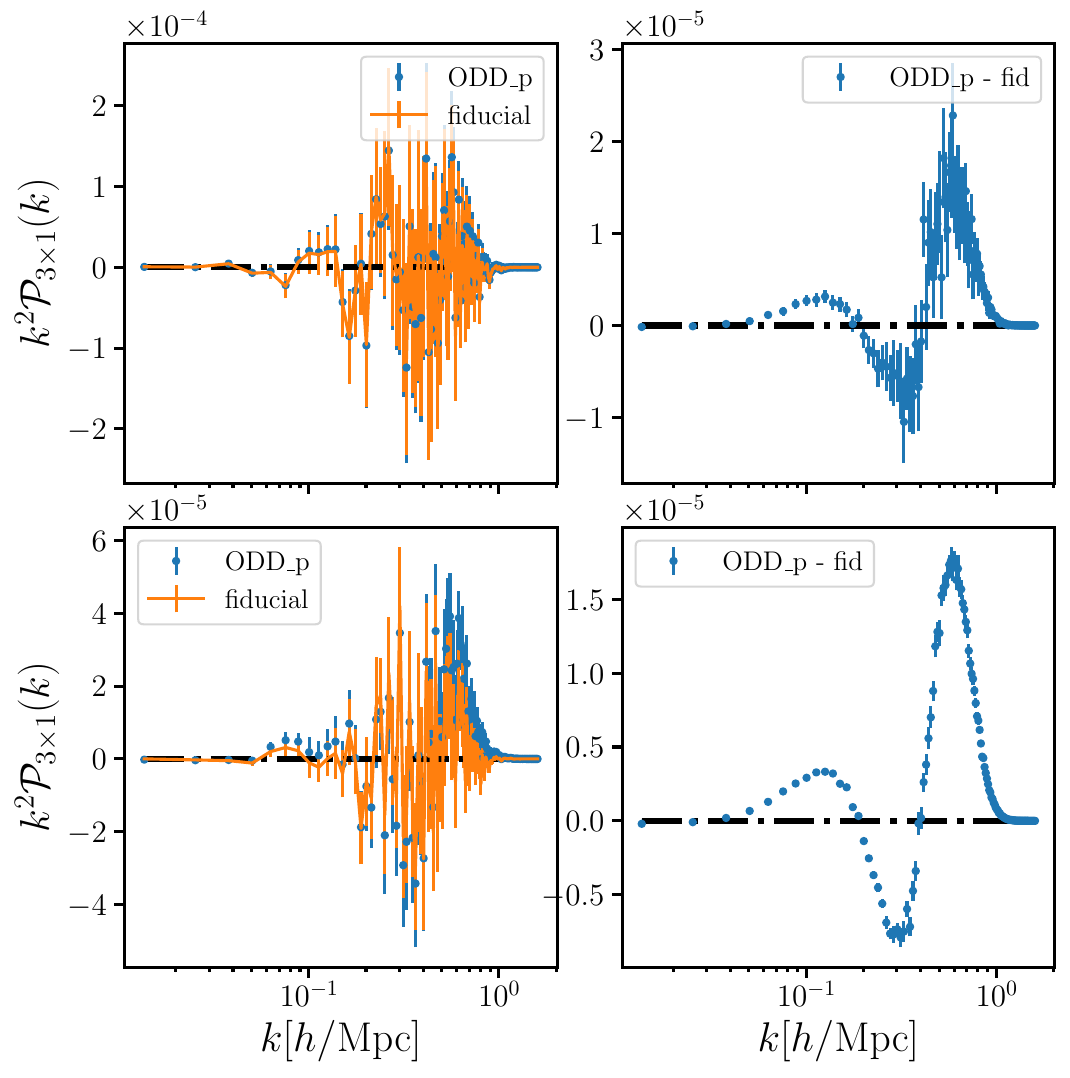}
            \caption{$R_G = 5\,\mathrm{Mpc}/h$}
            \label{fig:matter_3x1_RG5}
        \end{subfigure}
        \begin{subfigure}[b]{0.49\textwidth}
            \includegraphics[width=\linewidth]{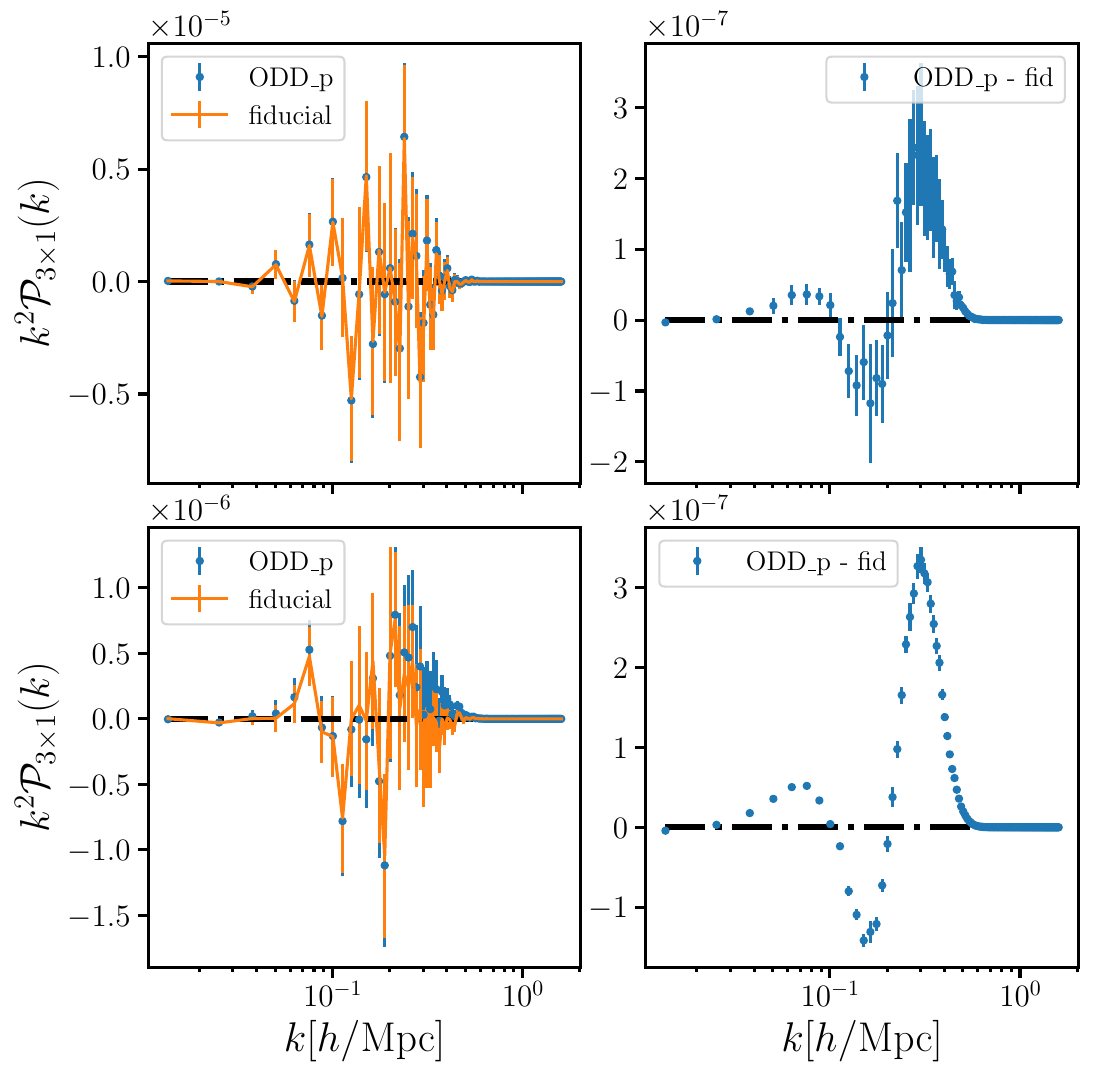}
            \caption{$R_G = 10\,\mathrm{Mpc}/h$}
            \label{fig:matter_3x1_RG10}
        \end{subfigure}
        \caption{Real-space matter kurto spectrum $\cP_{3\times1}$ at $z=1$ from \texttt{Quijote} snapshots for $R_G=5 \ {\rm Mpc}/h$ (figure~\ref{fig:matter_3x1_RG5}) and $R_G=10\ {\rm Mpc}/h$ (figure~\ref{fig:matter_3x1_RG10}). All spectra are multiplied by $k^{2}$ for visual clarity. Top row: single realization. Bottom row: average over 40 realizations. Left panels: \texttt{Quijote} ODD\_p (blue dots) and matched fiducial simulation (orange solid). Right panels: the difference of kurto spectra on \texttt{ODD\_p} and fiducial simulations (blue dots).}\vspace{-0.1in}
        \label{fig:Q_3x1_matter_RG5_RG10}
\end{figure}

We show the measured $\cP_{2 \times 2}^{\rm PO}$ and $\cP_{3 \times 1}^{\rm PO}$ for the \texttt{ODD\_p} and phase-matched fiducial DM fields (left panels), together with the per-realization difference $\Delta \cP$(right panels), at 
$z=1$ in figures~\ref{fig:Q_2x2_matter_RG5_RG10} and \ref{fig:Q_3x1_matter_RG5_RG10}, respectively. In each figure, subfigures (a) and (b) show the measurement with smoothing scales of $R_G=5\, \mathrm{Mpc}/h$ and $R_G=10\, \mathrm{Mpc}/h$, respectively, while the top row show measurements from a single realization and the bottom row show averages over 40 realizations.\footnote{This averaging approximates the noise reduction expected for a survey volume of $40 \ (\mathrm{Gpc}/h)^3$.}

In the left panels of figures \ref{fig:matter_2x2_RG5} and \ref{fig:matter_2x2_RG10}, in addition to the measured spectra on \texttt{ODD\_p} (blue dots) and the corresponding fiducial \texttt{Quijote} simulations (orange line), we also show the EPT \texttt{tot13} measurement (dashed green). In the right panel, we show the difference kurto spectrum (blue dots) and and the LO parity-odd signal, $\langle 3111 \rangle$ correlator, measured on the EPT field. In both panels, the black solid line shows the theoretical prediction of parity-odd signal from eqs. \eqref{eq:P_2x2_type1} and \eqref{eq:P_2x2_type2}, respectively. The error bars correspond to the variance estimated from square of measured kurto spectra from 500 fiducial simulations. In the left panels of figures \ref{fig:matter_2x2_RG5} and \ref{fig:matter_2x2_RG10}, we do not show the error bars since otherwise the comparison between the three measurements would not be visible. In figures \ref{fig:matter_3x1_RG5} and \ref{fig:matter_3x1_RG10}, the left panels show the $\cP_{3\times1}^{\rm PO}$ on \texttt{ODD\_p} and fiducial simulations, while the right panels show their difference. 

In figure \ref{fig:Q_2x2_matter_RG5_RG10}, for both smoothing scales, the single-realization measurements of $\cP_{2 \times 2}^{\rm PO}$ on \texttt{ODD\_p} and fiducial DM snapshots, as well as the EPT \texttt{tot13} contribution, are very noisy. As noted in section~\ref{subsubsec:ept_field}, although the $\mathcal{P}_{2\times 2}$ estimator for a parity-even field has zero mean, its variance is large and can bury the target parity-odd non-zero signal. Averaging over 40 realizations, reduces the noise in EPT \texttt{tot13} measurement---driven by parity-even correlations between the linear and cubic fields—so that the parity-odd signal becomes apparent and broadly matches the theoretical prediction for the smaller smoothing scale. For \texttt{ODD\_p} and fiducial DM snapshots, averaging over several realizations reduces the scatter by roughly an order of magnitude; however, the residual scatter is still large enough that the parity-odd signal in the \texttt{ODD\_p} measurement is not clearly identifiable.  We also observe that for $R_G=10\,\mathrm{Mpc}/h$ the EPT \texttt{tot13} amplitude can exceed the \texttt{ODD\_p} signal in the averaged measurements.  

In the right columns of figures \ref{fig:matter_2x2_RG5} and \ref{fig:matter_2x2_RG10}, the $\langle 3111 \rangle$ term from the EPT field (orange dashed line) has a slightly larger amplitude than the total parity-odd $\cP_{2\times 2}$ measured from the \texttt{Quijote-ODD} simulations (blue dots). This difference originates from higher-order parity-odd correlators present in the \texttt{Quijote-ODD} simulations. Increasing the smoothing scale reduces the amplitudes of both signal and noise, as expected, and shifts the signal peak to larger scales. Moreover, the difference between the \texttt{Quijote-ODD} and EPT measurements shrinks with increasing smoothing scale, indicating that filtering out smaller scales reduces the impact of higher-order parity-odd correlators. A similar behavior is expected for the $\cP_{3\times 1}$ estimator, even though, we dont make an explicit comparison in that case.

The measurements of $\cP_{3\times 1}$ (figure \ref{fig:Q_3x1_matter_RG5_RG10}) on a single-realization are noisy, but averaging reduces the scatter and makes the shape of the parity-odd signal identifiable by eye. Subtracting the fiducial measurement from \texttt{ODD\_p} clearly brings out the signal, even for a single realization. As with $\cP_{2\times 2}$, increasing the smoothing scale shifts the peak and trough to larger scales and lowers the noise level. Nevertheless, even the averaged $\cP_{3\times 1}$ measurements retain substantial scatter, which is considerably reduced when subtracting the fiducial measurement. Comparing figures \ref{fig:Q_2x2_matter_RG5_RG10} and \ref{fig:Q_3x1_matter_RG5_RG10}, the $\cP_{2\times 2}$ signal peaks at larger scales, whereas $\cP_{3\times 1}$ peaks at smaller scales. This does not by itself imply that the former is more affected by cosmic variance and the latter by small-scale nonlinearity: the trispectrum-like estimator couples small and large scales of the overdensity field. Furthermore, the number of $k$-modes contributing to $\cP_{3\times 1}$ exceeds that for $\cP_{2\times 2}$; therefore, we expect a higher detectability for the $\cP_{3\times 1}$ estimator (see section~\ref{subsec:cov_SNR_detectability}).

\begin{figure}[t]
  \begin{subfigure}{0.49\textwidth}
    \hspace{-0.1in}\includegraphics[width=\linewidth]{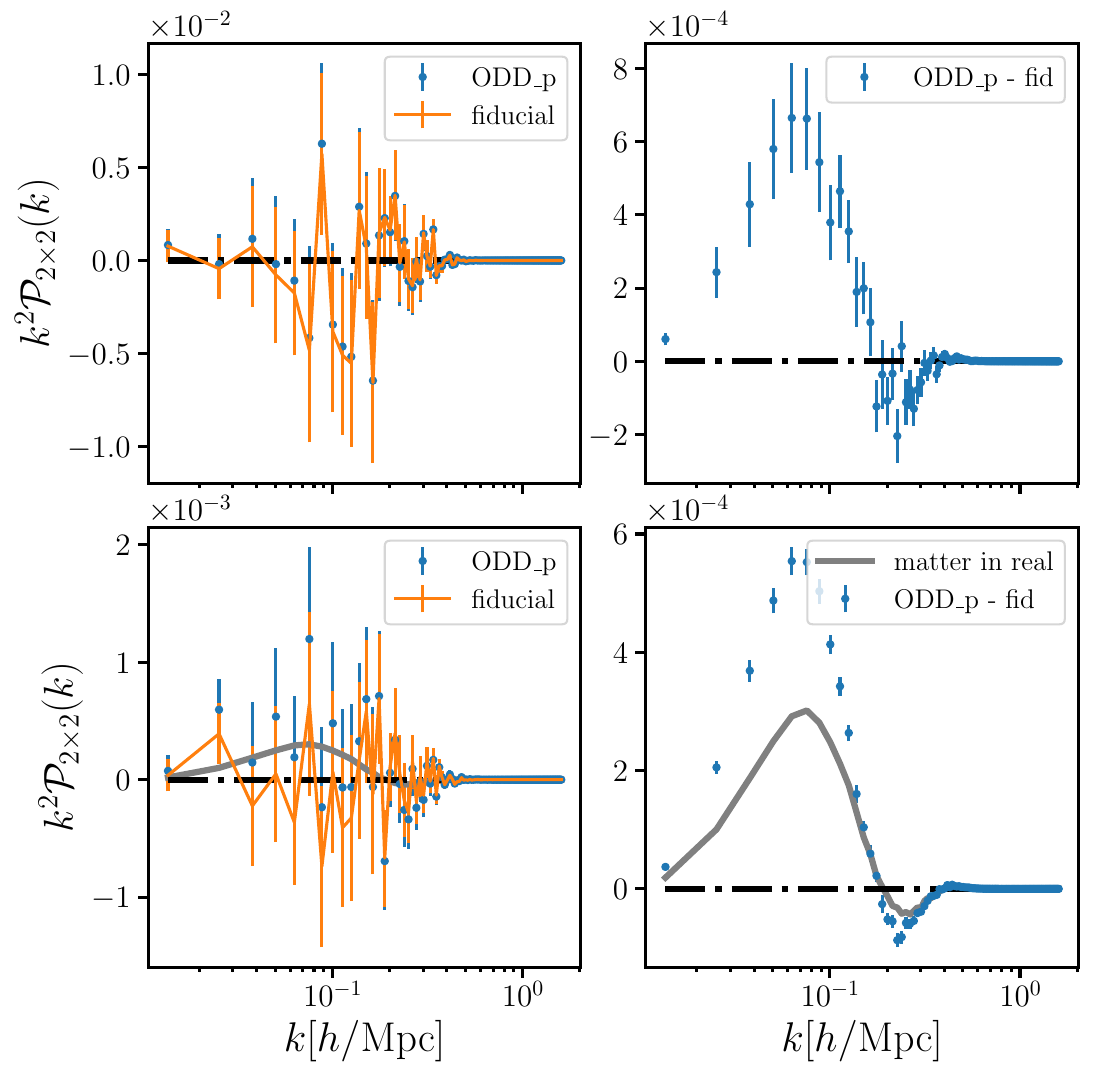}
    \caption{$R_G = 5\,\mathrm{Mpc}/h$}
    \label{fig:matter_RSD_2x2_RG5}
  \end{subfigure}\hfill
  \begin{subfigure}{0.49\textwidth}
    \hspace{-0.1in}\includegraphics[width=\linewidth]{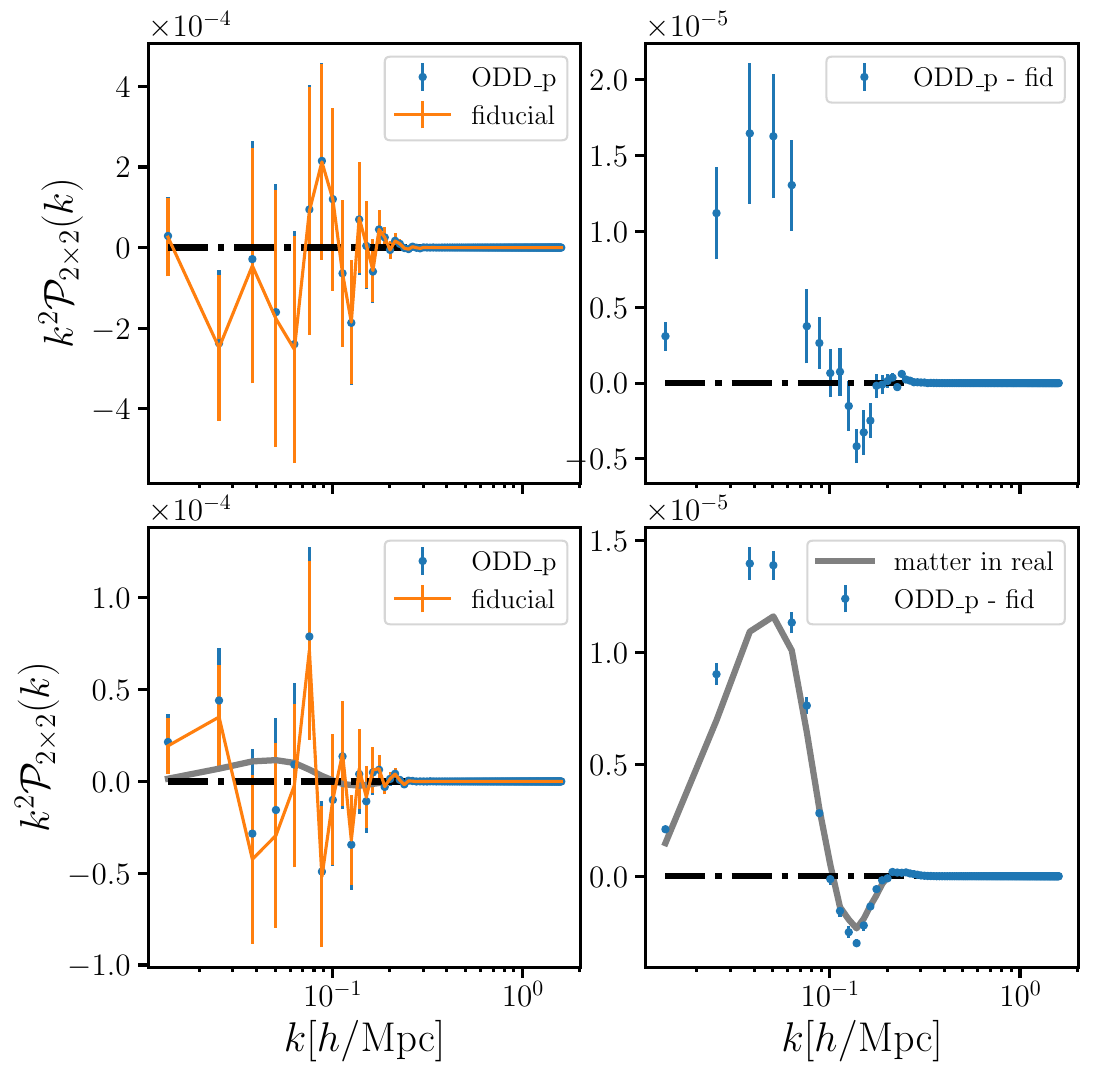}
    \caption{$R_G = 10\,\mathrm{Mpc}/h$}
    \label{fig:matter_RSD_2x2_RG10}
  \end{subfigure}\hfill
  \caption{Redshift-space matter kurto spectrum $\mathcal{P}_{2\times 2}$ at $z=1$ from \texttt{Quijote} snapshots for $R=5 \ {\rm Mpc}/h$ (figure~\ref{fig:matter_RSD_2x2_RG5}) and $R=10 \ {\rm Mpc}/h$ (figure~\ref{fig:matter_RSD_2x2_RG10}). All spectra are multiplied by $k^2$ for visual clarity. Top row: single realization. Bottom row: average over 40 realizations. Left panels: \texttt{Quijote ODD\_p} (blue dots) and matched fiducial simulation (orange solid). Right panels: the difference of kurto spectra on \texttt{ODD\_p} and fiducial simulations (blue dots), with the gray lines being the measured real-space difference.}
  \label{fig:Q_matter_2x2_RSD_RG5_RG10}
    \vspace{0.2in}
  \begin{subfigure}{0.49\textwidth}
    \includegraphics[width=\linewidth]{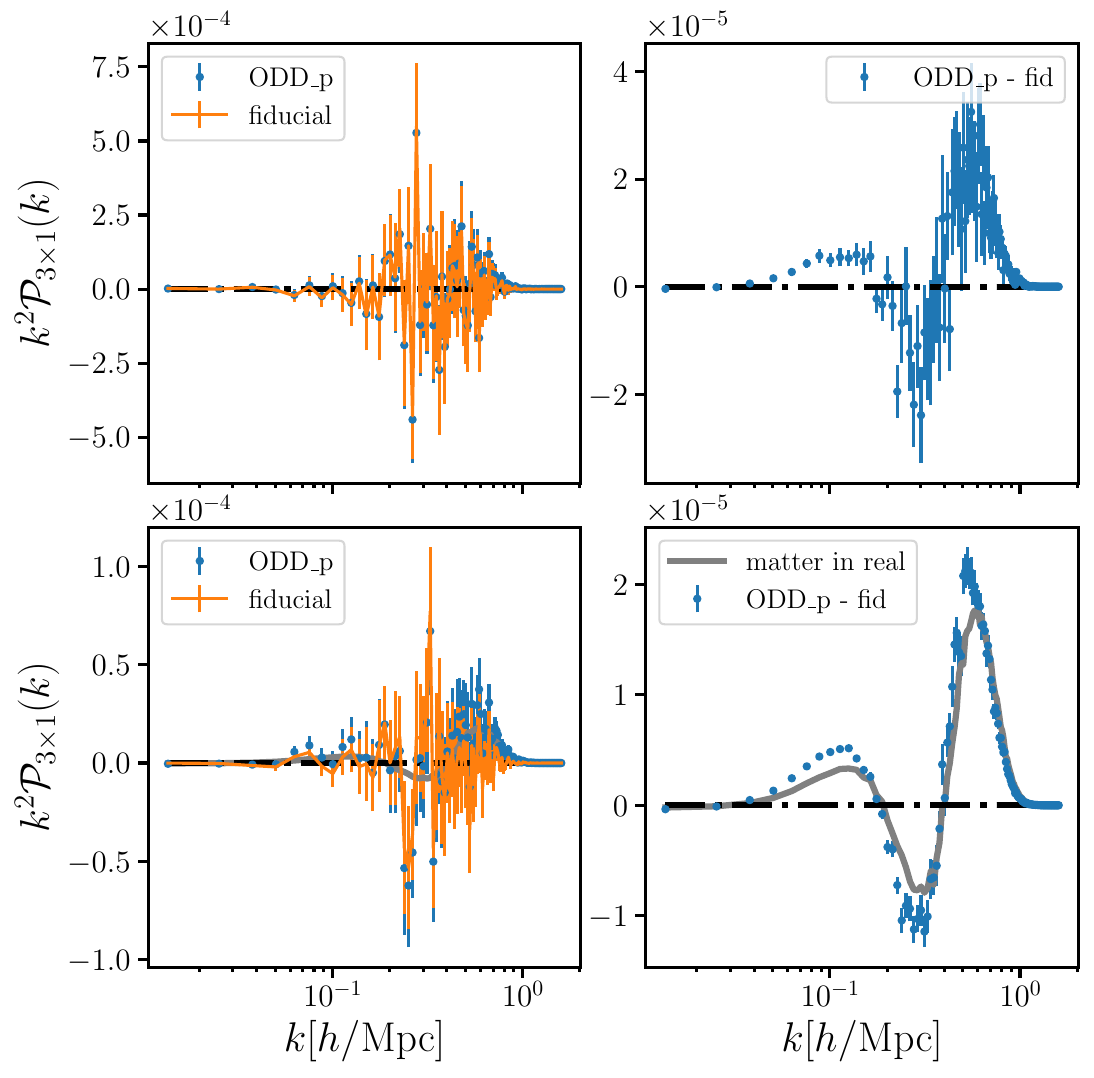}
    \caption{$R_G = 5\,\mathrm{Mpc}/h$}
    \label{fig:matter_RSD_3x1_RG5}
  \end{subfigure}\vspace{0.2in}
  \begin{subfigure}{0.49\textwidth}
    \includegraphics[width=\linewidth]{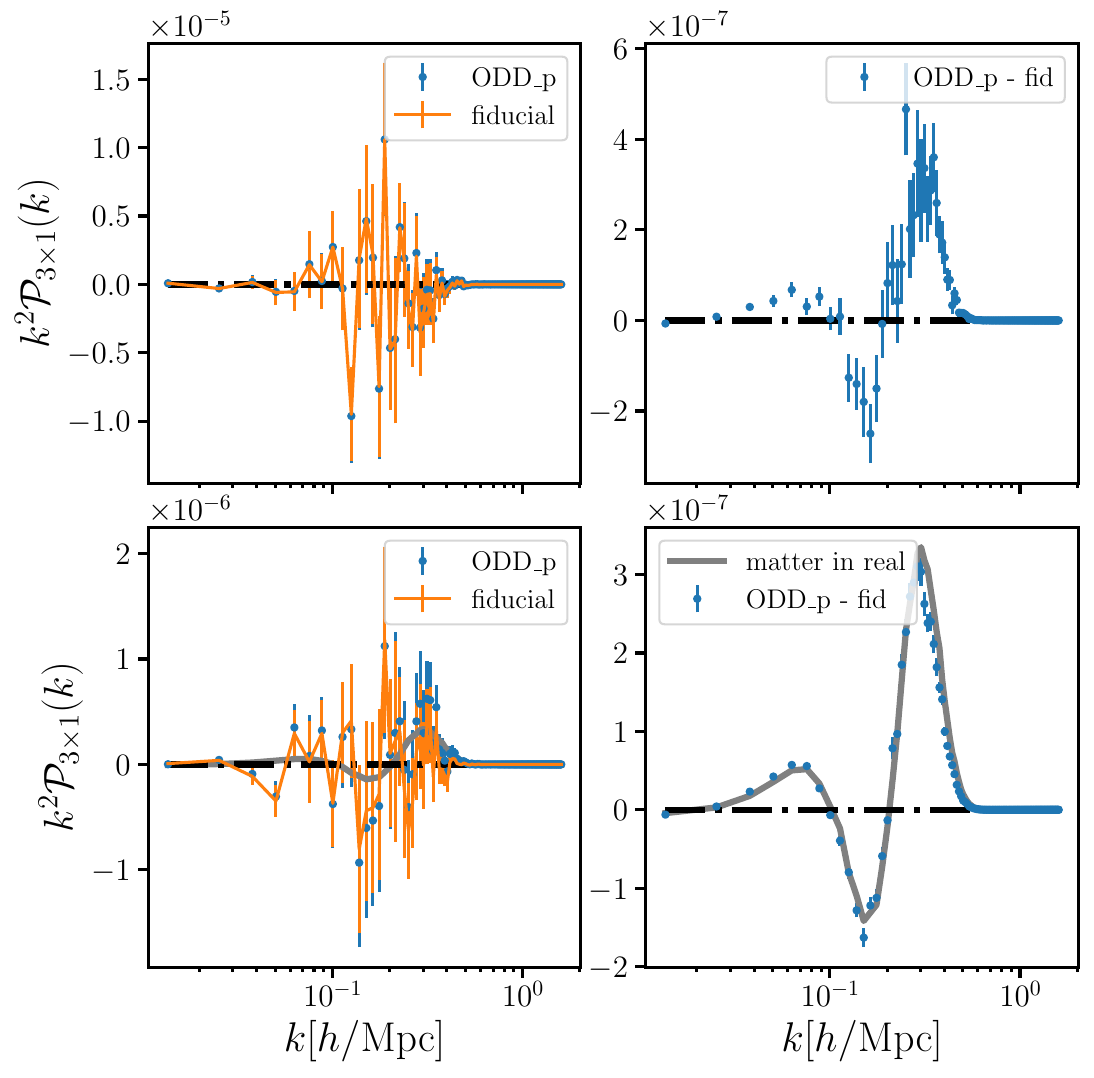}
    \caption{$R_G = 10\,\mathrm{Mpc}/h$}
    \label{fig:matter_RSD_3x1_RG10}
  \end{subfigure}\vspace{-0.1in}
  \caption{Same as figure \ref{fig:Q_matter_2x2_RSD_RG5_RG10}, but for $\cP_{3\times1}$ kurto spectrum.}
  \label{fig:Q_matter_3x1_RSD_RG5_RG10}
\end{figure}

Thus far we have shown the kurto spectrum measurements in real space. To assess the impact of redshift-space distortions (RSD), figures~\ref{fig:Q_matter_2x2_RSD_RG5_RG10} and \ref{fig:Q_matter_3x1_RSD_RG5_RG10} present redshift-space measurements of $\cP_{2\times2}$ and $\cP_{3\times1}$ from \texttt{ODD\_p} and fiducial DM snapshots at $z=1$, for Gaussian smoothing scales of $R_G=\{5,10\}\,\mathrm{Mpc}/h$. The overall shapes are similar to the real-space case, but the amplitudes are boosted, especially for $\cP_{2\times 2}^{\rm PO}$. For the larger smoothing scale of $R_G=10\ {\rm Mpc}/h$, the relative enhancement of redshift-space kurto spectra compared to real-space is less significant. This is an indication of the contribution of small-scale velocities to kurto spectra peak on large-scales, which is suppressed when smoothing scale is larger.

\subsection{Halo field}\label{subsec:Halo}

\begin{figure}[htbp!]
  \centering
  \begin{subfigure}[b]{0.49\textwidth}
    \hspace{-0.15in}\includegraphics[width=\linewidth]{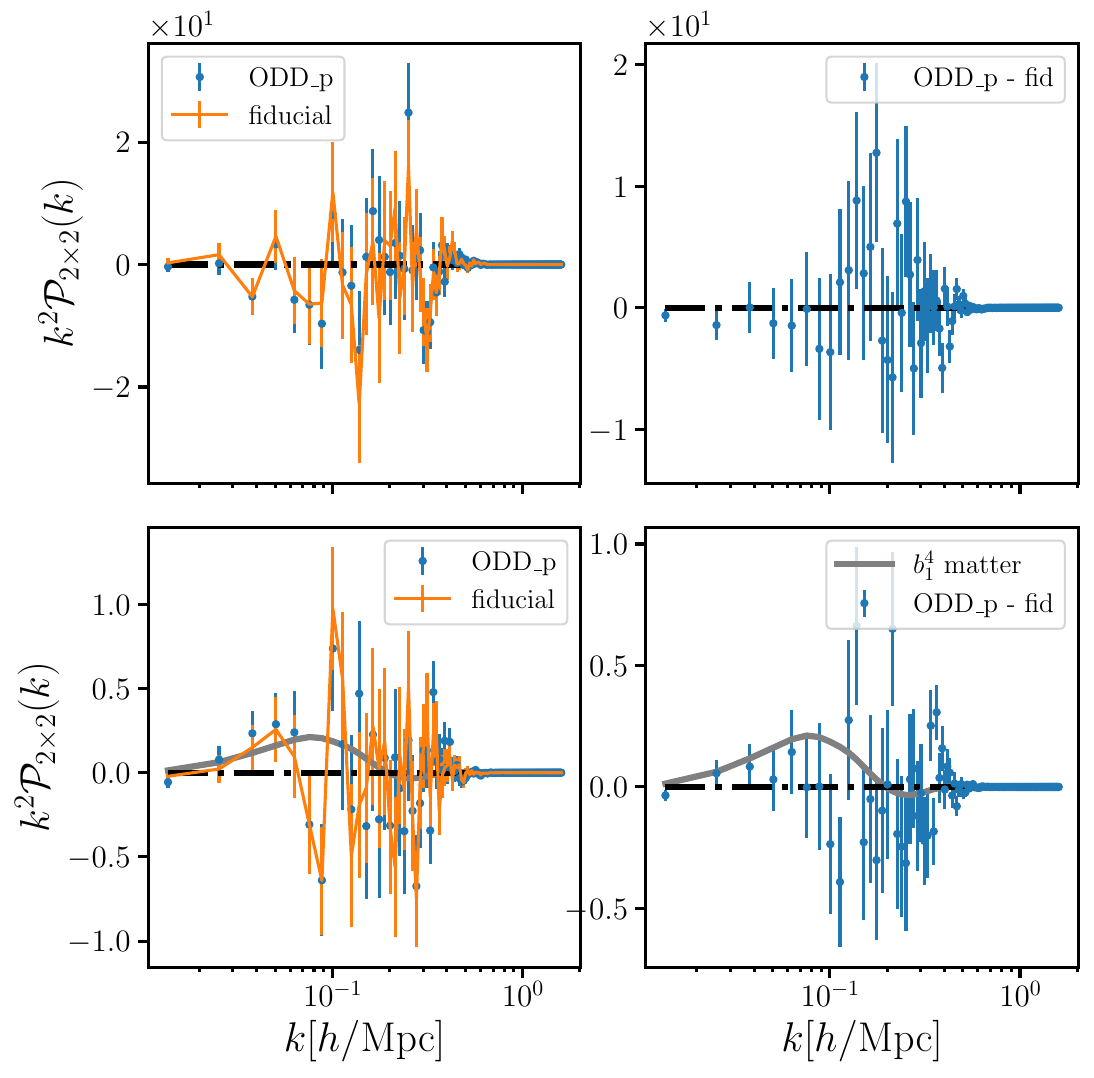}
    \caption{$R_G = 5\,\mathrm{Mpc}/h$}
    \label{fig:halo_2x2_RG5}
  \end{subfigure}
  \begin{subfigure}[b]{0.49\textwidth}
    \includegraphics[width=\linewidth]{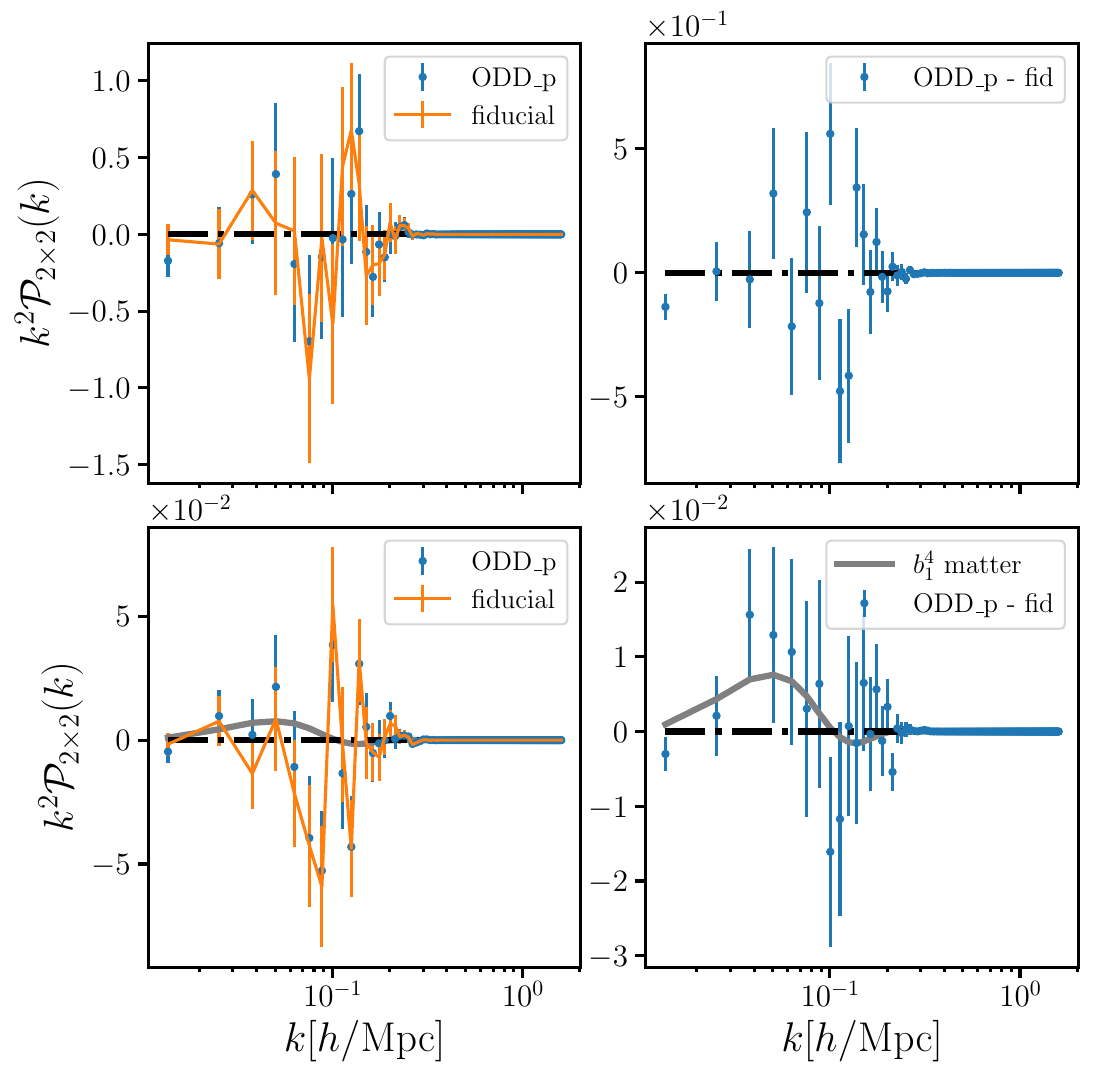}
    \caption{$R_G = 10\,\mathrm{Mpc}/h$}
    \label{fig:halo_2x2_RG10}
  \end{subfigure}
  \caption{Real-space halo kurto spectrum $\mathcal{P}_{2\times2}$ at $z=1$ from \texttt{Quijote} simulations for Gaussian smoothing scales $R_G=5\ {\rm Mpc}/h$ (figure~\ref{fig:halo_2x2_RG5}) and $R_G=10 \ {\rm Mpc}/h$ (figure~\ref{fig:halo_2x2_RG10}). All spectra are multiplied by $k^{2}$ for visual clarity. Top row: single realization. Bottom row: average over 500 realizations. Left panels: \texttt{Quijote ODD\_p} (blue dots) and matched fiducial simulation (orange solid). Right panels: the total parity-odd signal from the fiducial-subtracted simulations (blue dots). The gray lines are the measured kurto spectra of the fiducial-subtracted matter field, scaled by a factor of $b_1^4$.}
  \label{fig:Q_halo_2x2_RG5_RG10}
  \vspace{0.1in}
  \begin{subfigure}[b]{0.49\textwidth}
    \hspace{-0.15in}\includegraphics[width=\linewidth]{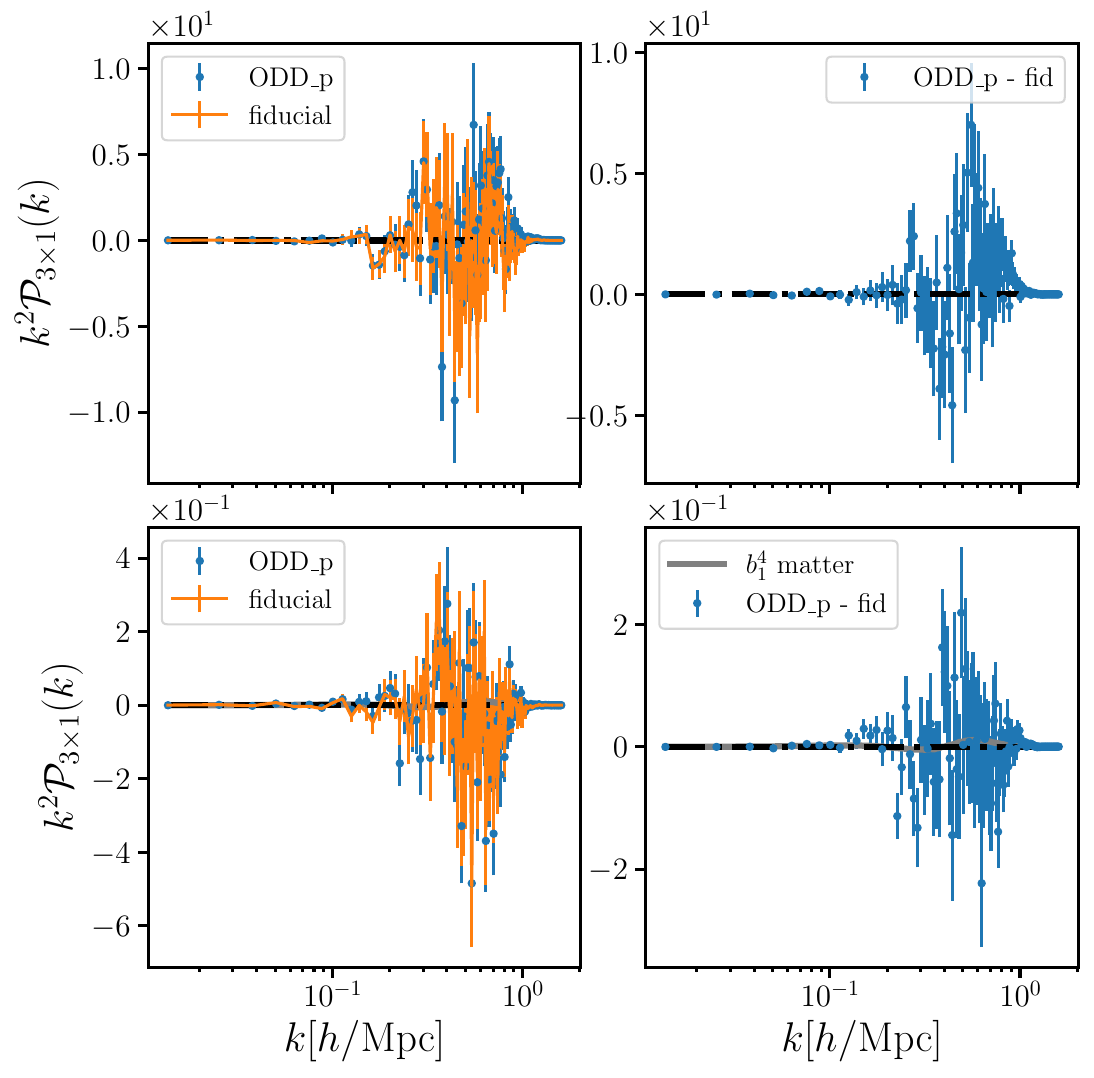}
    \caption{$R_G = 5\,\mathrm{Mpc}/h$}
    \label{fig:halo_3x1_RG5}
  \end{subfigure}
  \begin{subfigure}[b]{0.48\textwidth}
    \includegraphics[width=\linewidth]{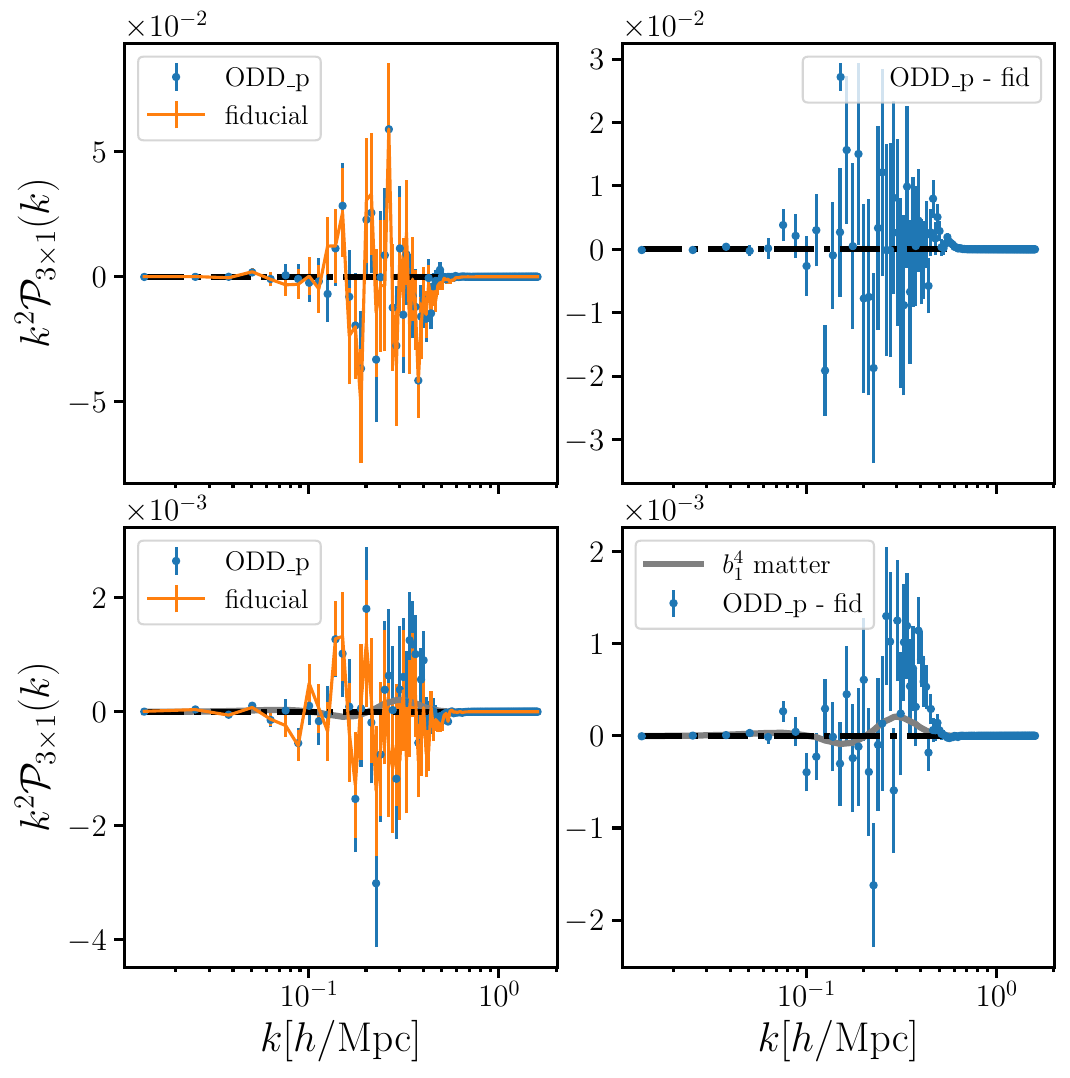}
    \caption{$R_G = 10\,\mathrm{Mpc}/h$}
    \label{fig:halo_3x1_RG10}
  \end{subfigure}
  \caption{Same as figure \ref{fig:Q_halo_2x2_RG5_RG10}, but for $\cP_{3\times1}$ kurto spectrum.} 
  \label{fig:Q_halo_3x1_RG5_RG10}
\end{figure}

In figures~\ref{fig:Q_halo_2x2_RG5_RG10} and \ref{fig:Q_halo_3x1_RG5_RG10}, we present measurements of $\cP_{2\times 2}$ and $\cP_{3\times 1}$ on \texttt{Quijote-ODD} \texttt{ROCKSTAR} halo catalogs at $z=1$, selecting halos in the mass range $5.5\times 10^{13}\, \mathrm{M}_\odot/h$–$3\times 10^{14}\, \mathrm{M}_\odot/h$. There are $\sim 13,800$ halos in this mass bin per simulation box, corresponding to a number density $\bar{n}\sim 1.38\times 10^{-5} \lb h/\mathrm{Mpc}\rb^3$. For the theory curve we adopt linear biasing from the matter field, i.e., $\mathcal{P}^{\rm halo} = b_1^4 \mathcal{P}^{\rm matter}$, neglecting higher–order bias contributions as a first approximation. Fitting the measured halo power spectrum on large scales with linear prediction $P_h(k) = b_1^2 P_m(k)$, we obtain obtain $b_1=4.9$. Because the parity-odd kurto spectra of the halo field are particularly noisy, here we average over 500 realizations (instead of 40 used for DM field).

For a smoothing scale $R_G=5\,\mathrm{Mpc}/h$, in contrast to DM case (figures \ref{fig:matter_2x2_RG5} and \ref{fig:matter_3x1_RG5})---where subtracting fiducial and parity-odd simulations clearly isolated the parity-odd signal---the halo measurements do not show a clear parity-odd detection even for the \texttt{ODD\_p} minus fiducial difference and after averaging over 500 realizations (see figures \ref{fig:halo_2x2_RG5} and \ref{fig:halo_3x1_RG5}). This suggests that the parity-odd signature affects halo formation, introducing stochasticity that differs from the fiducial cosmology. Consequently, the difference between fiducial and \texttt{ODD\_p} halo fields no longer isolates only the biased parity-odd component but also includes additional parity-even (stochastic) noise that dominates over the signal. We note that this stochasticity is still dominated by that from parity-even components, while the \texttt{ODD\_p} and fiducial ones fail to cancel. Using a larger smoothing scale, $R_G=10\,\mathrm{Mpc}/h$, the $\cP_{2\times 2}$ and $\cP_{3\times 1}$ measurements with fiducial subtraction show a bump in the noisy data that tracks the expected signal peak (see figures \ref{fig:halo_2x2_RG10} and \ref{fig:halo_3x1_RG10}); the nonzero signal is more apparent for the former. This indicates that the additional stochasticity is more influential on smaller scales of the overdensity field. The detectability in these two cases is quantified in Tab.~\ref{tab:detectability_match_filter}. 

\begin{figure}[t]
  \centering
  \begin{subfigure}[b]{0.49\textwidth}
    \hspace{-0.15in}\includegraphics[width=\linewidth]{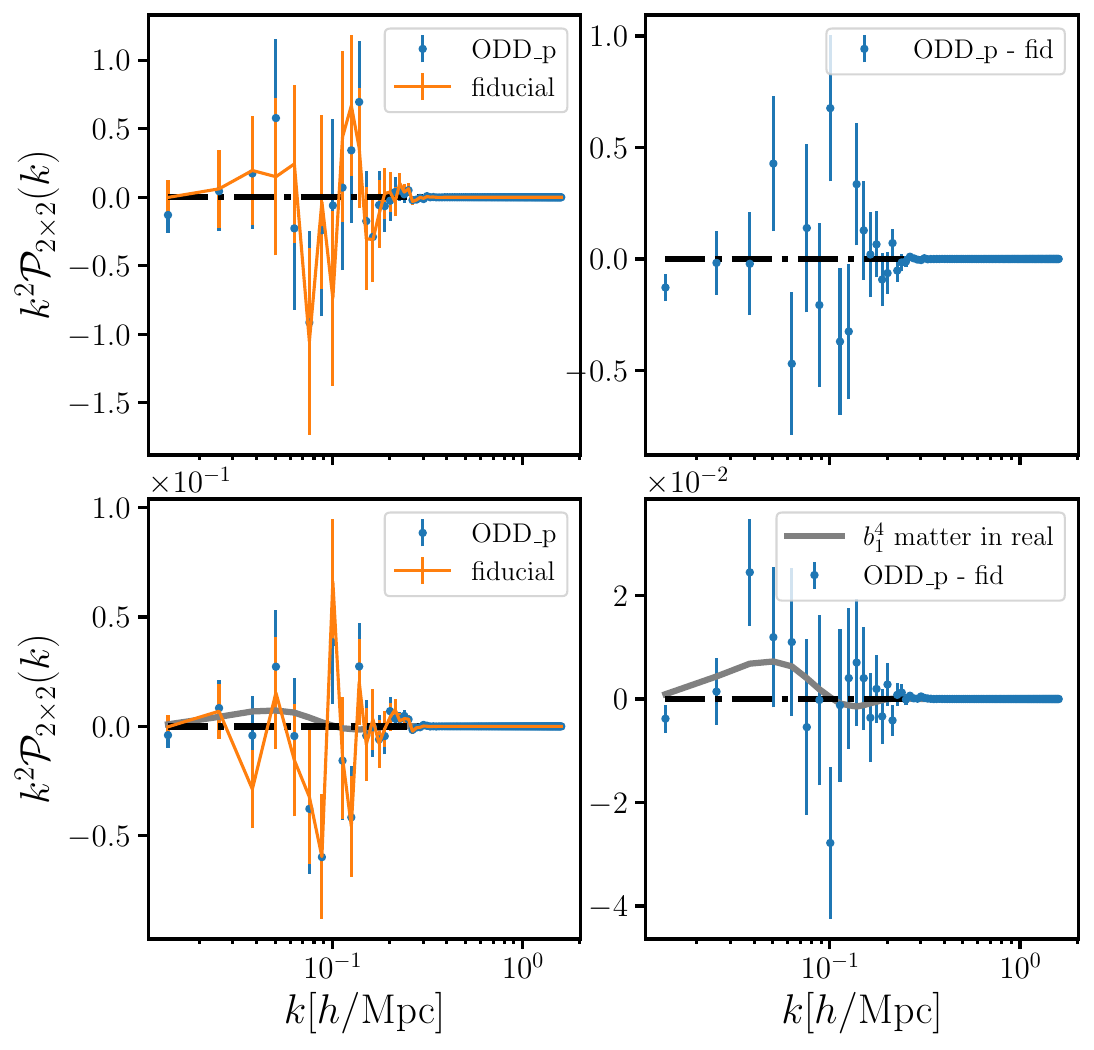}
    \caption{$R_G = 10\,\mathrm{Mpc}/h$}
    \label{fig:halo_RSD_2x2_RG10}
  \end{subfigure}
  \begin{subfigure}[b]{0.49\textwidth}
    \includegraphics[width=\linewidth]{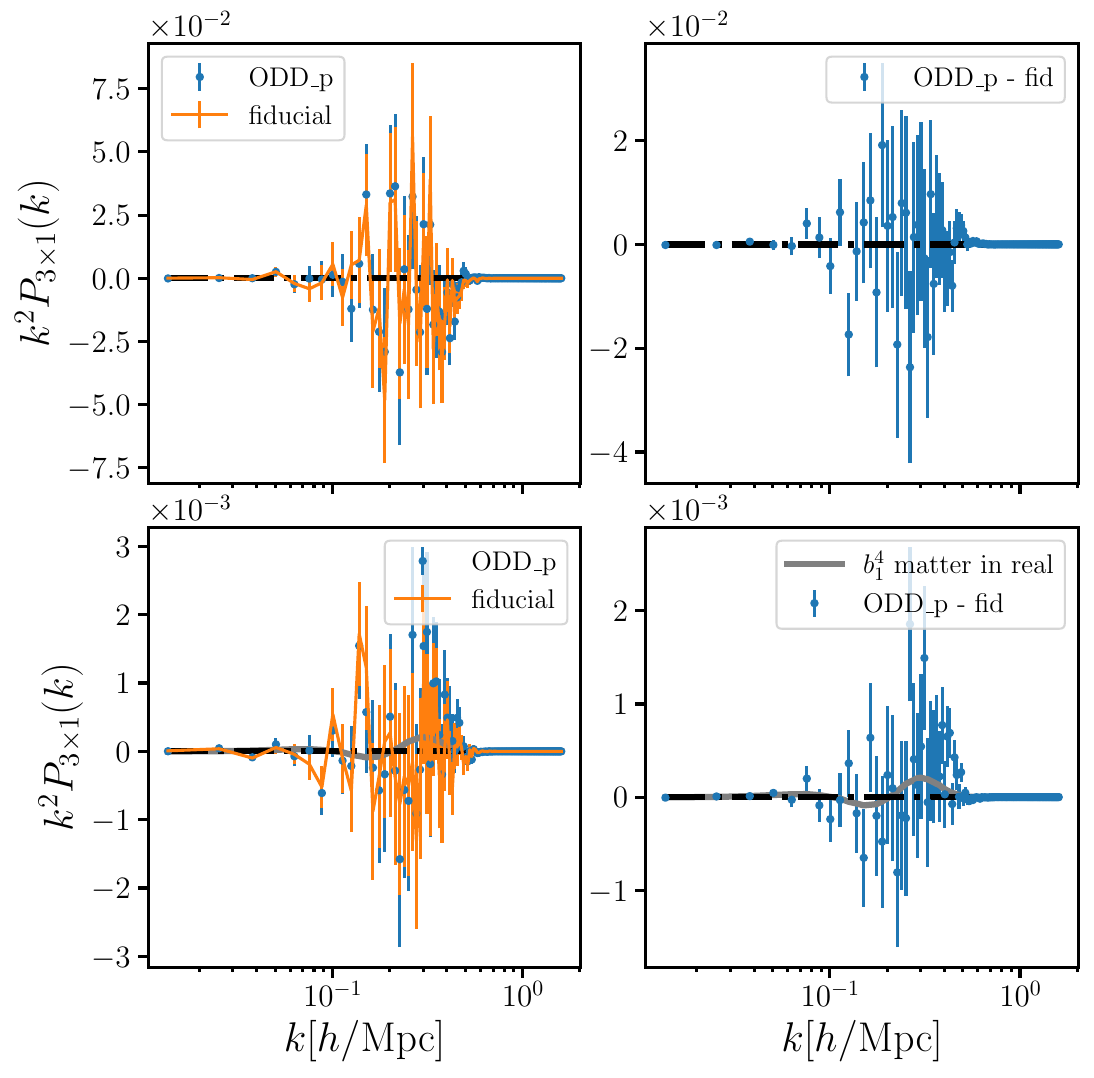}
    \caption{$R_G = 10\,\mathrm{Mpc}/h$}
    \label{fig:halo_RSD_3x1_RG10}
  \end{subfigure}
  \caption{Redshift-space halo kurto spectra, $\mathcal{P}_{2\times2}$ (figure~\ref{fig:halo_RSD_2x2_RG10}) and $\mathcal{P}_{3\times1}$ (figure~\ref{fig:halo_RSD_3x1_RG10}), from the \texttt{Quijote} halo catalogs at $z=1$ with Gaussian smoothing $R_G=10\ {\rm Mpc}/h$. All spectra are multiplied by $k^{2}$ for visual clarity. Top row: single realization. Bottom row: average over 500 realizations.  Left panels: \texttt{Quijote\ ODD\_p} (blue dots) and matched fiducial simulation (orange solid). Right panels: the total parity-odd signal from the fiducial-subtracted simulations (blue dots), and the matter measurement in real space (grey solid) scaled by a factor of $b_1^4$.}
  \label{fig:Q_halo_RSD_RG5_RG10}
\end{figure}

In figure~\ref{fig:Q_halo_RSD_RG5_RG10}, we show the redshift-space measurements of the halo kurto spectra for $R_G=10\,\mathrm{Mpc}/h$ to assess the impact of peculiar velocities on clustering. The fiducial redshift-space measurements exhibit larger noise than their real-space counterparts. After subtracting the fiducial measurement, the results remain dominated by stochasticity, with little discernible difference between redshift- and real-space cases. 

In practice, reaching a geometric comoving volume of 
$\sim 500 \ ({\rm Gpc}/h)^3$, would require roughly half-sky coverage extending to $z_{\rm max}\sim 8$, well beyond Stage-IV galaxy redshift surveys and likely achievable only with future wide-field intensity-mapping programs (e.g., \cite{PUMA:2019jwd,Karkare:2022bai}). The results of this section and the further investigation of section ~\ref{subsubsec:halo_HiddenValley} illustrates that aside from the finite simulation volume, the halo–mass resolution of the \texttt{Quijote} simulations is the major contributor to the large noise in our halo measurements. Increasing the tracer number density (e.g., by lowering the mass threshold or using higher-resolution mocks) reduces shot noise and boosts the effective volume $V_{\rm eff}(k)$, thereby lowering the geometric volume required for a significant detection of the parity-odd kurto spectra. For a survey with a fixed number density, weighting galaxies by some of their appropriate properties can also alleviate the the stochasticity \cite{Cai:2010gz,Fang:2023ypv}. We leave investigation of this point to a future work.

\subsection{Identifying the origin of the scatter in halo statistics}\label{subsec:scatter}

In this section, we investigate the sources of scatter in the halo statistics from three angles. First, we apply the $\cP_{2\times2}$ estimator to the \texttt{Hidden Valley} (HV) simulations, where no parity-odd signal is present, and show that increasing the halo number density substantially reduces the variance contributed by the parity-even field. Second, we quantify how the noise level scales with the mean tracer density by randomly down-sampling dark-matter particles in the \texttt{Quijote} snapshots. Third, we measure the halo–matter cross-correlation, which mitigates the stochastic (shot-noise) contribution introduced by the halo field.

\subsubsection*{Null measurement on Hidden Valley 
simulation}\label{subsubsec:halo_HiddenValley}

The HV simulations \cite{Modi:2019ewx} \faGlobe{\footnote{\url{http://cyril.astro.berkeley.edu/HiddenValley/}}} are approximate Nbody simulations that are run with FastPM code \cite{Feng:2016yqz} \faGithub{\footnote{\url{https://github.com/fastpm/fastpm}}}, and follow the dynamics of $10240^3$ particles in periodic cubic boxes of size $L_{\rm box} = 1024 \ {\rm Mpc}/h$. This gives a particle mass resolution of $M_{\rm min} = 8.57 \times 10^7 \ \mathrm{M_\odot}/h$ resolving halos with minimum mass of $\sim 10^9\ \mathrm{M_\odot}/h$. Thus, the HV simulations have a smaller shot noise and lower value of linear halo bias (considering all halos) compared to \texttt{Quijote\ ODD\_p} suite ($b_1^{\rm HV} \approx 1.3$, while $b_1^{\rm Quijote} \approx 4.9$). We specifically use the \texttt{HV10240-R} halo catalog for the measurement.

\begin{figure}[htbp!]
  \centering
  \begin{subfigure}[b]{0.37\textwidth}
    \hspace{-0.1in}\includegraphics[width=\linewidth]{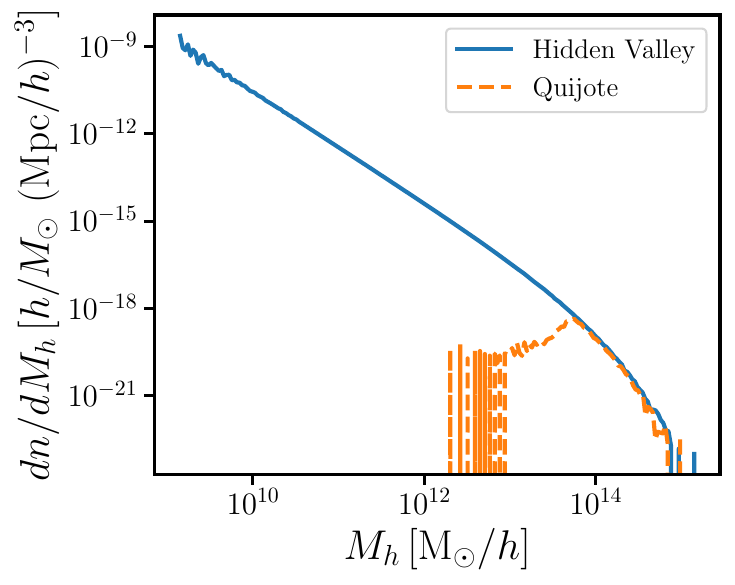}
    \caption{Halo mass functions}
    \label{fig:HMF_HV_Quijote}
  \end{subfigure} 
  \hfill 
  \begin{subfigure}[b]{0.62\textwidth}
    \includegraphics[width=\linewidth]{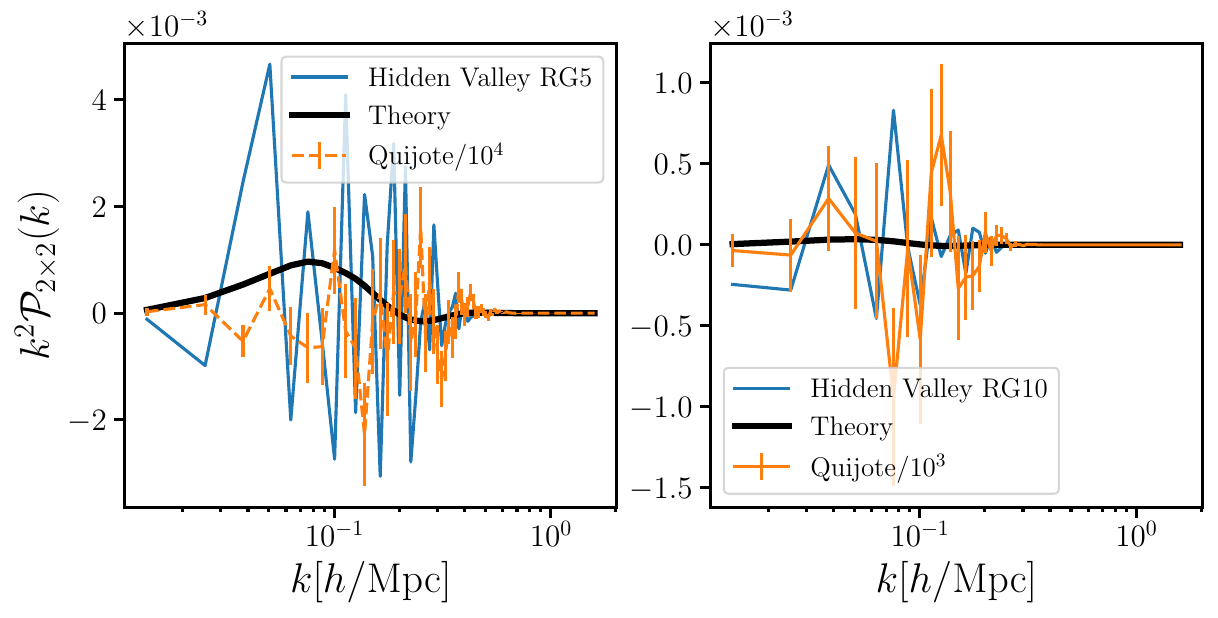}
    \caption{Parity-even contributions for two choices of $R_G$}
    \label{fig:22_HV_Quijote}
  \end{subfigure}
  \label{fig:HV_quijot} \vspace{-0.15in}
  \caption{Comparison between HV (blue) and \texttt{Quijote} fiducial (orange) simulations at $z=1$. figure~\ref{fig:HMF_HV_Quijote} shows the halo mass function while figure~\ref{fig:22_HV_Quijote} shows the parity-even contribution to $\mathcal{P}_{2\times2}^{\rm fid}$ (multiplied by $k^{2}$ for visual clarity), with $R_G=5\,\mathrm{Mpc}/h$ on the left and $R_G=10\,\mathrm{Mpc}/h$ on the right. The black curve is the theoretical prediction of parity-odd signal, obtained by scaling the matter kurto spectrum with the fourth power of the estimated linear biases. Each panel shows a single realization. }
\end{figure}

Even though HV and \texttt{Quijote} (fiducial) simulations have comparable volumes, the higher mass resolution—and thus much larger halo number density—in HV yields substantially lower (parity-even) stochasticity, which acts as a noise limiting detectability of parity-odd kurto spectra. To illustrate the differences in halo abundance and clustering, figure \ref{fig:HMF_HV_Quijote} shows the halo mass functions (\texttt{Quijote} in orange, HV in blue), and figure \ref{fig:22_HV_Quijote} shows the measured $\cP_{2\times2}^{\rm fid}$ for the two suites (same color scheme). In figure~\ref{fig:22_HV_Quijote}, the theory curves are obtained by scaling the matter kurto spectrum by $b_1^4$, and the left and right panels correspond to smoothing scales $R_G=5\,\mathrm{Mpc}/h$ and $R_G=10\,\mathrm{Mpc}/h$, respectively. In both cases we use a single realization. We find that the $\cP_{2\times2}^{\rm fid}$ of \texttt{Quijote} can be $4$ ($3$) orders of magnitude larger than that measured from HV. Although the expected theoretical $\cP_{2\times2}^{\rm PO}$ in the HV suite (if was injected in teh initial conditions) is about $200$ times smaller than in \texttt{Quijote\ ODD\_p} (due to the lower halo bias), the parity-odd kurto spectrum should be detectable at higher significance in HV thanks to its much lower stochasticity.

These results imply that forthcoming Stage-IV spectroscopic surveys---such as DESI and Euclid---should be capable of a high-significance detection of the parity-odd kurto spectra (if present), provided observational systematics are controlled. Their geometric volumes are $\mathcal{O}(40\text{--}50) \\ (\mathrm{Gpc}/h)^3$ (i.e., $\sim 40\times$ the HV volume), and their mean comoving number densities for the primary tracers are typically $\bar n \sim 10^{-4}\!-\!10^{-3}\,h^{3}\mathrm{Mpc}^{-3}$. Both factors suppress the parity-even stochastic contribution and boost the effective volume $V_{\rm eff}(k)$, thereby improving detectability relative to the HV and \texttt{Quijote} setups discussed above.

\begin{figure}[htbp!]
    \centering
    \includegraphics[width=0.45\linewidth]{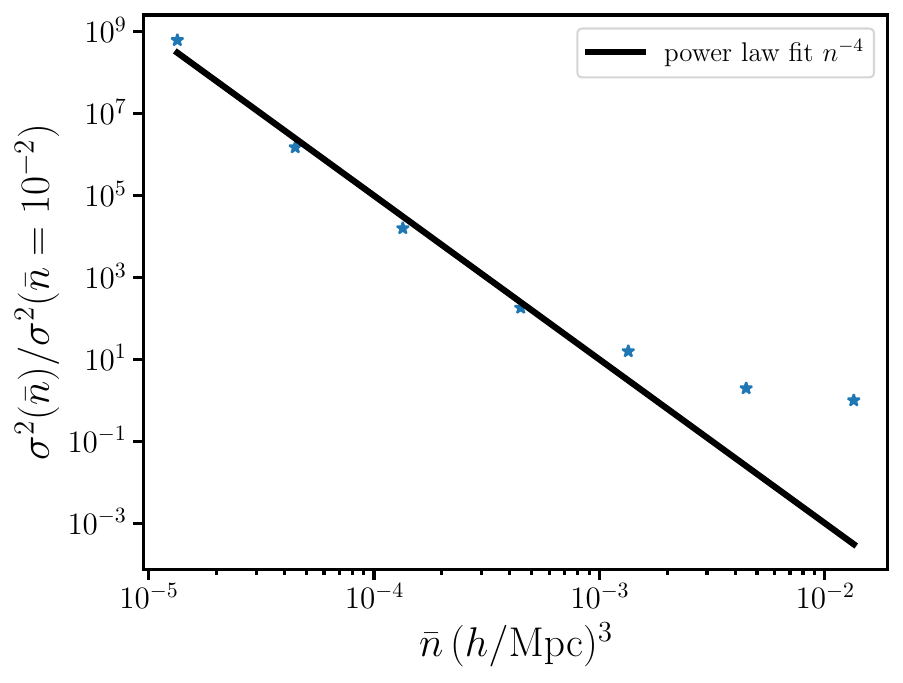}
    \caption{The scattering of $\mathcal{P}^{\rm fid}_{2\times 2}$ on the down-sampled DM field as a function of mean number density.}
    \label{fig:Q_DM_down_sample}
\end{figure}

\paragraph{Down-sampling the DM field} To quantify how the parity-even stochastic (shot-noise) contribution to the halo kurto spectrum scales with the tracer density, we down-sample the \texttt{Quijote} fiducial DM particles. The native mean number density is $\bar n_0 = 1.34\times10^{-1} \ (h/{\rm Mpc})^3$.
We randomly select DM particles to construct seven diluted catalogs with
$\bar n = \{1.34\times10^{-2},4.02\times 10^{-2},1.34\times 10^{-3},4.02\times 10^{-3},1.34\times 10^{-4},4.02\times 10^{-4},1.34\times 10^{-5}\}\, \ (h/{\rm Mpc})^3$.
For each \(\bar n\), we measure the parity-even contribution to  $\mathcal{P}^{\rm fid}_{2\times2}$ kurto spectrum and estimate its variance from 500 realizations,
$ \sigma^2(\bar n)\;\equiv\;
\mathrm{diag}\!\left\{\,\mathrm{Cov}\!\big[\mathcal{P}^{\rm fid}_{2\times2}(k)\big]\,\right\}$,
and we display the ratio relative to the highest-density sample,
$R(\bar n)\equiv\sigma^2(\bar n)/\sigma^2(\bar n_{\rm ref})$, with
$\bar n_{\rm ref}=10^{-2}\,h^{3}\mathrm{Mpc}^{-3}$,
in figure~\ref{fig:Q_DM_down_sample}.

At low number densities (\(\bar n\lesssim 10^{-3}\,h^{3}\mathrm{Mpc}^{-3}\)) the measurements follow a steep power law,
\be
\sigma^2(\bar n)\ \propto\ \bar n^{-\alpha},
\qquad \alpha\simeq 4,
\label{eq:pow_scaling}
\ee
consistent with expectations for kurto spectrum whose variance is dominated by shot-noise terms of the schematic form
\(\big[P(k)+1/\bar n\big]^p\!/V\) with \(p\simeq4\).
In the regime \(1/\bar n \gg P(k)\), eq.~\eqref{eq:pow_scaling} approaches \(\alpha\approx4\).

At high number densities (\(\bar n\gtrsim 10^{-3}\,h^{3}\mathrm{Mpc}^{-3}\)) the curve begins to flatten, indicating the emergence of an \(\bar n\)-independent floor set by the finite number of modes (cosmic variance), the chosen smoothing/assignment scheme, and the intrinsic non-Gaussianity of the field. A convenient approximation that captures both regimes is
\be
\sigma^2(\bar n)\;=\;\sigma^2_{\rm CV}\;+\;A\,\bar n^{-\alpha},
\qquad \alpha\simeq 3.8,
\label{eq:two_comp_model}
\ee
for which the plotted ratio becomes
\be
\frac{\sigma^2(\bar n)}{\sigma^2(\bar n_{\rm ref})}
\;=\;
\frac{\sigma^2_{\rm CV}+A\,\bar n^{-\alpha}}
     {\sigma^2_{\rm CV}+A\,\bar n_{\rm ref}^{-\alpha}}.
\label{eq:ratio_model}
\ee
If \(\bar n_{\rm ref}\) already lies in the \(nP\gg1\) regime, then \(\sigma^2(\bar n_{\rm ref})\approx\sigma^2_{\rm CV}\) and the denominator is nearly constant, which explains the apparent saturation.

Given the trend observed for the down-sampled dark-matter field, $\sigma^2(\bar n)\propto \bar n^{-\alpha}$ with $\alpha\simeq3.8$, we scale the \emph{rms} parity-even noise from simulations to surveys as
\be
\label{eq:survey_noise_level_estimate}
\mathcal{\sigma}^{\rm fid}_{\rm survey} = \mathcal{\sigma}^{\rm fid}_{\rm sim} \frac{\sqrt{\bar{n}_{\rm sim}^\alpha V_{\rm sim}}}{\sqrt{\bar{n}_{\rm survey}^\alpha V_{\rm survey}}} 
\ee
which is effectively bias-independent in the shot-noise regime ($nP\ll1$). If $nP\gg1$, an additive, $\bar n$–independent floor $\sigma_{\rm CV}$ should be included. For the signal we adopt the leading-order scaling
\be
\mathcal{P}^{\rm sig}_{\rm survey}
\simeq
\left(\frac{b_{\rm survey}}{b_{\rm sim}}\right)^{4}
\mathcal{P}^{\rm sig}_{\rm sim}.
\ee

\begin{figure}[t]
    \centering
    \begin{subfigure}[b]{0.485\textwidth}
    \hspace{-0.15in}\includegraphics[width=\linewidth]{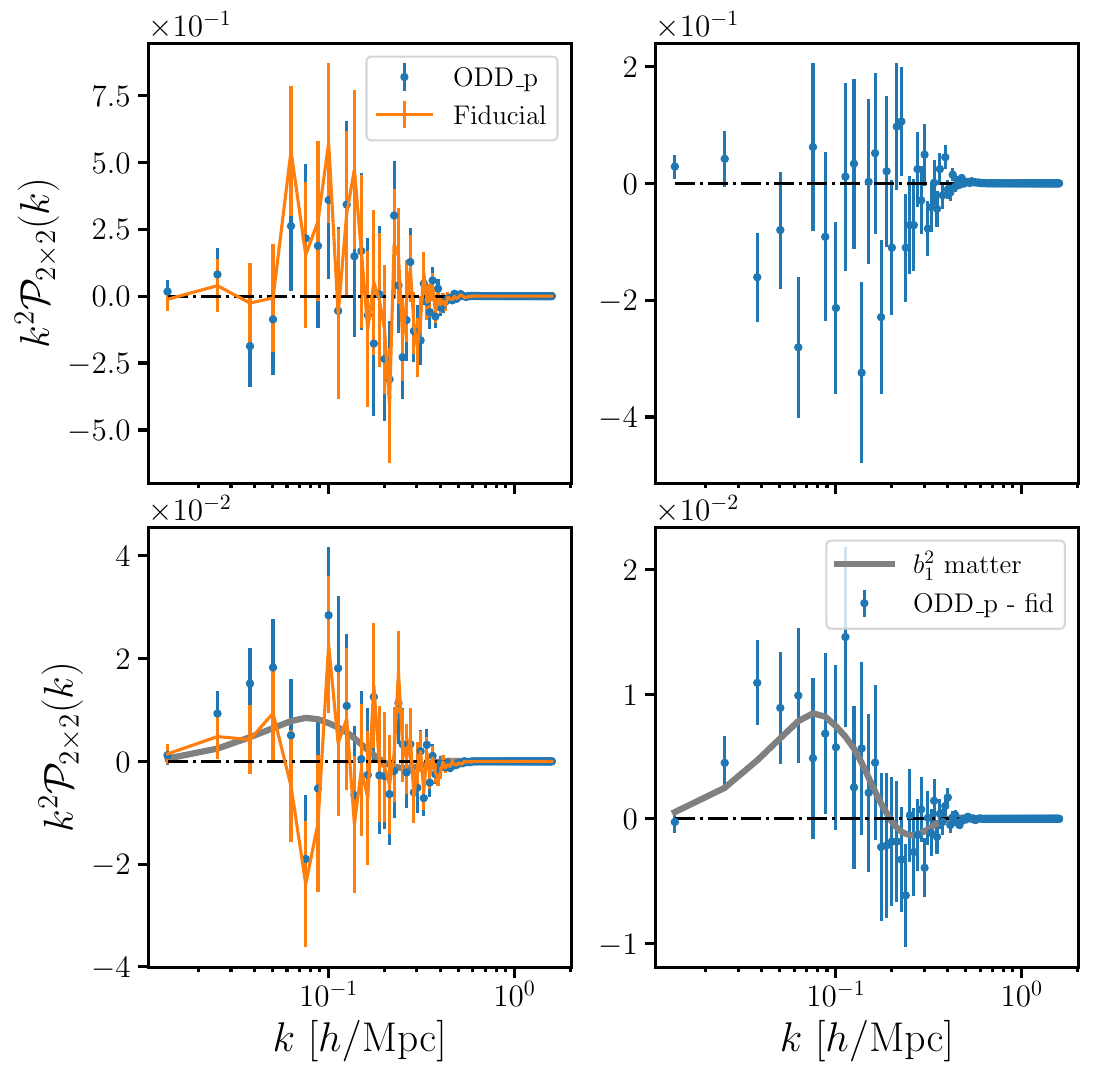}
    \caption{$R_G = 5\,\mathrm{Mpc}/h$}
    \label{fig:hhmm_2x2_RG5}
  \end{subfigure}
  \begin{subfigure}[b]{0.485\textwidth}
    \includegraphics[width=\linewidth]{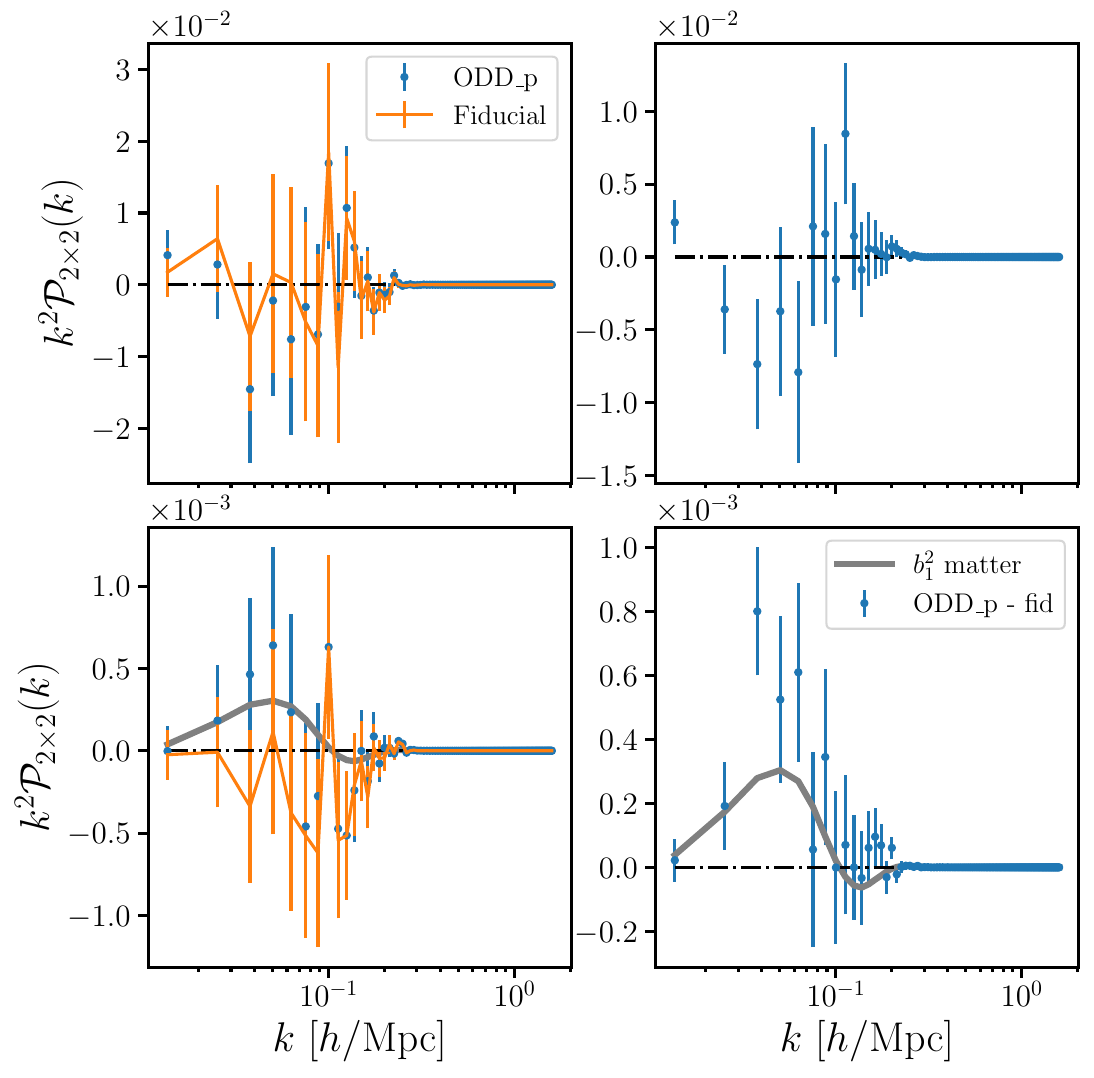}
    \caption{$R_G = 10\,\mathrm{Mpc}/h$}
    \label{fig:hhmm_2x2_RG10}
  \end{subfigure}
    \caption{Real-space halo-matter cross kurto spectrum $\cP_{2\times 2}$ from \texttt{Quijote} simulations, scaled with a factor of $k^2$, for $R_{G}=5\, \mathrm{Mpc}/h$ (figure \ref{fig:hhmm_2x2_RG5}) and $R_{G}=10\, \mathrm{Mpc}/h$ (figure \ref{fig:hhmm_2x2_RG10}). Top row: single realization. Bottom row: average over 500 realizations.  Left panels: \texttt{Quijote ODD\_p} (blue dots) and matched fiducial simulation (orange solid). Right panels: the total parity-odd signal from the fiducial-subtracted simulations (blue dots). The gray lines are the the matter measurement scaled by a factor of $b_1^2$.}
    \label{fig:hhmm_RG5_RG10}
\end{figure}

\paragraph{Halo and matter field cross correlation} Next, we measure the cross parity-odd kurto spectra of halo and matter fields, focusing on $\cP_{2\times 2}$ spectrum. There are two options to perform cross-correlation: (a) construct the $\vec V $ and $\vec A$ fields purely from halo DM and halo fields, respectively, then correlate them. We note this as $\langle hhmm\rangle$; (b) construct each of the two vectorial fields from one halo and one DM field, noted as $\langle hmhm \rangle$. As expected, the level of stochastic noise is lower in the first type since the leading-order shot noise doesn't contribute, while in the second type there is a contribution from shot noise from correlation of the two halo fields in each of the composite fields. Thus, we only demonstrate the first type of cross correlation. Again, we consider $500$ realizations of \texttt{Quijote ODD\_p} and fiducial simulations and compare the measurements for the two smoothing scales of $R_{G}=5\, \mathrm{Mpc}/h$ (figure~\ref{fig:hhmm_2x2_RG5}) and $R_{\rm G}=10\, \mathrm{Mpc}/h$ (figure~\ref{fig:hhmm_2x2_RG10}). The difference between fiducial and parity-odd simulation starts to show non-zero signature when averaged over $500$ realizations, while the \texttt{ODD\_p} measurement without subtraction is still noisy. 

\begin{figure}[t]
    \centering
    \begin{subfigure}[b]{0.485\textwidth}
    \hspace{-0.15in}\includegraphics[width=\linewidth]{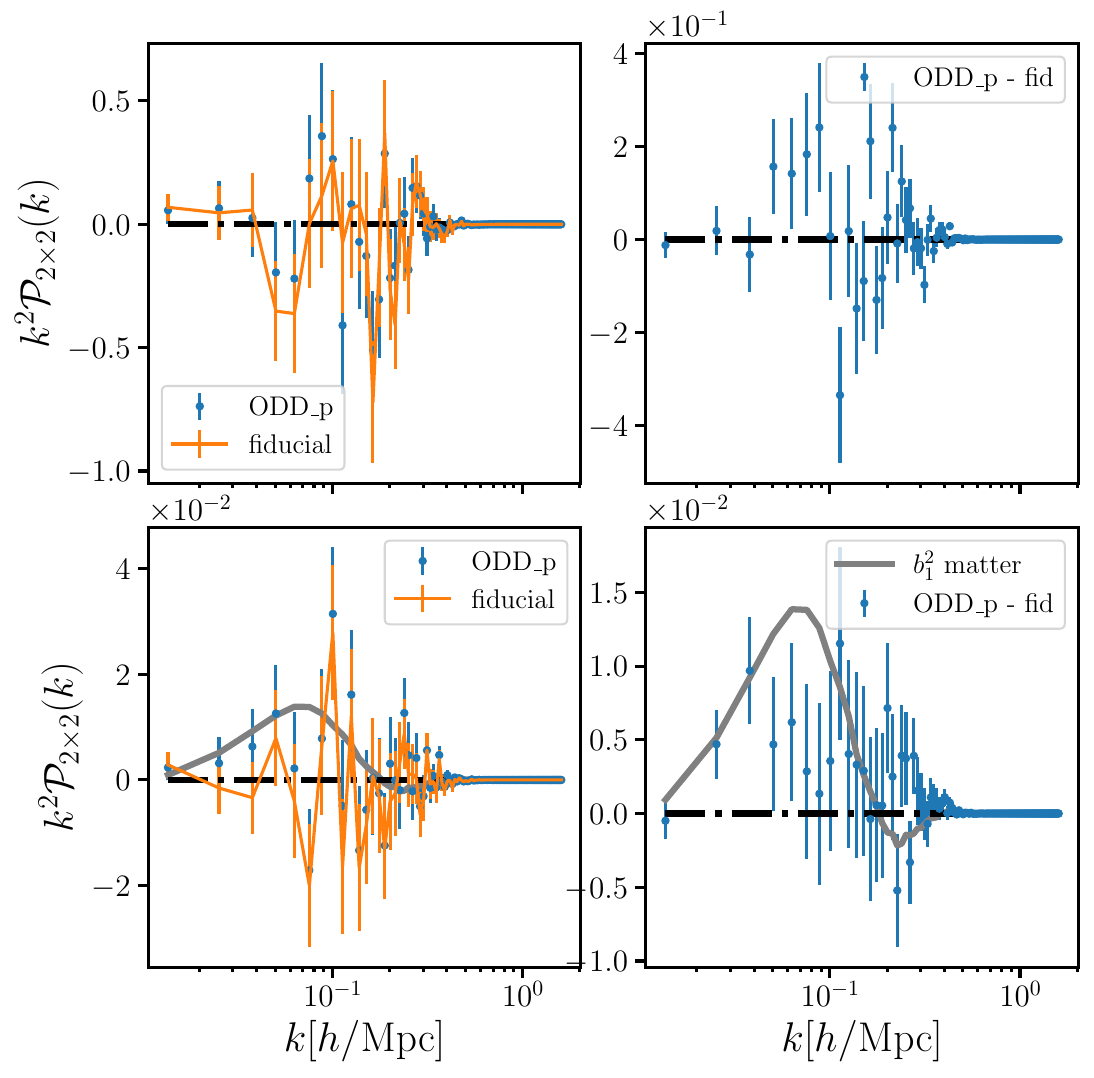}
    \caption{$R_G = 5\,\mathrm{Mpc}/h$}
    \label{fig:hhmm_RSD_2x2_RG5}
  \end{subfigure}
  \begin{subfigure}[b]{0.485\textwidth}
    \includegraphics[width=\linewidth]{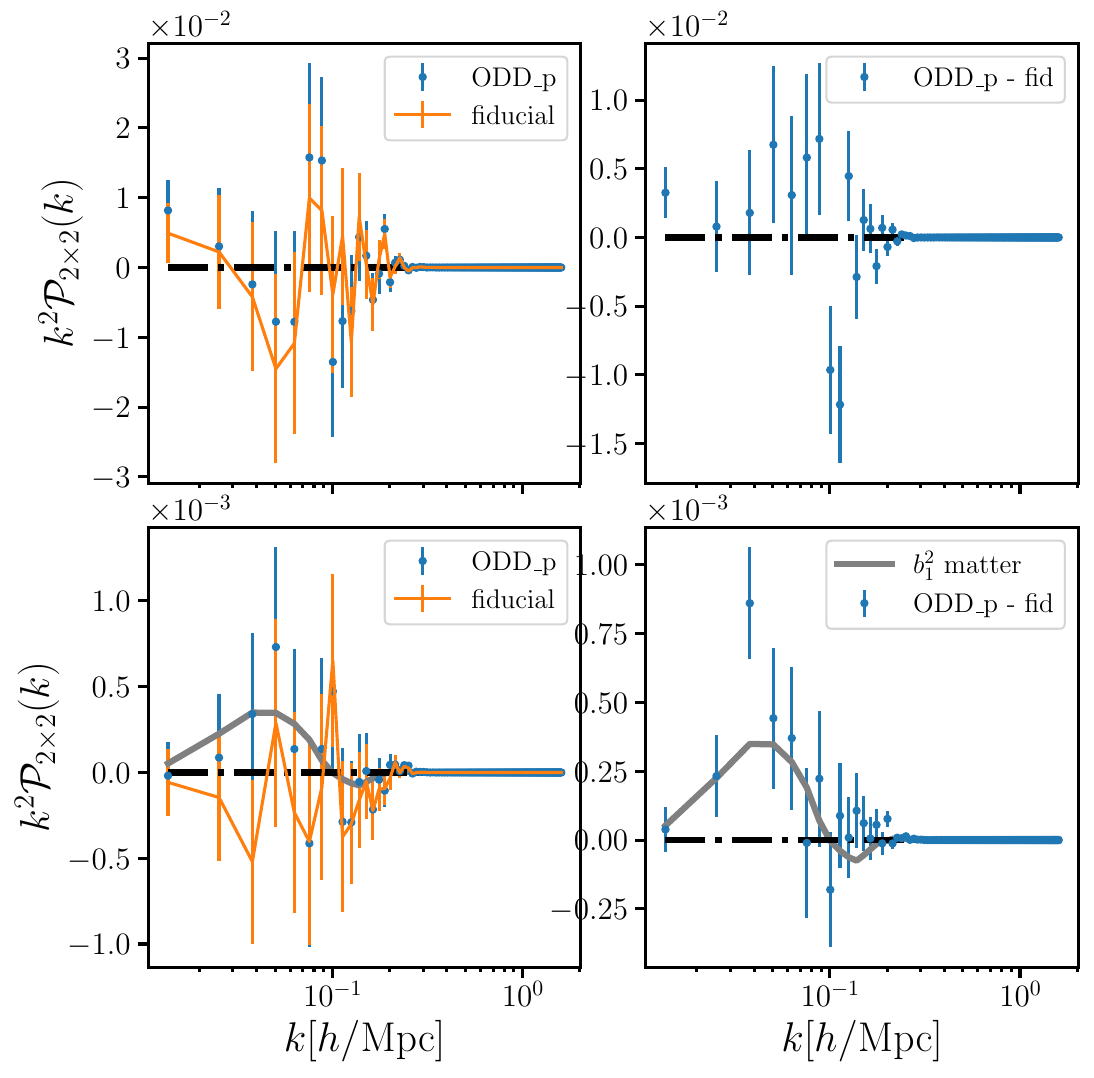}
    \caption{$R_G = 10\,\mathrm{Mpc}/h$}
    \label{fig:hhmm_RSD_2x2_RG10}
  \end{subfigure}
    \caption{The same setup as figure~\ref{fig:hhmm_RG5_RG10}, only with RSD effect.}
    \label{fig:hhmm_RSD_RG5_RG10}
\end{figure}

We show the redshift-space measurement of the halo-matter cross spectrum in figure \ref{fig:hhmm_RSD_RG5_RG10} for two smoothing scales of $R=5 \ {\rm Mpc}/h$ and $R=10 \ {\rm Mpc}/h$. For filtering length of $R_G = 5\, \mathrm{Mpc}/h$ the cross-correlation in real-space is less noisy than that with RSD, while in the $R_G = 10\, \mathrm{Mpc}/h$ case, the two are similar. This demonstrate that the RSD increase the stochasticity in halo field and that smoothing the field on larger scales which wash out the impact of small-scale velocities, suppresses the redshift-space kurto spectra on relatively large scales at its peak. 

\subsection{Kurto spectra with optimal weighting}\label{sec:opt_filter}

In this section, we apply the optimal weighting described in section \ref{subsec:skew_kurto} to parity-odd kurto spectra of matter field and halo catalogs of \texttt{Quijote} simulations and show that this weighting increase the detectability of the parity-odd signal. Additionally, we also replace the Gaussian smoothing filter with a (smooth) Heaviside-step-function like filter, as
\be
W(k)=\frac{1}{2}\Bigl[1-\tanh\!\Bigl(\alpha\,\ln\!\frac{k}{k_{\max}}\Bigr)\Bigr],
\ee
where we choose $\alpha=50$ and $k_{\rm max}=0.45\, h/\mathrm{Mpc}$ to match the scale cut-off similar to the Gaussian filtering with $R_G=5\, \mathrm{Mpc}/h$.

To match the parity-odd kurto spectra to the ML estimator for the trispectrum amplitude $g_{\rm NL}^{\rm PO}$, we adopt the optimal (inverse-variance) weighting for the separable template in eq.~\eqref{eq:Tg_lin} (with kernel in eq.~\eqref{eq:Bsep}). The resulting cubic field entering $\cP_{3\times1}$ in eq.~\eqref{eq:D_3_beta} is
\begin{align}
\mathcal K_\beta(\vk)
&= \epsilon_{ijk} \int_{\bq_1}\!\int_{\bq_2}\,
A_i(\bq_1)\,B_j(\bq_2)\,C_k(\vk-\bq_1-\bq_2)\non\\
&\hspace{1.2em}\times\;
\frac{\delta_{\rm obs}(\bq_1)}{\mathcal{M}(q_1)}\frac{P_L(q_1)}{P_{\rm obs}(q_1)}\,
\frac{\delta_{\rm obs}(\bq_2)}{\mathcal{M}(q_2)}\frac{P_L(q_2)}{P_{\rm obs}(q_2)}\,
\frac{\delta_{\rm obs}(\vk-\bq_1-\bq_2)}{\mathcal{M}(|\vk-\bq_1-\bq_2|)} \frac{P_L(|\vk-\bq_1-\bq_2|)}{P_{\rm obs}(|\vk-\bq_1-\bq_2|)} ,
\end{align}
where $A,B,C$ are the kernel factors from eq.~\eqref{eq:K3_Coulton}, $\delta_{\rm obs}$ is the observed (matter or halo) overdensity field, and $P_{\rm obs}$ includes noise. The two quadratic composite fields entering $\cP_{2\times2}$ are
\begin{align}
\mathcal S_{\beta,12}(\vk)
&= \epsilon_{ijk}\int_{\bq}\,
A_i(\bq)\,B_j(\vk-\bq)\,
\frac{\delta_{\rm obs}(\bq)\,P_L(q)}{\mathcal{M}(q)\,P_{\rm obs}(q)}\,
\frac{\delta_{\rm obs}(\vk-\bq)\,P_L(|\vk-\bq|)}{\mathcal{M}(|\vk-\bq|)\,P_{\rm obs}(|\vk-\bq|)} ,\\
\mathcal S_{\beta,34}(\vk)
&= \int_{\bq}\,
C_k(\bq)\,\mathcal{M}(|\vk-\bq|)\,
\frac{\delta_{\rm obs}(\bq)\,P_L(q)}{\mathcal{M}(q)\,P_{\rm obs}(q)}\,
\frac{\delta_{\rm obs}(\vk-\bq)}{P_{\rm obs}(|\vk-\bq|)} .
\end{align}

\begin{figure}[htbp!]
    \centering
    \begin{subfigure}[b]{0.485\textwidth}
    \hspace{-0.15in}\includegraphics[width=\linewidth]{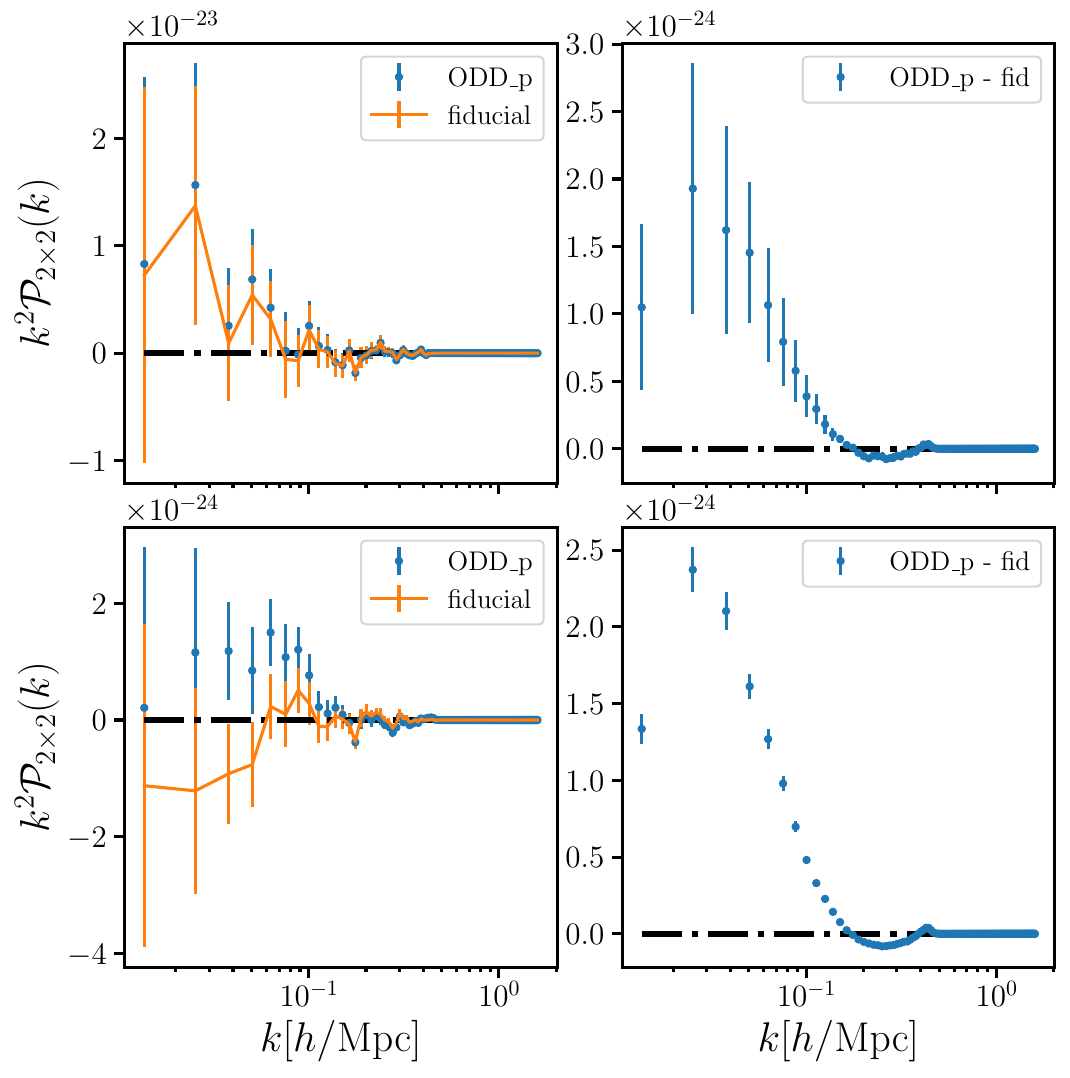}
    \caption{matter (40)}
    \label{fig:tanh_2x2_matter}
  \end{subfigure}
  \begin{subfigure}[b]{0.485\textwidth}
    \includegraphics[width=\linewidth]{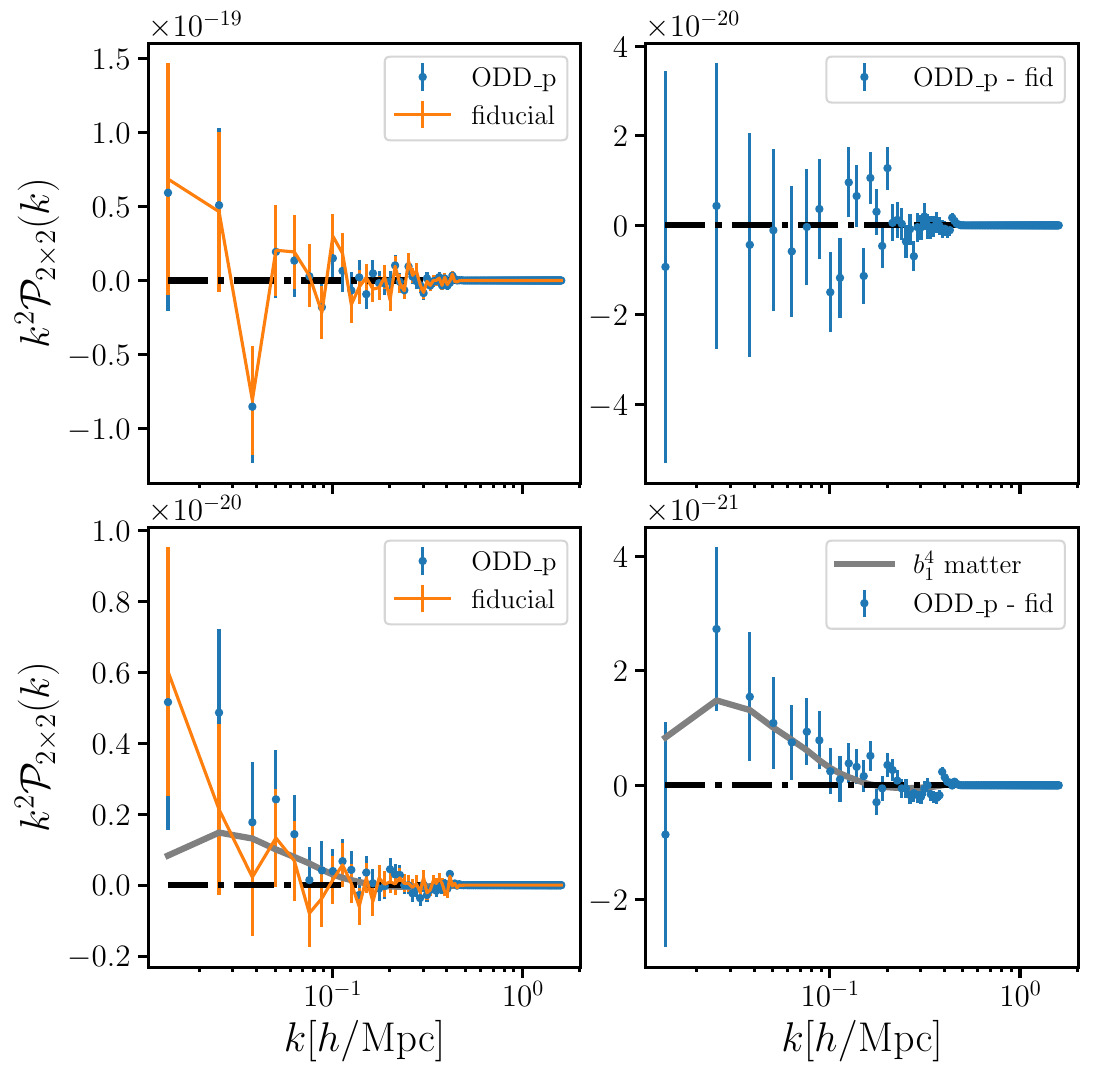}
    \caption{halo (500)}
    \label{fig:tanh_2x2_halo}
  \end{subfigure}
    \caption{Real-space kurto spectrum $\mathcal{P}_{2\times 2}$ of matter (figure \ref{fig:tanh_2x2_matter}) and halos (figure \ref{fig:tanh_2x2_halo}) using the optimal weighting (to match ML estimator) and tanh smoothing kernel (compatible to smoothing with a Gaussian filter with $R_G=5 \ {\rm Mpc}/h$). Top row: single realization. Bottom row: average over 500 realizations. Left panels: \texttt{Quijote ODD\_p} (blue dots) and matched fiducial simulation (orange solid). Right panels: the total parity-odd signal from the fiducial-subtracted simulations (blue dots). The gray line is the theoretical predictions obtained by $b_1^4 \mathcal{P}^{\rm po - fid}$ (linear scaling of blue dots in figure \ref{fig:tanh_2x2_matter}).}
    \label{fig:tanh_2x2}
\end{figure}

\begin{figure}
    \begin{subfigure}[b]{0.485\textwidth}
    \hspace{-0.15in}\includegraphics[width=\linewidth]{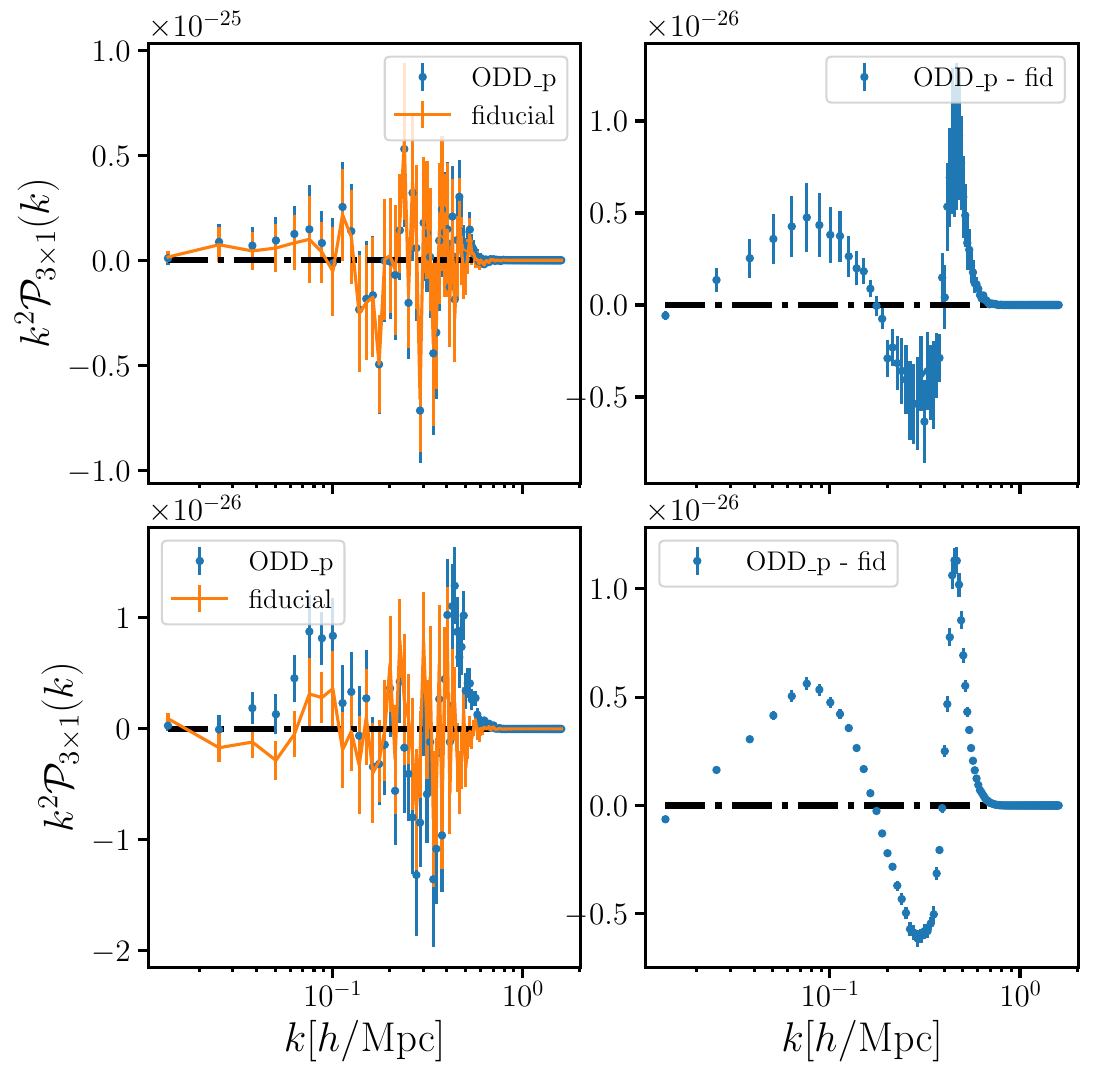}
    \caption{matter (40)}
    \label{fig:tanh_3x1_matter}
  \end{subfigure}
  \begin{subfigure}[b]{0.485\textwidth}
    \includegraphics[width=\linewidth]{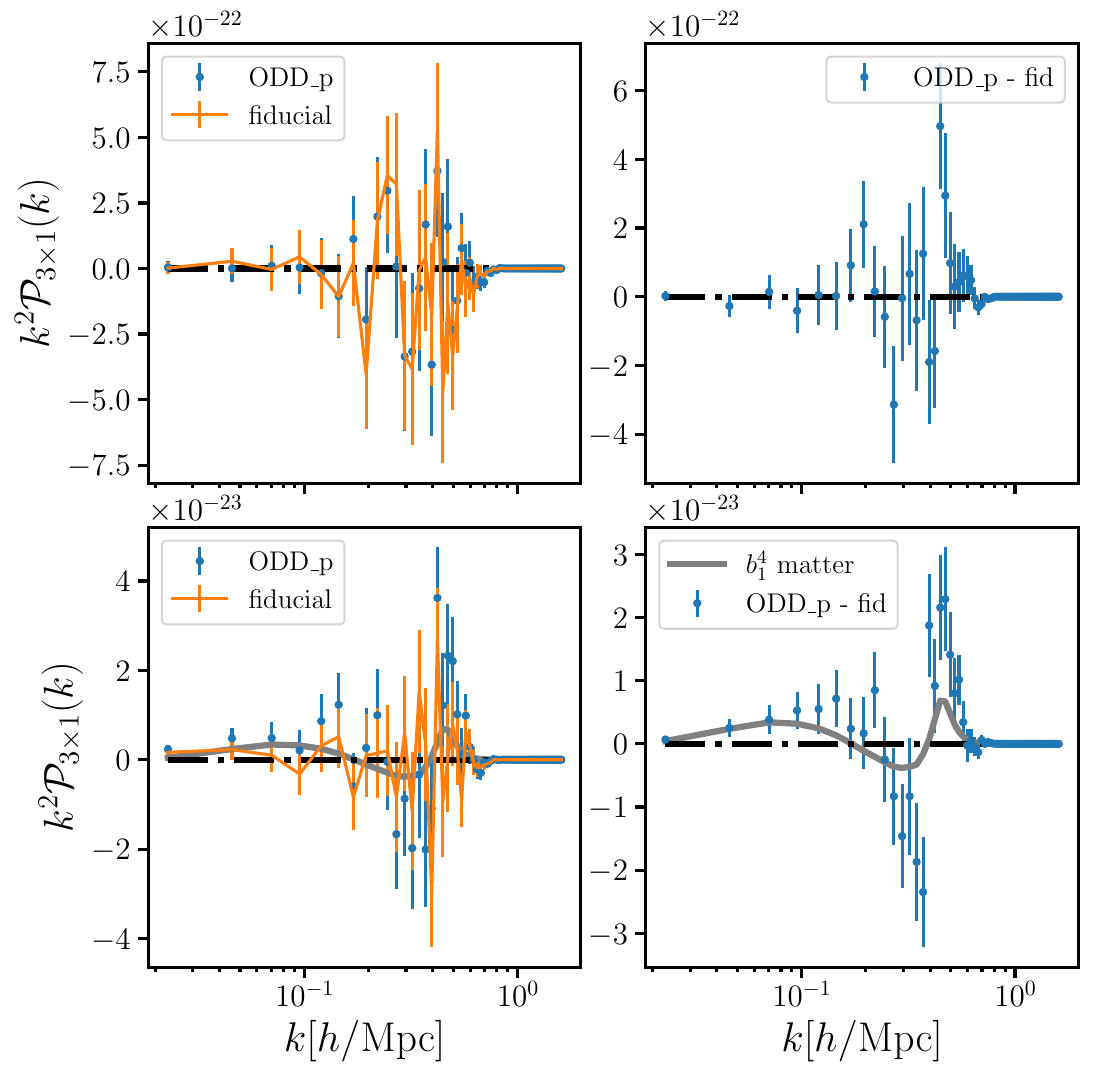}
    \caption{halo (500)}
    \label{fig:tanh_3x1_halo}
  \end{subfigure}
    \caption{Same as figure \ref{fig:tanh_2x2} but for $\mathcal{P}_{3\times 1}$ kurto spectrum.}
    \label{fig:tanh_3x1}
\end{figure}

In figures \ref{fig:tanh_2x2} and \ref{fig:tanh_3x1}, we show the real-space measured $\cP_{2\times 2}$ and $\cP_{3\times 1}$ spectra on fiducial and parity-odd \texttt{Quijote} matter field and halo catalogs with optimal weighting and tanh smoothing kernel (matching the scales included using the Gaussian smoothing scale of $R=5 \ {\rm Mpc}/h$). Again, the top row in each plot are measurements on a single realizations while the bottom rows are averaged over 40 (500) realizations for matter (halos). Compared with figures \ref{fig:halo_2x2_RG5} and \ref{fig:halo_3x1_RG5}, we see that the optimal weighting clearly brings out the parity-odd signal both for matter and halos. For matter, the signal is even detectable by eye without subtracting the fiducial simulations.

We present a quantitative comparison of the effects of optimal weighting and filter choice in Table~\ref{tab:detectability_match_filter} (Sec.~\ref{subsec:cov_SNR_detectability}). The results show that the SNR gain arises from a \emph{combination} of the optimal weighting and the smooth, nearly top-hat (tanh) window. The optimal estimator yields the largest improvement for the difference statistic $\Delta\cP$, whereas replacing the Gaussian filter with the tanh window enhances both $\Delta\cP$ and the direct \texttt{ODD\_p} measurement. A plausible interpretation is that the optimal weights primarily \emph{boost the signal} along the target template, while the tanh window \emph{reduces mode coupling}, limiting leakage of small-scale stochasticity into large-scale cosmological modes.

\subsection{Covariance matrices, signal-to-noise ratio, and detectability}\label{subsec:cov_SNR_detectability}

In this section, we (i) present covariance matrices for the two parity-odd kurto spectra, (ii) define the cumulative signal-to-noise ratio (SNR) and a matched-filter \emph{detectability} statistic, and (iii) provide an illustrative Euclid-like forecast for the expected SNR and detectability.

\begin{figure}[htbp!]
    \centering
    \includegraphics[width=\linewidth]{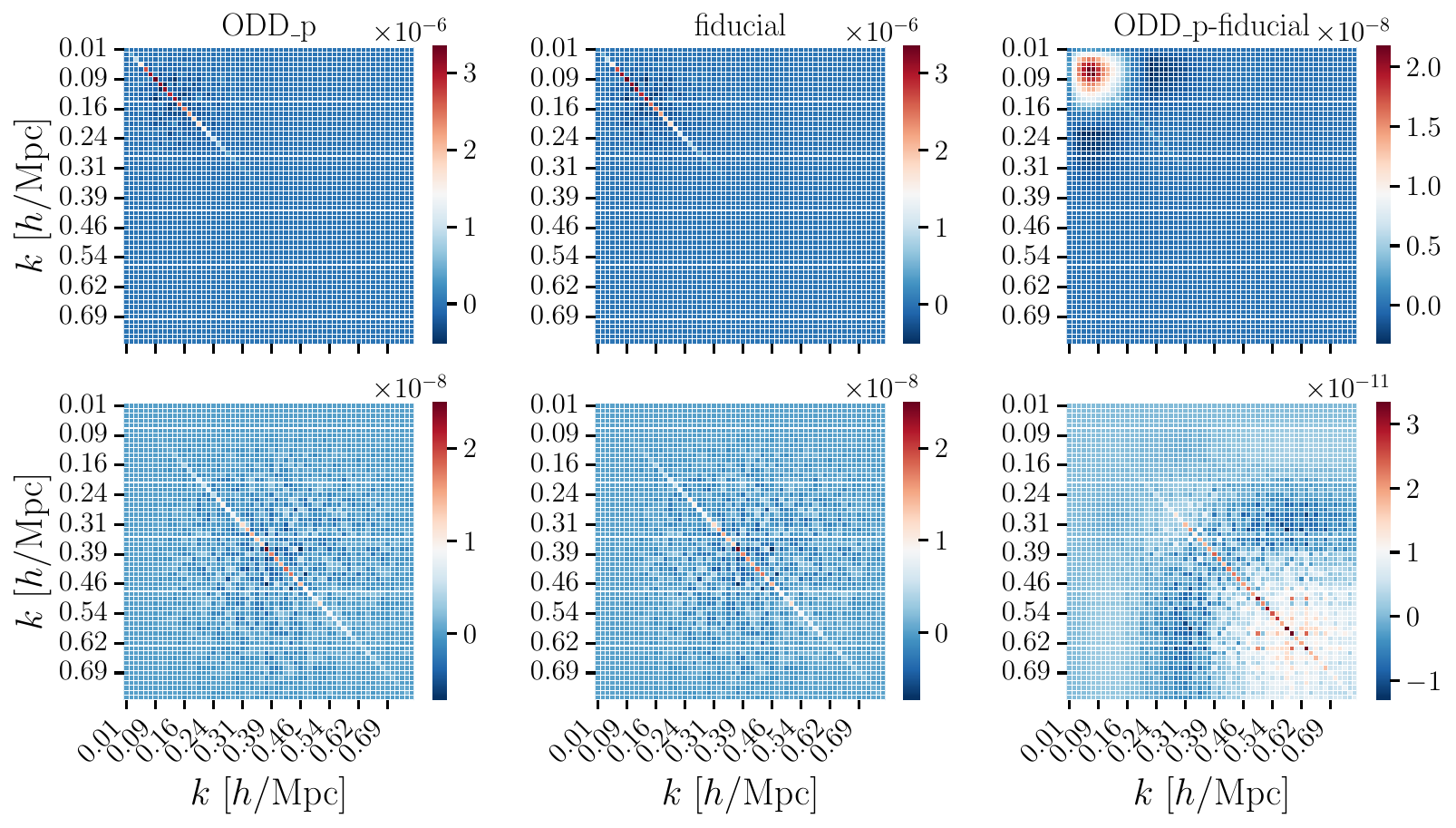}
    \caption{Covariance matrices of the real-space matter kurto spectra constructed from $500$ simulations at $z=1$ with $R_G=5\ {\rm Mpc}/h$. First row: $k^2\cP_{2\times 2}$; second row: $k^2\cP_{3\times 1}$. Columns (left to right): \texttt{ODD\_p}, fiducial, and fiducial-subtracted ($k^2\Delta\mathcal{P}$).}\label{fig:cov_2x2_3x1_RG5_m}
    \end{figure}

\begin{figure}
    \includegraphics[width=\linewidth]{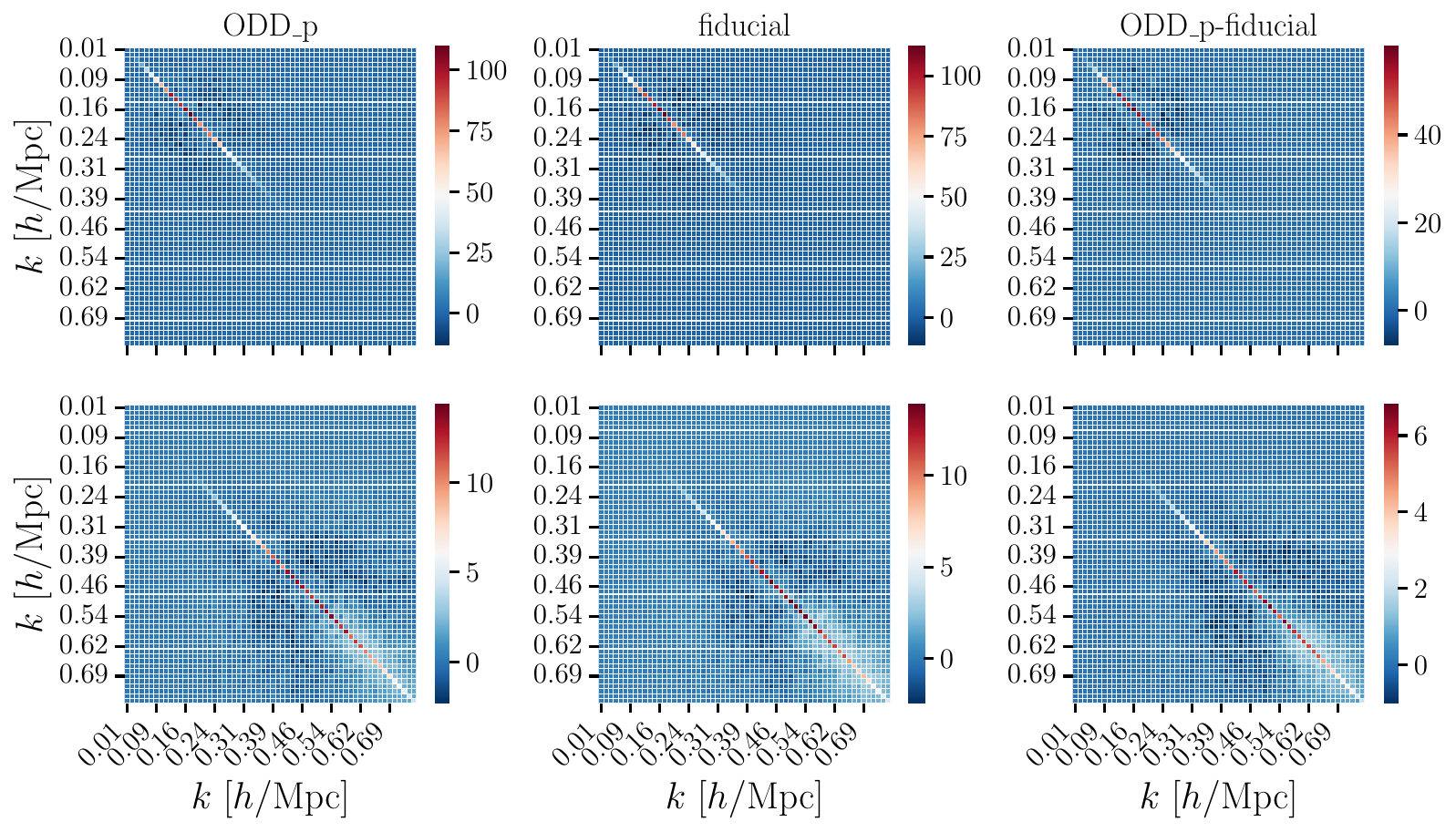}
    \caption{Same as figure \ref{fig:cov_2x2_3x1_RG5_m}, but for halos.}
    \label{fig:cov_2x2_3x1_RG5_h}
\end{figure}

For a data vector $\mathbf{d}=\{d_i\}$ (with $i$ indexing $k$-bins), we estimate the covariance with the unbiased sample estimator,
\be
\label{eq:cov_matrix}
\mathbf{C}_{ij}
= \frac{1}{N-1}\sum_{n=1}^{N}\big(d^{(n)}_i - \bar{d}_i\big)\big(d^{(n)}_j - \bar{d}_j\big),
\qquad
\bar{d}_i \equiv \frac{1}{N}\sum_{n=1}^{N} d^{(n)}_i .
\ee
For realization-averaged measurements, the per-realization estimate is scaled by the number of realizations used in averaging. We use two signal data vectors: (a) the ensemble mean of the per–realization difference $\Delta\cP(k)\equiv \cP(\texttt{ODD\_p})-\cP(\texttt{fid})$, and (b) the ensemble means $\cP$(\texttt{ODD\_p}),  $\cP$(\texttt{fid}). The error bars in previous figures correspond to the \emph{diagonal} of covariance matrices estimated from $N=500$ realizations of the \texttt{Quijote} fiducial and \texttt{ODD\_p} simulations.  In figure~\ref{fig:cov_2x2_3x1_RG5_m}, we show the full covariance matrices of the real-space matter $k^2\cP_{2\times2}$ (first row) and $k^2\cP_{3\times1}$ (second row) at $z=1$; columns correspond to \texttt{ODD\_p}, fiducial, and the difference $k^2\Delta\cP$. All results use Gaussian smoothing $R_G=5\,\mathrm{Mpc}/h$. The non-subtracted covariances are strongly diagonal, with off-diagonal correlations becoming more prominent near the scales where the signal peaks. For the covariance of subtracted signals, their overall amplitudes track the signal envelopes: $k^2\cP_{2\times2}$ peaks at larger scales, $k^2\cP_{3\times1}$ at smaller scales. \texttt{ODD\_p} and fiducial covariances are similar in amplitude (and much larger than the subtracted case), indicating that parity-even fluctuations dominate the noise. In the case with dominant parity-even noise, the shape covariance depends on the choice of kurto spectra and filtering functions. In contrast, the covariance of $k^2\Delta\cP$ shows pronounced off-diagonal structure near the peak, reflecting non-Gaussian mode coupling in the parity-odd signal. The corresponding halo covariances (figure~\ref{fig:cov_2x2_3x1_RG5_h}) are comparable across \texttt{ODD\_p}, fiducial,  $k^2\Delta\cP$, consistent with stochastic (shot-noise–like) contributions dominating in all cases.


\begin{figure}[t]
    \centering
    \hspace{-0.1in}\begin{subfigure}[b]{0.45\textwidth}
    \includegraphics[width=0.94\linewidth]{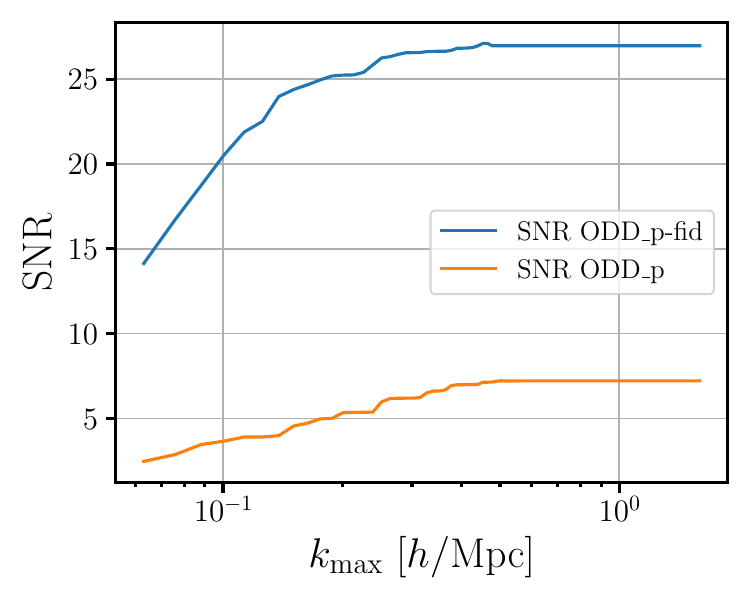}
    \caption{matter}
    \end{subfigure}
    \begin{subfigure}[b]{0.45\textwidth}
    \includegraphics[width=0.94\linewidth]{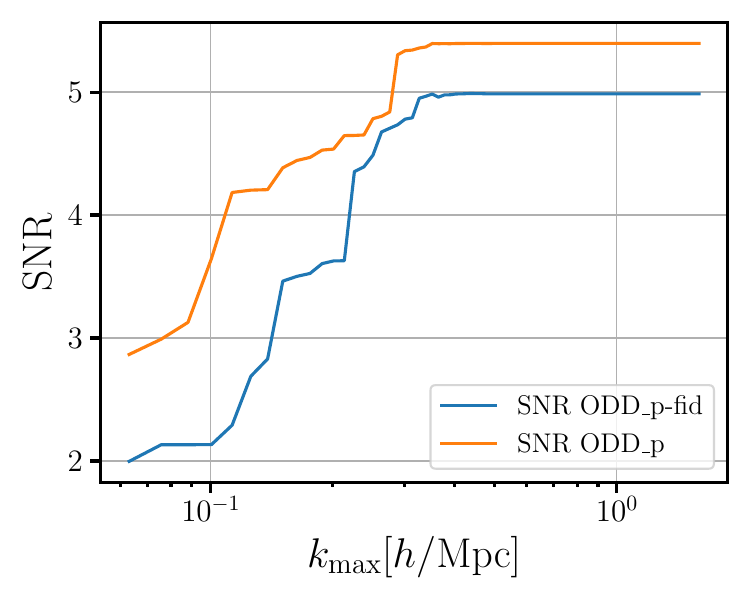}
    \caption{halo}
    \end{subfigure}
    \caption{Cumulative SNR for the real-space matter (left) and halo (right) $\mathcal{P}_{2\times2}$ as a function of the maximum wavenumber $k_{\max}$, using a Gaussian smoothing scale $R_G=5 \ {\rm Mpc}/h$. Two signal definitions are shown: (blue) $\cP^{\rm PO}_{2\times2}(k)-\cP^{\rm fid}_{2\times2}(k)$, and (orange) $\cP^{\rm PO}_{2\times2}(k)$. The covariance is estimated from 500 fiducial and \texttt{ODD\_p} realizations.}
    \label{fig:SNR_2x2_RG5}
    \vspace{0.2in}
    \hspace{-0.1in}\begin{subfigure}[b]{0.45\textwidth}
    \includegraphics[width=0.94 \linewidth]{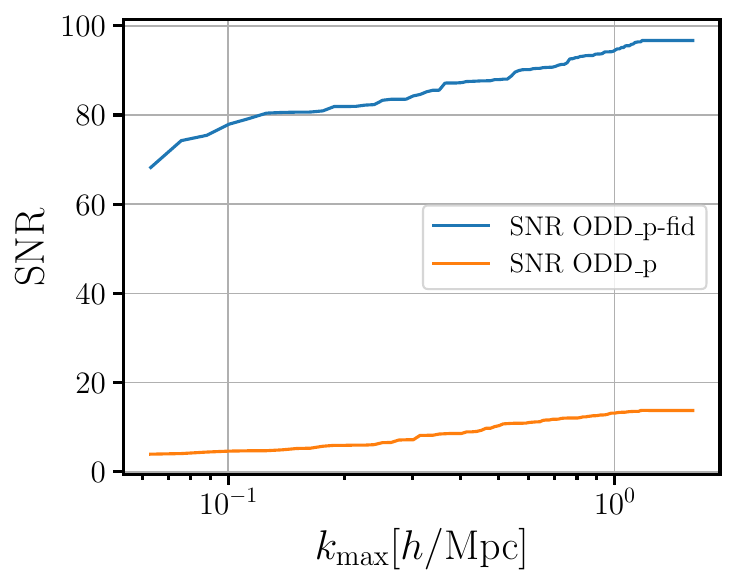}
    \caption{matter}
    \end{subfigure}
    \begin{subfigure}[b]{0.45\textwidth}
    \includegraphics[width= 0.94 \linewidth]{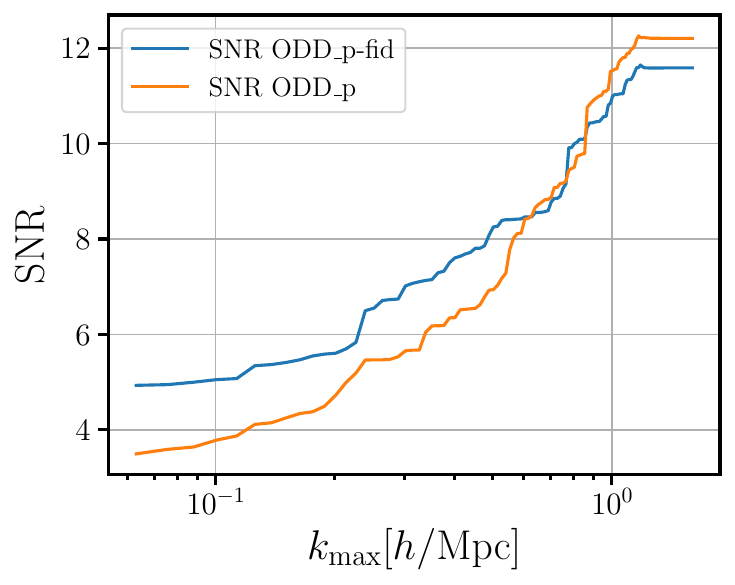}
    \caption{halo}
    \end{subfigure}
    \caption{Same as figure \ref{fig:SNR_2x2_RG5} but for $\mathcal{P}_{3\times 1}$.}
    \label{fig:SNR_3x1_RG5}
\end{figure}

Next, we compute the cumulative SNR up to $k_{\max}$ using
\be
\label{eq:SNR}
\mathrm{SNR}^2(k_{\max}) \;=\; \sum_{k_i,k_j\le k_{\max}} d_i\,\mathbf{C}^{-1}_{ij}\,d_j,
\ee
for the two data vectors defined above. The covariance is inverted via an SVD with a small–eigenvalue cutoff to stabilize the inversion at high $k$. The resulting SNR curves for $R_G=5\,\mathrm{Mpc}/h$ are shown in figure~\ref{fig:SNR_2x2_RG5} for $\cP_{2\times2}$ and figure~\ref{fig:SNR_3x1_RG5} for $\cP_{3\times1}$. For DM field, as expected, the subtracted data vector $\Delta\cP$ yields considerably higher SNR than the direct parity-odd measurement. For halos, the SNR is considerably lower than the DM and subtraction of fiducial kurto spectra does not systematically improve the SNR due to dominant stochasticity. In both DM an halo, $\mathcal{P}_{3\times1}$ attains higher SNR than $\mathcal{P}_{2\times2}$, reflecting the larger number of contributing $k$-modes.

In addition to SNR, we quantify the detectability of a fixed template $\mathbf f$ (from a primordial trispectrum model) using a matched filter~\cite{Zubeldia:2021mzx}. We model the measured data vector as
\be
\mathbf d \;=\; s_0\,\mathbf f \;+\; \mathbf n,\qquad \mathbf n\sim\mathcal N(\mathbf 0,\mathbf C),
\ee
with full covariance $\mathbf C$. The maximum–likelihood amplitude and its $1\sigma$ uncertainty are
\be
\hat s_0 \;=\; \frac{\mathbf f^{\top}\mathbf C^{-1}\mathbf d}{\mathbf f^{\top}\mathbf C^{-1}\mathbf f},
\qquad
\sigma_{s_0} \;=\; \big[\mathbf f^{\top}\mathbf C^{-1}\mathbf f\big]^{-1/2}.
\ee
The corresponding detection statistic (a $z$–score) is
\be
\nu \;\equiv\; \frac{\hat s_0}{\sigma_{s_0}}
\;=\; \frac{\mathbf f^{\top}\mathbf C^{-1}\mathbf d}{\sqrt{\mathbf f^{\top}\mathbf C^{-1}\mathbf f}}\,.
\ee
Under the null (no signal) with Gaussian noise, $\nu\sim\mathcal N(0,1)$, so $|\nu|$ gives the significance in “$1\sigma$” units and $\Delta\chi^2\equiv\nu^2\sim\chi^2_{(1)}$. For intuition, $\nu=0$ is noise-consistent; $|\nu|=1$ is a common $1\sigma$ fluctuation; $|\nu|=3$ is rare. The
sign indicates alignment $(+)$ or anti-alignment $(-)$ with the template. 

We report $\nu$ for matter, halos, and their cross kurto spectra (various smoothing scales) in Table~\ref{tab:detectability_match_filter}, always using the \emph{full} covariance $\mathbf C$. When averaging over $N_{\rm real}$ realizations, we use the covariance of the mean $\mathbf C_{\rm mean}=\mathbf C/N_{\rm real}$. Note that the plotted SNR uses the quadratic norm
$\mathrm{SNR}^2=\bar{\mathbf d}^{\top}\mathbf C_{\rm mean}^{-1}\bar{\mathbf d}$, which collects \emph{all} power in the averaged data (signal \emph{and} noise) and therefore grows with the number of $k$-bins even under the null. The matched filter, $\nu$, projects only onto the template and returns its amplitude in $1\sigma$ units. Writing $\bar{\mathbf d}=a\,\mathbf f+\mathbf r$ with $\mathbf f^{\top}\mathbf C_{\rm mean}^{-1}\mathbf r=0$ gives
$\mathrm{SNR}^2=\nu^2+\mathbf r^{\top}\mathbf C_{\rm mean}^{-1}\mathbf r$, hence
$\mathrm{SNR}\ge |\nu|$; the gap is power orthogonal to the template (e.g.\ stochasticity or
template mismatch).

For the matter \texttt{ODD\_p} field, $\cP_{3\times1}$ outperforms $\cP_{2\times2}$; increasing the Gaussian smoothing $R_G$ reduces detectability. In redshift space, the signal is enhanced but additional velocity-induced variance lowers the net significance. For halos, detectability is reduced for both kurto spectra, with $\cP_{3\times1}$ generally higher than $\cP_{2\times2}$. Increasing $R_G$ \emph{increases} detectability for the difference statistic $\Delta\cP$ but \emph{decreases} it for the direct $\cP(\texttt{ODD\_p})$; $\sim 1\sigma$ shifts should be regarded as statistical fluctuations. In redshift space, $\cP_{2\times2}$ degrades while $\cP_{3\times1}$ changes little—consistent with its greater weight on small-scale configurations already curtailed by smoothing. For the halo–matter cross, detectability improves markedly relative to the halo auto (reflecting lower stochasticity). With optimal weights and a tanh window, we find $\hat{g}_{\rm NL}^{\rm PO}=(1.7\pm 0.6)\times 10^6$ from $\cP_{2\times2}(\texttt{ODD\_p})$ and $(1.8\pm 0.5)\times 10^6$ from $\cP_{3\times1}(\texttt{ODD\_p})$ (1$\sigma$ errors).

\begin{table}[t]
    \centering
    \begin{tabularx}{\textwidth}{l|XXXXXX}
    \hline
         & po & po-fid & po \ \ \ (opt.) & po-fid (opt.) & po (opt.+tanh) & po-fid (opt.+tanh) \\
    \hline
    \multicolumn{7}{c}{$\mathcal{P}_{2\times 2}$} \\
    \hline
    m RG5 ($40$) & 4.2 & - & - & - & - & -\\
    m RSD RG5 ($40$) & 2.5 & - & - & - & - & -\\
    m RG10 ($40$) & 0.0 & - & - & - & - & -\\
    m RSD RG10 ($40$) & 1.5 & - & - & - & - & -\\
    h RG5 ($500$)  & 1.6 & 0.4 & 1.5 & 2.9 & 3.0 & 3.6\\
    h RG10 ($500$) & 0.5 & 2.6 & - & - & - & - \\
    h RSD RG10 ($500$)& 0.0 & 1.9 & - & - & - & - \\
    h x m RG5 ($500$) & 2.5 & 4.6 & - & - & - & -\\
    h x m RSD RG5 ($500$) & 1.2 & 3.0 & - & - & - & -\\
    h x m RG10 ($500$) & 2.5 & 4.2 & - & - & - & -\\
    h x m RSD RG10 ($500$) & 2.2 & 4.2 & - & - & - & -\\
    \hline 
    \multicolumn{7}{c}{$\mathcal{P}_{3\times 1}$} \\
    \hline 
    m RG5 ($40$) & 8.0 & - & - & - & - & - \\
    m RSD RG5 ($40$) & 4.0 & - & - & - & - & -\\
    m RG10 ($40$) & 5.6 & - & - & - & - & -\\
    m RSD RG10 ($40$) & 3.2 & - & - & - & - & -\\
    h RG5 ($500$)  & 1.7 & 2.3 & 2.0 & 3.6 & 3.5 & 4.2\\
    h RG10 ($500$) & 0.6 & 2.5 & - &- & - &- \\
    h RSD RG10 ($500$)& 0.2 & 2.8 & - & - & - & - \\
    \hline
    \end{tabularx}
    \caption{Detectability of the template amplitude in units of its 1$\sigma$ error in \texttt{Quijote-ODD} simulations with $g_{\rm NL}^{\rm PO}=10^6$. Names `po' and `fid' stand for the kurto spectra measured from Quijote simulation with PO initial condition and Gaussian initial condition. The `opt' means applying the optimal estimator, while `tanh' means that we use the smooth Heaviside function (as described in~\ref{subsec:numerical_implementation}) to filter the fields. Row-wise, `m' and `h' stand for matter and halo fields. `RG' stands for the typical scale for the Gaussian filtering function. The number in the brackets represent the number of simulations we averaged over.}
    \label{tab:detectability_match_filter}
\end{table}

\begin{figure}[htbp!]
    \centering
    \includegraphics[width=0.6\linewidth]{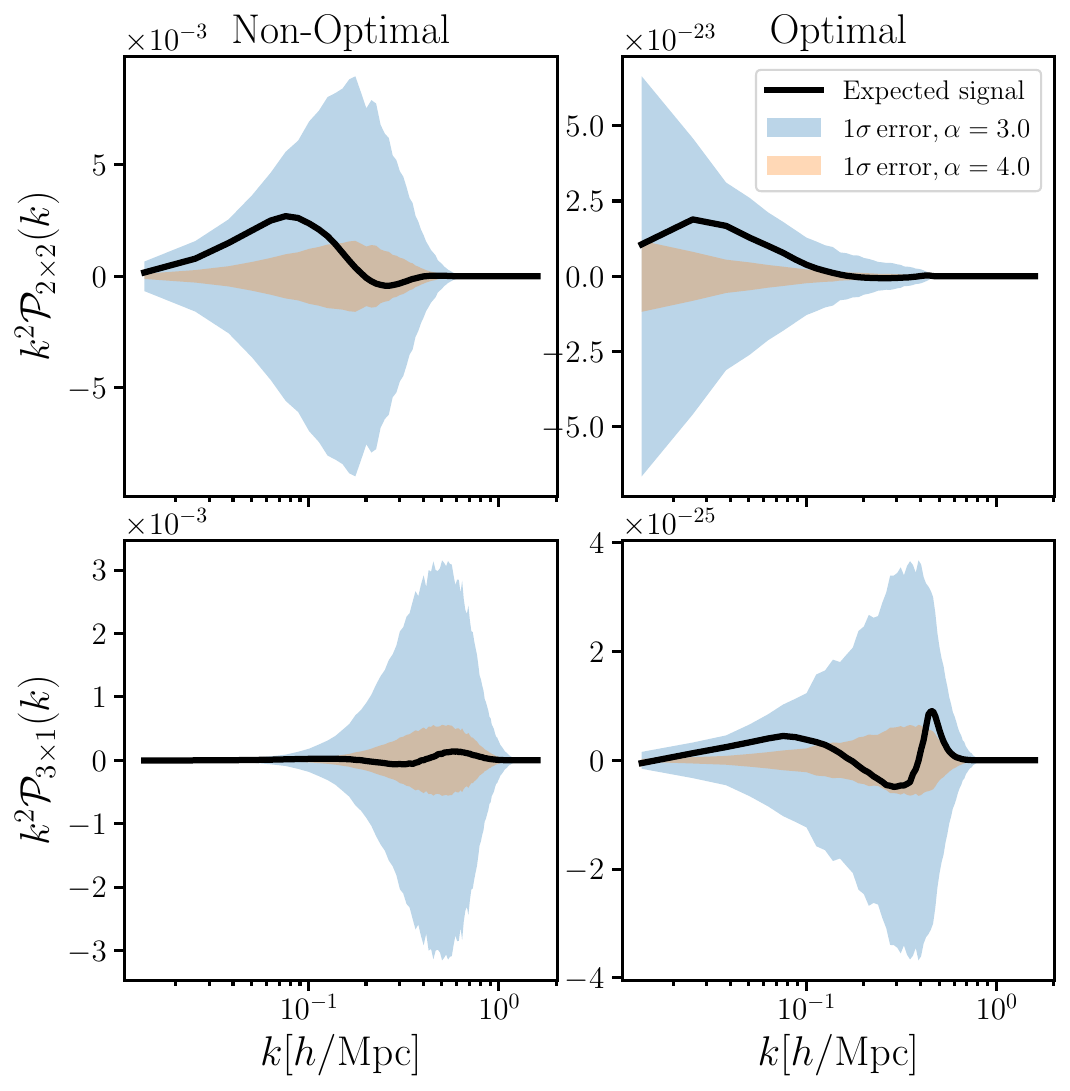}
    \caption{Expected signal (bias-scaled theory) and $1\sigma$ noise bands for the combined Euclid spectroscopic sample. Rows: $\mathcal{P}_{2\times 2}$ (top) and $\mathcal{P}_{3\times 1}$ (bottom). Columns: non-optimal (left) and optimal (right) estimators. Blue and orange bands correspond to the stochasticity–density slope choices $\alpha=3.0$ and $\alpha=4.0$ in eq.~\eqref{eq:survey_noise_level_estimate}.}
    \label{fig:Euclid_full_survey}
\end{figure}

We close this section by presenting an illustrative forecast of the parity-odd detectability for a survey with specified volume $V$, mean number density $\bar n$, and linear bias $b_1$ in figure \ref{fig:Euclid_full_survey}. The signal template is obtained by scaling our theory curves with $b_1^4$, and the noise is propagated using the empirical relation of eq.~\eqref{eq:survey_noise_level_estimate}. We assume the power-law scaling of the parity-even stochasticity with $\bar n$ applies to both kurto spectra and to the optimal estimators. Adopting Euclid-like values---total spectroscopic volume $V=43.36 \ (\mathrm{Gpc}/h)^3$, mean density $\bar n=4.39\times 10^{-4} \ (h/\mathrm{Mpc})^3$, and bias $b_1=1.68$~\citep{Euclid:2019clj} (averaged over the whole spectroscopic redshift range)---the resulting theory curves and $1\sigma$ noise bands are shown in figure~\ref{fig:Euclid_full_survey}. If the noise-number density slope is $\alpha=4$, for both non-optimal and optimal $\cP_{2\times 2}$ measurement, the Euclid full-volume survey is able to detect the PO signal introduced in the \texttt{Quijote ODD} suite. For $\cP_{3\times 1}$, only by using optimal estimator (with $tanh$ window function) can the survey detect such a signal. On the other hand, changing the slope results in a drastic change of the error band. We quote also the SNR here. For $\cP_{2\times 2}$, taking $k_{\rm max}=0.38\, h/\mathrm{Mpc}$ ($N_k=30$), we obtain the SNR for Euclid full survey as $\mathrm{SNR}=8$ (non-optimal) and $\mathrm{SNR}=7.3$ (optimal). For $\cP_{3\times 1}$, taking $k_{\rm max}=0.64\, h/\mathrm{Mpc}$ ($N_k=50$), we have $\mathrm{SNR}=3.6$ (non-optimal) and $\mathrm{SNR}=8.9$ (optimal). We note again that for the optimal estimator, noise-number density scaling relation is sample-dependent and should be re-examined from realistic mocks of Euclid data given that the shot noise of the field is taken account by the $P_{\rm obs}(k)$ in the estimator, therefore we might be underestimating the SNR for the optimal estimator. Besides, we note that the error band is independent of the parity-odd signal of the universe, but depends only on the survey specifications and the form of kurto-spectrum. Therefore, testing these CF-spectrum estimator on realistic survey simulation will serves as a more robust way to obtain the error band, which can help us further to forecast the constraining power of a survey for various parity-odd models. 

Let us emphasize that this forecast inherits uncertainties from (i) extrapolating the $\bar n$–scaling measured from down-sampled \emph{matter} to halos and to optimized estimators, (ii) the approximate $V^{-1}$ mode-counting, and (iii) neglect of survey systematics, projection, and RSD. A more robust assessment of detectability/SNR should employ survey-specific mocks to quantify the error budget accurately, accounting for all relevant observational systematics.

\section{Conclusions}\label{sec:conclusions}

In this work we investigated the detection of \emph{primordial scalar} parity violation using composite–field spectra (``kurto spectra'') that compress the galaxy trispectrum into power–spectrum–like, one–dimensional statistics. Because parity–odd contributions first appear at leading order in the \emph{four–point} function for scalar observables, we constructed estimators that are explicitly parity–odd---i.e., they \emph{vanish} for parity–even fields by symmetry. For a given scalar field one can form two kurto spectra: a vector–pseudovector spectrum, $\cP_{2\times2}$, and a scalar–pseudoscalar spectrum, $\cP_{3\times1}$. We also developed an FFTLog–based pipeline that enables fast and accurate theory predictions for these statistics; the method applies generally to separable kernels (applicable to scalar/vectorial/tensorial observables).

For the bulk of this paper, we tested the kurto spectra on $N$-body and perturbative simulations of late-time dark-matter and halo fields. Our main empirical finding is that, although the parity–even kurto spectra vanish in expectation, they generate a variance that typically exceeds the amplitude of the parity–odd signal.

To isolate the impact of gravitational nonlinearity, we constructed EPT fields in a
$1\,(\mathrm{Gpc}/h)^3$ box and decomposed the kurto spectra into their perturbative pieces. Both the linear and parity–even nonlinear contributions produce stochastic variance that is $\mathcal O(10)$ larger than the leading, nonzero parity–odd signal. This same dominance of parity-even variance is directly seen in our measurements from the \texttt{Quijote} simulations. Including second-order kernels allows a measurement of the NLO signal and simultaneously reveals the limitations of standard EPT: accurate NLO predictions require proper IR resummation.

We then turned to the \texttt{Quijote} simulations with parity-violating initial conditions. For each scenario we measured $\cP(\texttt{ODD\_p})$, the phase-matched Gaussian counterpart $\cP(\texttt{fid})$, and their difference $\Delta\cP\equiv \cP(\texttt{ODD\_p})-\cP(\texttt{fid})$. Covariances were estimated from $500$ realizations, and we evaluated cumulative SNRs as well as a matched-filter detectability statistic. For the \emph{matter} field, $\Delta\cP$ cleanly reveals the parity-odd signal, whereas the direct measurement is noise-dominated by roughly an order of magnitude. We therefore take $\Delta\cP$ as the fiducial signal for matter, and—assuming leading bias scaling—use $b_1^4\,\Delta\cP$ as a proxy for the halo auto case and $b_1^2\,\Delta\cP$ for the halo–matter cross (reflecting the reduced number of biased legs). We also find that $\cP_{3\times 1}$ achieves higher detectability than $\cP_{2\times 2}$, owing to its larger number of contributing small-scale modes. For the \emph{halo} field, even after averaging over $500$ realizations of simulation boxes with a volume of $1\ {\rm (Gpc}/h)^3$ (representative of a survey with volume of $\sim 500\,(\mathrm{Gpc}/h)^3$), $\Delta\cP$ does not yield an unambiguous nonzero detection; the matched-filter significance remains $<3\sigma$. This indicates that parity-odd initial conditions affect halo formation in a way that introduces additional (parity-even) stochasticity relative to Gaussian initial conditions, which dominates the error budget for halo-based estimators. This is an indication of the stochastic operators in the parity-odd sector of EFT/bias expansion. 

We investigated the origin of the large stochastic variance in the \texttt{Quijote} \emph{halo} measurements. A null test on the \texttt{Hidden Valley} simulations showed that the noise scales \emph{inversely} with tracer number density. Halo–matter cross correlations further demonstrated that stochasticity is the dominant noise source in the kurto spectra. By down-sampling the matter field (randomly selecting subsets of DM particles), we measured a power-law relation between the parity-even variance and the mean number density. This empirical scaling enables forecasts given a survey’s volume, tracer bias, and number density. Applying it to an Euclid-like spectroscopic sample, we find that the \texttt{Quijote-ODD}–type signal should be detectable with the full survey. These forecasts, however, rely on extrapolating the measured scaling and neglect detailed survey systematics; a more robust assessment requires Euclid-like mocks with injected parity-odd signals.

Several directions naturally follow from this work. \emph{(i) Reducing stochasticity:} explore additional weighting schemes in the composite fields (e.g., mass– or environment–dependent halo weights, multi–tracer combinations, and targeted mode filters/deprojections) and leverage cross–correlations with low–stochasticity tracers (e.g., lensing) to suppress halo shot noise while preserving the parity–odd signal. \emph{(ii) Survey realism:} test the estimators on Euclid/DESI–like mocks that include realistic galaxy populations, masks, selection and completeness, fiber assignment, imaging systematics, and redshift–space/AP effects, and validate with end–to–end null tests. \emph{(iii) New observables:} extend the formalism to vector/tensor fields---most immediately weak–lensing shear and galaxy shape correlations with explicit E/B and parity separation---and to lensing–galaxy cross kurto spectra. \emph{(iv) Angular implementation:} develop full–sky, tomographic harmonic–space parity–odd kurto spectra (including beyond–Limber terms and odd–$L$ selection rules) for direct application to wide surveys. \emph{(v) Theory and inference:} improve NLO modeling (IR resummation, bias expansion), build a bank of separable parity–odd templates (e.g., massive–spin exchange, axion–gauge scenarios, helical PMFs), and identify which models are best constrained by current and future data, as well as how 2D vs. 3D parity-odd kurto spectra and full trispectra complement one another. These steps will clarify how far parity–odd kurto spectra can push beyond current trispectrum analyses in forthcoming large-scale structure observables.

\section*{Acknowledgments}

We thank Antonio Riotto for many insightful discussions on theoretical models that can generate parity-odd signatures in the scalar and tensor sectors, Dennis Linde for discussions related to the construction of the EPT field using GridSPT algorithm, Emiliano Sefusatti for discussions on stochasticity, and Alex Barreira for an early discussion on parity-odd estimators. The work of ZG and AMD is supported by the \emph{Agence Nationale de la Recherche} (ANR) under grant No. ANR-23-CPJ1-0160-01. Z.V. acknowledges support from the Kavli Foundation. All measurements on simulations were performed using the computing and storage resources of GENCI at IDRIS, thanks to allocation 2025-AD010416295 on the CSL partition of the Jean Zay supercomputer.

\appendix

\appendix
\appendix
\section{Helicity Basis and Tensor Harmonics}
\label{append:helicity_basis}

In this appendix we summarize the helicity basis and irreducible tensor decomposition used in section ~\ref{sec:parity-CFs}. This provides the technical backbone for the classification of parity-odd spectra.

\paragraph{Helicity basis for vectors}
The helicity basis $\{\hat{\vk}, \vec e^+, \vec e^-\}$ is constructed as follows.  
Given a wavevector $\vk$ and an arbitrary auxiliary direction $\hat{\vec n}$ not collinear with $\hat{\vk}$, we define
\be
\vec e_k = \hat{\vk}, \qquad 
\vec e_1 = \frac{\vk \times \hat{\vec n}}{|\vk \times \hat{\vec n}|}, \qquad 
\vec e_2 = \hat{\vk}\times \vec e_1.
\ee
This gives an orthonormal triad $\{\vec e_k, \vec e_1, \vec e_2\}$.  

The vector field can then be decomposed as
\be
\label{Eq:vector_decomposition}
\vec V(\vk) = v^{(0)}(\vk)\,\vec e_k + v^{(1)}(\vk)\,\vec e_1 + v^{(2)}(\vk)\,\vec e_2.
\ee
Introducing helicity eigenvectors
\be
\label{Eq:pm_basis}
\vec e_\pm = \mp \frac{1}{\sqrt{2}}\left(\vec e_1 \pm i\vec e_2\right),
\ee
we can equivalently write
\be
\label{Eq:pm_component}
V_i(\vk) = v^{(0)}(\vk)\,\hat k_i + v^{(+1)}(\vk)\,e^+_i + v^{(-1)}(\vk)\,e^-_i,
\ee
with
\be
v^{(-1)} = \tfrac{1}{\sqrt{2}}(v^{(1)}+iv^{(2)}), 
\qquad
v^{(+1)} = -\tfrac{1}{\sqrt{2}}(v^{(1)}-iv^{(2)}).
\ee

\paragraph{Vector correlator decomposition} 
Statistical isotropy implies that correlators are diagonal in helicity,
\be
\langle v^{(m)}(\vk)\,v^{(m')}(\vk')\rangle 
= (2\pi)^3 \delta_D(\vk+\vk')\,\delta^K_{mm'}\,P^{(m)}(k),
\ee
with $m=0,\pm1$. The most general isotropic correlator of two vectors is then
\be
\langle V_i(\vk)V_j(\vk')\rangle = (2\pi)^3 \delta_D(\vk+\vk')
\Big[-P_0(k)\hat k_i\hat k_j + P_+(k)(\delta^K_{ij}-\hat k_i\hat k_j) + iP_\times(k)\epsilon_{ijl}\hat k_l\Big].
\ee
Here $P_0$ and $P_+$ are parity-even, while $P_\times$ is parity-odd, corresponding to the difference between the $+1$ and $-1$ helicity spectra.

\paragraph{Rank-2 tensor harmonics} 
A symmetric rank-2 tensor $\Pi_{ij}(\vk)$ can be decomposed into irreducible representations of SO(3):
\be
\Pi_{ij}(\vk) = \tfrac{1}{3}\delta^K_{ij}\,\Pi^{(0)}_0(\vk) 
+ \sum_{m=-2}^2 \Pi^{(m)}_2(\vk)\,Y^{(m)}_{ij}(\hat k).
\ee
The scalar part $\Pi^{(0)}_0$ is the trace ($\ell=0$), while $\Pi^{(m)}_2$ ($\ell=2$) are the traceless components expanded in the tensor harmonics
\begin{align}
Y^{(0)}_{ij} &= \sqrt{\tfrac{3}{2}}\Big(\hat k_i\hat k_j - \tfrac{1}{3}\delta^K_{ij}\Big), \\
Y^{(\pm1)}_{ij} &= \sqrt{\tfrac{1}{2}}(\hat k_i e^\pm_j + \hat k_j e^\pm_i), \\
Y^{(\pm2)}_{ij} &= e^\pm_i e^\pm_j.
\end{align}

\paragraph{Transformation properties} 
The harmonics obey
\be
Y^{(m)}_{ij}(\hat k)^* = (-1)^m Y^{(-m)}_{ij}(\hat k), 
\qquad
Y^{(m)}_{ij}(-\hat k) = (-1)^{\ell+m} Y^{(-m)}_{ij}(\hat k).
\ee
Parity $\mathbb P:\vk\mapsto -\vk$ acts as
\be
\mathbb P: \hat k \to -\hat k, \quad e^\pm \to -e^\mp, 
\quad Y^{(m)}_{ij}(\hat k) \to Y^{(m)*}_{ij}(\hat k).
\ee

\paragraph{Parity-odd tensor spectra} 
The most general isotropic two-point function of tensor components is
\be
\langle \Pi^{(m)}_\ell(\vk)\,\Pi^{(m')}_{\ell'}(\vk')\rangle
= (2\pi)^3\delta_D(\vk+\vk')\,\delta^K_{mm'}\,P^{(m)}_{\ell\ell'}(k).
\ee
Parity invariance enforces $P^{(+m)}_{\ell\ell'}(k)=P^{(-m)}_{\ell\ell'}(k)$. Violations of this equality define parity-odd spectra.  
For $\ell=2$ there are two such independent channels:  
\be
P_d^\times(k) \;\;\leftrightarrow\;\; m=\pm1, 
\qquad
P_e^\times(k) \;\;\leftrightarrow\;\; m=\pm2,
\ee
the natural tensorial analogs of $P_\times(k)$ in the vector case.  

\medskip
This formalism makes explicit that parity violation is always encoded in the antisymmetric differences between opposite-helicity modes. The parity-odd kurto spectra in the main text are scalar projections designed to isolate precisely these helicity-odd sectors.


\bibliographystyle{utphys}
\bibliography{ms}

\end{document}